\documentclass[iop,apjl,numberedappendix]{emulateapj}
\usepackage{natbib}
\usepackage{graphicx}
\usepackage{latexsym}
\usepackage{amsmath,amsfonts}
\usepackage{longtable}
\usepackage{multirow,booktabs}
\usepackage{hyperref}
\usepackage{afterpage} 
\usepackage[normalem]{ulem}
\usepackage{color}

\hypersetup{colorlinks=false,pdfborder={0 0 0}}

\slugcomment{Accepted for publication in ApJ, August 25, 2018}

\setcitestyle{notesep={; }}

\def\sgra{Sgr~A$^{\ast}$}
\newcommand{\rin}{r_{\rm in}} 

\def\lsim{\mathrel{\raise.3ex\hbox{$<$\kern-.75em\lower1ex\hbox{$\sim$}}}}
\def\gsim{\mathrel{\raise.3ex\hbox{$>$\kern-.75em\lower1ex\hbox{$\sim$}}}}
\def\gtwid{\mathrel{\raise.3ex\hbox{$>$\kern-.75em\lower1ex\hbox{$\sim$}}}}
\def\proptwid{\mathrel{\raise.3ex\hbox{$\propto$\kern-.75em\lower1ex\hbox{$\sim$}}}}

\begin{document}

\title{ The Scattering and Intrinsic Structure of Sagittarius A$^\ast$ at Radio Wavelengths }
\shorttitle{The Scattering and Intrinsic Structure of \sgra}

\author{Michael D.~Johnson\altaffilmark{1}, 
Ramesh Narayan\altaffilmark{1}, 
Dimitrios Psaltis\altaffilmark{2}, 
Lindy Blackburn\altaffilmark{1}, 
Yuri Y.\ Kovalev\altaffilmark{3,4,5}, 
Carl R.\ Gwinn\altaffilmark{6}, 
Guang-Yao Zhao\altaffilmark{7}, 
Geoffrey C.\ Bower\altaffilmark{8}, 
James M.\ Moran\altaffilmark{1}, 
Motoki Kino\altaffilmark{9,10}, 
Michael Kramer\altaffilmark{5,11}, 
Kazunori Akiyama\altaffilmark{12}, 
Jason Dexter\altaffilmark{13}, 
Avery E.\ Broderick\altaffilmark{14,15},
and Lorenzo Sironi\altaffilmark{16}}
\shortauthors{Michael D.~Johnson et al.}
\altaffiltext{1}{Harvard-Smithsonian Center for Astrophysics, 60 Garden Street, Cambridge, MA 02138, USA}
\altaffiltext{2}{Astronomy Department, University of Arizona, 933 N. Cherry Ave, Tucson, AZ 85721}
\altaffiltext{3}{Astro Space Center of Lebedev Physical Institute, Profsoyuznaya 84/32, 117997 Moscow, Russia}
\altaffiltext{4}{Moscow Institute of Physics and Technology, Institutsky per.~9, Dolgoprudny 141700, Russia}
\altaffiltext{5}{Max-Planck-Institut f\"ur Radioastronomie, Auf dem H\"ugel 69, 53121 Bonn, Germany}
\altaffiltext{6}{Department of Physics, University of California, Santa Barbara, CA 93106, USA}
\altaffiltext{7}{Korea Astronomy and Space Science Institute, Daedeok-daero 776, Yuseong-gu, Daejeon 34055, Korea}
\altaffiltext{8}{Academia Sinica Institute of Astronomy and Astrophysics, 645 N. A{'o}hoku Place, Hilo, HI 96720, USA}
\altaffiltext{9}{Kogakuin University of Technology \& Engineering, Academic Support Center, 2665-1 Nakano, Hachioji, Tokyo 192-0015, Japan}
\altaffiltext{10}{National Astronomical Observatory of Japan, 2-21-1 Osawa, Mitaka, Tokyo, 181-8588, Japan}
\altaffiltext{11}{Jodrell Bank Centre for Astrophysics, The University of Manchester, M13 9PL, United Kingdom}
% \altaffiltext{9}{National Astronomical Observatory of Japan, 2-21-1 Osawa, Mitaka, Tokyo 181-8588, Japan}
\altaffiltext{12}{Massachusetts Institute of Technology, Haystack Observatory, Route 40, Westford, MA 01886, USA}
\altaffiltext{13}{Max Planck Institute for Extraterrestrial Physics, Giessenbachstr.\ 1, 85748 Garching, Germany}
\altaffiltext{14}{Perimeter Institute for Theoretical Physics, 31 Caroline Street North, Waterloo, ON N2L 2Y5, Canada}
\altaffiltext{15}{Department of Physics and Astronomy, University of Waterloo, 200 University Avenue West, Waterloo, ON N2L 3G1, Canada}
\altaffiltext{16}{Department of Astronomy, Columbia University, 550 W 120th St, New York, NY 10027, USA}
\email{mjohnson@cfa.harvard.edu } 

\keywords{ radio continuum: ISM -- scattering -- ISM: structure -- Galaxy: nucleus -- techniques: interferometric --- turbulence \vspace{4cm}}

\begin{abstract}
Radio images of the Galactic Center supermassive black hole, Sagittarius~A$^\ast$ (\sgra), are dominated by interstellar scattering. 
Previous studies of \sgra\ have adopted an anisotropic Gaussian model for both the intrinsic source and the scattering, and they have extrapolated the scattering using a purely $\lambda^2$ scaling to estimate intrinsic properties. However, physically motivated source and scattering models break all three of these assumptions. 
They also predict that refractive scattering effects will be significant, which have been ignored in standard model fitting procedures. 
We analyze radio observations of \sgra\ using a physically motivated scattering model, and we develop a prescription to incorporate refractive scattering uncertainties when model fitting. 
We show that an anisotropic Gaussian scattering kernel is an excellent approximation for \sgra\ at wavelengths longer than 1\,cm, with an angular size of $(1.380 \pm 0.013) \lambda_{\rm cm}^2\,{\rm mas}$ along the major axis, $(0.703 \pm 0.013) \lambda_{\rm cm}^2\,{\rm mas}$ along the minor axis, and a position angle of $81.9^\circ \pm 0.2^\circ$. 
We estimate that the turbulent dissipation scale is at least $600\,{\rm km}$, with tentative support for $r_{\rm in} = 800 \pm 200\,{\rm km}$, suggesting that the ion Larmor radius defines the dissipation scale. We find that the power-law index for density fluctuations in the scattering material is $\beta < 3.47$, shallower than expected for a Kolmogorov spectrum ($\beta=11/3$), and we estimate $\beta = 3.38^{+0.08}_{-0.04}$ in the case of $r_{\rm in} = 800\,{\rm km}$. We find that the intrinsic structure of \sgra\ is nearly isotropic over wavelengths from 1.3\,mm to 1.3\,cm, with a size that is roughly proportional to wavelength: $\theta_{\rm src} \sim (0.4\,{\rm mas}) \times \lambda_{\rm cm}$. We discuss implications for models of \sgra, for theories of interstellar turbulence, and for imaging \sgra\ with the Event Horizon Telescope. 
\end{abstract}

\section{Introduction}

The compact radio source at the Galactic Center, Sagittarius A$^\ast$ (\sgra), was discovered in 1974 \citep{Balick_Brown_1974}. Within two years, observers had deduced that the radio image was dominated by scatter broadening caused by the ionized interstellar medium (ISM) based on an observed scaling of image size with the squared observing wavelength, $\theta \propto \lambda^2$ \citep{Davies_1976}. In the decades since the initial discovery of \sgra, knowledge of its scattering properties has continually improved, but scattering uncertainties remain the primary limitation in determining the intrinsic structure of \sgra\ at wavelengths longer than a few millimeters. 

Motivated by the $\theta \propto \lambda^2$ scaling and approximately Gaussian image, many observers have sought to accurately measure the image of \sgra\ at a wide range of radio wavelengths, seeking to constrain the scattering law at long wavelengths (where the scattering dominates) and then to deconvolve its effects at shorter wavelengths to estimate the intrinsic source parameters. An advantage of treating both the source and the scattering as Gaussian is that the scattered image is then also a Gaussian because the time-averaged scattering acts as a convolution \citep[see, e.g.,][]{Coles_1987,GoodmanNarayan89,Johnson_Gwinn_2015}. Consequently, many techniques have been developed to accurately estimate Gaussian image parameters for \sgra\ from interferometric data, including image-domain parameter estimation \citep{Bower_2006}, model fitting using only closure quantities \citep{Bower_2004,Shen_2005,Bower_2014b,Ortiz_2016,Brinkerink_2016}, and self-calibration \citep{Doeleman_2001,Lu_2011,Ortiz_2016}. In addition, many techniques have been applied to ensure conservative estimates of parameter uncertainty, including standard exploration of the chi-squared hypersurface \citep[e.g.,][]{Bower_2014b}, Monte Carlo approaches \citep{Ortiz_2016}, and bootstrap approaches using multi-epoch data \citep[e.g.,][]{Lu_2011}. Nevertheless, the reported sizes and position angles still have significant unresolved discrepancies \citep[see][]{Psaltis_2015}. 

In addition to the simplified scattering model, a major missing component from all these previous studies has been refractive scattering effects. Refractive scattering will distort the instantaneous image, giving systematic departures from the ensemble-average image that are independent of observing quality \citep{Blandford_Narayan_1985}. Refractive scattering also introduces substructure in the image, which contributes additional ``refractive noise'' to interferometric measurements on baselines that resolve the image \citep{NarayanGoodman89,GoodmanNarayan89,Johnson_Gwinn_2015}. Recently, refractive noise was discovered in 1.3\,cm observations of \sgra\ \citep{Gwinn_2014}, suggesting that it may contribute significantly to the error budget when fitting Gaussian models. Refractive noise is especially problematic because the longer baselines, which are most affected, are also the most sensitive to compact structure; their measurements are what dominate Gaussian model fits. Because refractive noise tends to bias long-baseline visibility amplitudes upward, detections interpreted without a noise budget for refractive substructure will tend to imply artificially compact structure \citep[see, e.g.,][]{Johnson_2016RA,Pilipenko_2018}. Thus, refractive scattering effects are essential to include when fitting models to interferometric data, and they contribute in multiple ways, both by modulating the ``true'' instantaneous image size and orientation and by adding a new type of ``noise'' to interferometric measurements.

Here, we analyze archival observations of \sgra\ at wavelengths from 1.3\,mm (EHT) to 30\,cm (VLA). We develop a framework to efficiently incorporate refractive noise into parametric model fitting, and we show how to isolate components of the refractive noise that may be absorbed into fitted model parameters (e.g., refractive flux modulation and image wander). We constrain a physically motivated scattering model \citep{Psaltis_2018}, which generically produces Gaussian scatter-broadening that scales as $\lambda^2$ in the limit $\lambda \rightarrow \infty$, but which differs at short wavelengths because of a finite inner scale $r_{\rm in}$ of the interstellar turbulence with an associated power-law index $\alpha$. In addition to these two parameters, the scattering model depends on the Gaussian scatter broadening in the long-wavelength limit, which we parameterize via the major axis full width at half maximum (FWHM) $\theta_{\rm maj,0}$, minor axis FWHM $\theta_{\rm min,0}$, and major axis position angle $\phi_{\rm PA}$, all specified at a reference wavelength $\lambda_0$ (we use $\lambda_0 \equiv 1\,{\rm cm}$). 

% , , and . These characterize 

We estimate uncertainties in our parameter estimates by fitting representative ensembles of synthetic datasets that match the baseline coverage and sensitivity of the observations. These synthetic datasets are created using numerical simulations of the scattering and also include wavelength-dependent systematic gain calibration uncertainties to simulate imperfect amplitude and phase calibration. This approach allows us to incorporate thermal noise, refractive uncertainties, and systematic calibration errors in the overall error budget, and to verify that our model fitting is not biased by any of these effects or by the anisotropic baseline coverage. Using our estimated scattering model, we compute the wavelength-dependent intrinsic size of \sgra.

We begin, in \S\ref{sec::Scattering_Model}, with a brief review of scattering theory. Next, in \S\ref{sec::Model_Fitting_Procedure}, we describe our procedure to fit individual observations and motivate how we can use the full set of observations to constrain the scattering model. In \S\ref{sec::Observations}, we provide details about the observations used to constrain the scattering model and give the results of Gaussian fits to each. In \S\ref{sec::Composite_Constraints}, we derive our parameter estimates and uncertainties for the scattering model, describe the expected scattering properties, and estimate the intrinsic source size of \sgra. 
In \S\ref{sec::Discussion}, we discuss implications for models of \sgra, implications for theories of interstellar turbulence, consequence of unmet assumptions in our approach, and prospects for continued study of \sgra. We summarize our findings in \S\ref{sec::Summary}.

\section{Scattering Model}
\label{sec::Scattering_Model}

% Note: most of this section should be referenced to . 

\subsection{Background}
\label{sec::Background}

The basic properties of interstellar scattering have been summarized in several reviews \citep[e.g.,][]{Rickett_1990,Narayan_1992,TMS}, and our specific scattering model is derived and discussed in detail in a companion paper \citep{Psaltis_2018}. Here, we will only briefly summarize the key properties that are of immediate relevance for the remainder of this paper.

Interstellar scattering and scintillation at radio wavelengths is caused by density inhomogeneities in the ionized ISM. Neglecting a weak birefringence from the magnetic field (which is negligible for the observing wavelengths we consider), the local index of refraction of the ISM is given by $n \approx 1 - \frac{1}{2} \left( \frac{\nu_{\rm p}}{\nu} \right)^2$, where $\nu$ is the wave frequency, $\nu_{\rm p} \approx 9.0 \times \sqrt{\frac{n_{\rm e}}{1\,{\rm cm}^{-3}}}\,{\rm kHz}$ is the plasma frequency, and $n_{\rm e}$ is the electron density \citep[see, e.g.,][]{TMS}. A density fluctuation $\delta n_{\rm e}$ along a path length $dz$ then introduces a corresponding phase change $\delta \phi = -r_{\rm e} \lambda \times dz \times \delta n_{\rm e}$, where $r_{\rm e} \approx 2.8 \times 10^{-13}~{\rm cm}$ is the classical electron radius and $\lambda$ is the wavelength. Note that this dispersion relation is quite general and is independent of a specific ISM scattering model or geometry. 

Along many lines of sight, the effects of scattering can be approximated via a single thin phase screen $\phi(\mathbf{r})$, where $\mathbf{r}$ is a two-dimensional vector transverse to the line of sight. Electron density fluctuations imprint their spectrum on the power spectrum $Q(\mathbf{q})$ of phase fluctuations, which are typically characterized by a single, unbroken power law between some outer ($r_{\rm out}$) and inner ($r_{\rm in}$) scales: $Q(\mathbf{q}) \propto \left|\mathbf{q}\right|^{-\beta}$. This description is expected for a top-down turbulent cascade between an injection scale and a dissipation scale, and a Kolmogorov spectrum of density fluctuations gives $\beta=11/3$ \citep{Goldreich_Sridhar_1995}. 
  
The effects of scattering on interferometric measurements are conveniently characterized using the phase structure function of the scattering screen, $D_{\phi}(\mathbf{r}) \equiv \left \langle \left[ \phi(\mathbf{r}' + \mathbf{r}) - \phi(\mathbf{r}') \right]^2 \right \rangle \propto \lambda^2$. In the ensemble-average scattering limit \citep[see][]{NarayanGoodman89,GoodmanNarayan89}, the effects of scattering are to convolve an unscattered image with a scattering kernel or, equivalently, to multiply unscattered interferometric visibilities by the appropriate Fourier-conjugate kernel. The Fourier-conjugate kernel is given by $\exp\left[ -\frac{1}{2} D_{\phi}\left( \mathbf{b}/(1+M) \right) \right]$, where $\mathbf{b}$ is the vector baseline of the interferometer and $M = D/R$ is the ``magnification'' of the scattering screen ($D$ is the observer-screen distance; $R$ is the source-screen distance). For spatial displacements smaller than $r_{\rm in}$, the phase fluctuations will be smooth, $\phi(\mathbf{r}' + \mathbf{r}) \approx \phi(\mathbf{r}') + \mathbf{r} \cdot \nabla \phi(\mathbf{r}')$. In this limit, $D_{\phi}(\mathbf{r}) \propto r^2 \lambda^2$ \citep{Tatarskii_1971}. This expression -- which makes no assumptions other than the cold plasma dispersion relation and smoothness of phase fluctuations below some scale -- shows that ensemble-average scatter-broadening should act as a (possibly anisotropic) Gaussian blurring that scales as $\theta_{\rm scatt} \propto \lambda^2$ for baselines $\mathbf{b} \lsim (1+M) r_{\rm in}$ (i.e., on angular scales $\theta \gsim \lambda/((1+M) r_{\rm in})$. Moreover, because the time-averaged scattering kernel from an ensemble of thin screens is determined by the cumulative convolution of all the individual screens, this generic asymptotic behavior is not limited to thin-screen scattering. At longer baselines, $D_{\phi}(\mathbf{r}) \propto \left| \mathbf{r} \right|^\alpha$, where $\alpha\equiv \beta-2$, and the corresponding image becomes non-Gaussian. In this regime, the angular broadening scales as $\theta_{\rm scatt} \propto \lambda^{1+\frac{2}{\alpha}}$ and the interferometric scattering kernel falls as $e^{-\left|\mathbf{b} \right|^{\alpha}}$. However, note that $D_{\phi}(\mathbf{r}) \propto \lambda^2$ regardless of $\alpha$ or the scattering model. 

While scatter broadening is produced by phase fluctuations on the diffractive scale\footnote{Formally, the diffractive scale is defined by $D_{\phi}(r_{\rm diff}) \equiv 1$.}  $r_{\rm diff} \sim \lambda/((1+M)\theta_{\rm scatt})$, refractive scintillation is dominated by modes that are comparable to the refractive scale (i.e., the projected size of angular broadening on the scattering screen): $r_{\rm ref} \sim \theta_{\rm scatt} D$. The Fresnel scale $r_{\rm F} \equiv \sqrt{\frac{D R}{D+R} \frac{\lambda}{2\pi}}$, which is defined entirely by geometrical parameters of the scattering, corresponds to the geometric mean of the diffractive and refractive scales. 
When $r_{\rm ref} > r_{\rm F} > r_{\rm diff}$, the scattering is said to be ``strong'' (e.g., the scattering of \sgra\ is strong for all frequencies lower than a few ${\rm THz}$). In this case, refractive effects are most naturally described using the power spectrum of phase fluctuations: $Q(\mathbf{q}) \equiv (2\pi)^2 \lambda^{-2}\int d^2\mathbf{r} \left \langle \phi(\mathbf{r}' + \mathbf{r}) \phi(\mathbf{r}) \right \rangle e^{-i \mathbf{q} \cdot \mathbf{r}}$. In this expression, the prefactor renders $Q(\mathbf{q})$ independent of wavelength. 

To describe a full scattering model then requires six parameters. Three are needed to characterize the long-wavelength behavior (Gaussian scatter broadening with a $\lambda^2$ dependence). As described before, we use the FWHM along the major and minor axes at a reference wavelength and the major axis position angle: $\theta_{\rm maj,0}$, $\theta_{\rm min,0}$, and $\phi_{\rm PA}$.  In addition, the power-law index $\alpha$, inner scale $r_{\rm in}$, and outer scale $r_{\rm out}$ are needed. \citet{Psaltis_2018} show how to compute the phase structure function, power spectrum, and scattering properties once these parameters are specified. 

We caution that the exact specification of these parameters is not unique, and the radio scattering literature is particularly inconsistent in defining the inner scale. For example, \citet{Rickett_1990} and \citet{Smirnova_2010} taper the power spectrum by $e^{-q^2 r_{\rm in}^2}$, \citet{Coles_1987} and \citet{Armstrong_1995} use $e^{-q^2 r_{\rm in}^2/2}$, \citet{Lambert_Rickett_1999} use $e^{-q^2 r_{\rm in}^2/4}$, and \citet{Spangler_Gwinn_1990} use $e^{-q r_{\rm in}/(2\pi)}$. We use a power spectrum taper of the form $e^{-q^2 r_{\rm in}^2}$.

% ?? define scattering model and its parameters ??

\subsection{Refractive Noise}
\label{sec::Refractive_Noise}

Refractive scattering modes introduce many types of stochastic effects. They modulate the total flux density of an image, displace its centroid, and distort the overall image. They also introduce image substructure, even on scales for which the unscattered source was smooth. All of these effects introduce a new type of ``noise'' for interferometric measurements. This refractive noise has a fractional bandwidth of order unity and varies slowly, on the refractive timescale $t_{\rm ref} = r_{\rm ref}/V_{\perp}$, where $V_{\perp}$ is the characteristic relative transverse velocity of the observer, scattering, and source. At centimeter wavelengths, \sgra\ has $r_{\rm ref} \approx (2 \times 10^{13}\,{\rm cm}) \times \lambda_{\rm cm}^2$. Taking $V_{\perp} \sim 50\,{\rm km/s}$ gives $t_{\rm ref} \sim (50~{\rm days})\times \lambda_{\rm cm}^2$.

\citet{Johnson_Gwinn_2015} and \citet{Johnson_Narayan_2016} provide expressions for how to compute properties of refractive noise, including the variance of refractive fluctuations of a complex visibility measured on a vector baseline $\mathbf{b}$: $\sigma_{\rm ref}^2(\mathbf{b}) \equiv \left \langle \left| \Delta V(\mathbf{b}) \right|^2 \right \rangle$. However, for short baselines, this variance is not the correct quantity to apply to standard VLBI analyses. Namely, part of the variance is due to variations in the total flux density, caused by refractive focusing, and part is due to image wander, caused by refractive deflections. Both of these effects would be eliminated in a typical VLBI analysis. The flux variations would be absorbed into the estimated total source flux density, and the image wander would be eliminated by centering the image (since VLBI has no concept of absolute position without absolute phase referencing). 

Appendix~\ref{sec::CalculatingRenormalizedRefractiveNoise} shows how to compute a {\it renormalized} visibility variance, $\hat{\sigma}_{\rm ref}(\mathbf{b})$, that eliminates these contributions. For example, the renormalized refractive noise is zero in the limit of zero baseline. On short baselines, it is dominated by changes in the overall image size from scattering -- a property that we utilize in \S\ref{sec::Image_Size_Fluctuations}. We will include renormalized refractive noise in the error budget for our model fitting.

\subsection{Assumed Scattering Properties of Sgr A*}  
\label{sec::Scattering_Geometry}  
  
We will use a few supplementary measurements and assumptions about the scattering of \sgra. The first is for the scattering geometry of \sgra. Because the Galactic Center magnetar lies only $2.4''$ from \sgra\ \citep{Bower_2015}, its radio pulsations permit an estimate of temporal broadening associated with the scattering toward \sgra. This measurement can be combined with the angular broadening to estimate the location of the scattering material \citep{Gwinn_Bartel_Cordes}. For instance, if the scattering is isotropic, then the pulse broadening function is exponential: $g(t) = e^{-t/\tau}$. This expression follows by relating a radial distance $r$ on the scattering screen to its corresponding geometric delay, $t(r) \approx \frac{r^2}{2 c}\left( \frac{1}{D} + \frac{1}{R} \right)$, and then expressing the brightness distribution on the sky in terms of $t$. 
The $1/e$ scale of the temporal broadening, $\tau$, is related to the FWHM angular size of the scattered image, $\theta_{\rm scatt}$, via \citep[e.g.,][]{Cordes_Lazio_1997}
\begin{align}
\label{eq::tau_theta_relationship}
 c \tau = \frac{ M d_{\rm src} }{8 \ln(2)} \theta_{\rm scatt}^2, 
\end{align}
where $d_{\rm src} = D+R$ is the distance from the observer to the source, and $M=D/R$. Because the magnetar shows angular broadening comparable to \sgra, the same scattering material is thought to dominate the angular broadening of each \citep{Bower_Magnetar_2014,Bower_2015}. The temporal broadening of the magnetar can then be combined with the angular broadening and distance to \sgra\ to estimate the location of the scattering material for both objects. 

For anisotropic scattering, the pulse broadening function is monotonically decreasing but not exponential. For a scattered image with FWHM $\theta_{\rm maj}$ and $\theta_{\rm min}$ along the major and minor axes, the pulse broadening function takes the form\footnote{\citet{Rickett_2009} and \citet{Gwinn_2016} derive similar expressions for $g(t)$. However, note that \citet{Rickett_2009} express their results in terms of the scattering angle of the screen $\theta_{\rm s}$ rather than the observed scattering angle $\theta = R (D+R)^{-1} \theta_{\rm s} = (1+M)^{-1} \theta_{\rm s}$.}
\begin{align}
g(t) &= I_0\left(  \frac{4\ln(2) c t}{M d_{\rm src} \theta_{-}^2}  \right) e^{- \frac{4 \ln(2) c t}{M d_{\rm src} \theta_{+}^2}},\\
\nonumber \theta_{\pm} &\equiv \frac{\theta_{\rm maj} \theta_{\rm min}} {\sqrt{\theta_{\rm maj}^2 \pm \theta_{\rm min}^2}}.
\end{align}
In this case, determining $\tau$ (i.e., solving $g(\tau) = g(0)/e$) must be done numerically. 

% so the generalization of Eq.~\ref{eq::tau_theta_relationship} to anisotropic scattering must be done numerically. 

We will assume a distance to \sgra\ of $8.1\,{\rm kpc}$ \citep{Gravity_2018}. Using our estimated scattering kernel parameters (see, e.g., Table~\ref{tab::GaussianParameters}), we obtain $\tau_{\rm 1\,GHz}/(1~{\rm s}) \approx  2.47 M$. Using the measured value $\tau_{\rm 1\,GHz} = 1.3\pm 0.2\,{\rm s}$ \citep{Spitler_2014} then gives $M = 0.53 \pm 0.08$. Note that this estimate differs slightly from the simpler approach of using the isotropic scattering result with the geometric mean of the major and minor scattering axes, which gives $\tau_{\rm 1\,GHz}/(1~{\rm s}) \approx  2.77 M$ \citep{Bower_Magnetar_2014}. Using the exact expression for anisotropic scattering, we obtain $D = 2.7\,{\rm kpc}$ and $R = 5.4\,{\rm kpc}$. 

There is now compelling evidence that at least some of the temporal broadening of the magnetar is local to the Galactic Center region \citep[see, e.g.,][]{Dexter_2017,Desvignes_2018}. Because angular broadening is more sensitive to scattering material closer to the observer, it is likely that the angular broadening and refractive effects are dominated by the scattering region that is distant from the Galactic Center. Because the temporal broadening caused by this material may be smaller than the value used above, the corresponding $M$ for the scattering responsible for the angular broadening may be somewhat \emph{lower} and the scattering material somewhat further from the Galactic center. However, our later results are insensitive to changes in $M$. Refractive noise scales as $\sigma_{\rm ref} \propto M^{-1+\frac{\alpha}{2}}$ (e.g., $\sigma_{\rm ref} \propto M^{-1/6}$ for a Kolmogorov spectrum), while our later inner scale constraint is proportional to $(1+M)^{-1}$. Thus, even a change in our assumed temporal broadening by a factor of 2 would not strongly affect our conclusions, and so we will work within the single-screen framework for the remainder of this paper. 

Our second assumption is that the outer scale for the scattering of \sgra\ is sufficiently large to be irrelevant for our calculations (specifically, we require $r_{\rm out} \gg 10\,{\rm AU}$). We will discuss the validity of this assumption in \S\ref{sec::Outer_Scale}.

\section{Scattering Model Fitting Procedure}
\label{sec::Model_Fitting_Procedure}

We now describe our procedure to fit observations, constrain the full scattering model, and estimate parameter uncertainties. Our fitting strategies are guided by synthetic datasets. We generated datasets with identical baseline coverage and sensitivity to our actual observations of \sgra, but with visibilities generated from simulated scattered images. We use a Monte Carlo approach to determine our uncertainties, fitting an ensemble of synthetic observations of scattered images. Thus, our reported uncertainties account for thermal noise, limitations of the fitting procedure, and systematic uncertainties from refractive scattering.

% Our fits incorporate refractive noise, but they do not include the full covariance of the noise among different baselines because doing so is computationally prohibitive. 

\subsection{Anisotropic Gaussian Model Assumption}
\label{sec::Gaussian_Model_Assumption}

One significant simplification in our model fitting approach is that we model the brightness distribution of the source on the sky as a wavelength-dependent anisotropic Gaussian. In \S\ref{sec::Scattering_Model}, we demonstrated that this assumption is well motivated for the shape of the scatter broadening because it corresponds to universal scattering behavior in the limit of long wavelength. Moreover, our approach to estimate parameter uncertainties uses the full, non-Gaussian scattering model, so our final error budget accounts for limitations in the assumption of Gaussian scatter broadening. However, the intrinsic source may be non-Gaussian, especially when the emission region becomes optically thin. Nevertheless, even in this regime, the Gaussian source assumption is well motivated for model fitting and has a meaningful associated FWHM, as we will now demonstrate.

Specifically, the interferometric visibility $\tilde{I}(\mathbf{u})$ on a short baseline $\mathbf{u}$ can be approximated as
\begin{align}
\label{eq::VCZ}
\nonumber \tilde{I}(\mathbf{u}) &= \int d^2 \mathbf{x}\, I(\mathbf{x}) \, e^{-2\pi i \mathbf{u} \cdot \mathbf{x}}\\
 &\approx \int d^2 \mathbf{x}\, I(\mathbf{x}) \left[ 1 - 2\pi i \mathbf{u} \cdot \mathbf{x} - 2\pi^2  \left( \mathbf{u} \cdot \mathbf{x} \right)^2 \right],
\end{align}
where $I(\mathbf{x})$ denotes the image, with $\mathbf{x}$ an angular coordinate on the sky \citep{TMS}. The term that is linear in $\mathbf{u}$ reflects an interferometric phase that is proportional to the image centroid projected along the baseline direction (from the Fourier shift theorem). Standard VLBI observations (including all those used in this paper) do not have absolute phase information, so we can set this term to zero (i.e., we use the image centroid to define the origin of the sky coordinates). The remaining terms in Eq.~\ref{eq::VCZ} show that the visibility amplitude decreases quadratically with baseline length for short baselines. The quadratic coefficient is proportional to the second moment of the image projected along the baseline direction. This second moment can then be used to define a characteristic image FWHM, using the relationship corresponding to a perfectly Gaussian image. For example, the major axis FWHM $\theta_{\rm maj}$ is given by 
\begin{align}
\label{eq::fwhm_maj}
\theta_{\rm maj} = \sqrt{-\frac{2 \ln(2)}{\pi^2 \tilde{I}(\mathbf{0})}\left. \nabla^2_{\mathbf{\hat{u}}_{\rm maj}} \tilde{I}(\mathbf{u})\right\rfloor_{\mathbf{u}=0}},
\end{align}
where $\tilde{I}(\mathbf{0})$ is the total flux density of the image, and $\nabla^2_{\mathbf{\hat{u}}_{\rm maj}}$ is the second directional derivative along the major axis direction. The three characteristic Gaussian parameters $\{ \theta_{\rm maj}, \theta_{\rm min}, \phi\}$ can be determined by diagonalizing the image covariance matrix. 

For this universal Gaussian behavior for the source visibility function to be applicable, the baselines must only marginally resolve the \emph{unscattered} source. For \sgra, this assumption can be assessed post hoc using the inferred intrinsic size. Using the characteristic size $\theta_{\rm src} \sim (0.4\,{\rm mas}) \times \lambda_{\rm cm}$ that we derive later (see \S\ref{sec::Scattering_Intrinsic}), we estimate that projected baselines must have a length of approximately $3000\,{\rm km}$ for the normalized visibility function of the intrinsic source to fall to $1/e$ (this length is independent of wavelength because the source grows linearly with wavelength while angular resolution scales inversely with wavelength). For all observations we analyze other than 1.3\,mm and 3.5\,mm, the longest baselines are significantly shorter than this limit (because longer baselines heavily resolve the \emph{scattered} source). Thus, for the wavelengths we analyze to estimate the scattering kernel ($\lambda \geq 7\,{\rm mm}$), the quadratic expansion of Eq.~\ref{eq::VCZ} and Gaussian approximation are well motivated for the intrinsic structure of \sgra.

\vspace{0.4cm}

\subsection{Anisotropic Gaussian Fitting Procedure}
\label{sec::Gaussian_Model_Fitting}

To estimate the scattering kernel of \sgra, we independently fit anisotropic Gaussian models to observations of \sgra\ at wavelengths from 1.3\,mm to 30\,cm. In principle, fitting a Gaussian to interferometric data is quite simple. In practice, the fitting is subtle and subject to many sources of noise and bias. These include thermal noise, station-based systematic errors in amplitude and phase, and refractive noise. We developed a simplified prescription for Gaussian model fitting that accounts for all these errors. Our prescription is motivated by tests on synthetic datasets (see \S\ref{sec::Synthetic_Observations}); it sacrifices some exactness for the sake of computational efficiency. Nevertheless, our approach provides a reliable error budget despite shortcomings in the model fitting procedure. 

For each observation, we began with complex visibilities that had a priori amplitude calibration applied but no self calibration. We first flagged all visibilities for which the elevation at either station was below $5^\circ$. Next, on a scan-by-scan basis, we flagged all stations that did not have a signal-to-noise (snr) of at least 12 on any baseline. Thus, at each time, a station was only included if it had at least one strong fringe detection. This station-based cut was chosen to avoid including visibilities with a noise bias from the fringe search; a baseline-based cut would also avoid spurious fringes but would potentially bias the set of unflagged, low-snr visibility amplitudes upward. Next, we computed the expected renormalized refractive noise (see \S\ref{sec::Refractive_Noise}) for each point, and we eliminated all visibilities for which the ensemble-average visibility function was less than four times the renormalized refractive noise. This cut eliminates visibilities that are dominated by refractive noise from the Gaussian model fits (we only performed this final cut for the Gaussian model fitting and include these visibilities for estimates of the long-baseline refractive noise). 

Next, we jointly fit for complex, time-dependent station gains at every site and the Gaussian image parameters (i.e., self-calibration to a model), seeking the maximum a posteriori estimate of all parameters. For this estimate, we used flat priors on the station phases and Gaussian priors on the logarithm of the gain amplitude, centered on a gain of amplitude of unity and with wavelength- and array-dependent spread. We used 5\% uncertainty for VLA data at 15-30\,cm, 5\% uncertainty at 3.6\,cm for VLBA data, and 10\% uncertainty at 1.3\,cm (VLBA) and 7\,mm (KaVA). At 3.5\,mm, the a priori calibration is sufficiently poor that we do not constrain the gain amplitudes (similar to an analysis using only closure quantities). These values can be validated after fitting the actual data and were guided by the expected performance for each wavelength/array combination. We assumed that the visibilities had complex Gaussian random noise, with standard deviation that was the quadrature sum of the measured thermal noise and the renormalized refractive noise. In this way, we included additional tolerance for visibility errors from refractive noise.

\subsection{Synthetic Observations for Monte Carlo Uncertainty Estimates}
\label{sec::Synthetic_Observations}

To estimate uncertainties for the fitted parameters, we used a Monte Carlo approach. Namely, for each dataset analyzed, we generated a representative ensemble of synthetic datasets and analyzed each using our procedure for the actual data. To create synthetic datasets, we scattered Gaussian source images using the \texttt{stochastic-optics} module of the \texttt{eht-imaging} library \citep{Chael_2016,Stochastic_Optics}. The source and scattering parameters were chosen to match the current best-estimates in our iterative fitting procedure (see \S\ref{sec::Fitting_Strategy}). We then sampled each scattered image on the observed $(u,v)$ coordinates and added complex Gaussian noise that was equal to the measured thermal noise. Next, we injected two types of gain uncertainty to the measurements: 1.) fluctuations of the station gains that were uncorrelated from scan to scan, and 2.) an overall uncertainty in each station gain that was constant over the entire observation but different among the different synthetic datasets. Each of these gain errors was a Gaussian random variable with unit mean and wavelength-dependent uncertainty, matching the values given \S\ref{sec::Gaussian_Model_Fitting}. 

As a concrete example, a single realization of the simulated 1.3\,cm VLBA data would have rapid jitter from thermal noise that was uncorrelated among all baselines and times, rapid station-based jitter from the gain errors (rms of $\sqrt{2} \times 10\%$ of each visibility amplitude), and a constant station-based error (rms of $\sqrt{2} \times 10\%$ of each visibility amplitude). For instance, all baselines to a particular antenna might be systematically underestimating the true flux density in one realization and over-estimating in another. Each realization also produced an image with FWHM fluctuations from refractive image distortions and with additional noise on long baselines from refractive substructure. 

To estimate our parameter uncertainties, we compute the rms of the parameter estimates from each simulated data set with respect to the true, ensemble-average parameter. Thus, our uncertainties account for thermal scatter in the model fitting, for systematic scatter from the refractive scattering, and for systematic errors and bias in the model fitting procedure.

Real data have additional imperfections that our simplified prescription does not capture, including bandpass errors, polarimetric leakage, and gain errors that are elevation dependent. However, the polarization of \sgra\ is negligible at cm wavelengths, and residual bandpass errors are small. As we will demonstrate, the dominant source of uncertainty for many of our measurements is refractive scattering, and our Monte Carlo approach fully accounts for this uncertainty.

\subsection{Overall Fitting Strategy}
\label{sec::Fitting_Strategy}
As described in the previous sections, we independently fit Gaussian models to data at multiple frequencies. However, these fits used refractive noise corresponding to the scattering properties that are estimated using the full multi-frequency dataset. Thus, our overall fitting procedure is iterative:
\begin{enumerate}
\item We fit Gaussian models to the 1.3\,cm and 3.6\,cm data. These fits require estimates of $r_{\rm in}$ and $\alpha$ to determine the refractive noise to include in the model fitting procedure and in the Monte Carlo uncertainty estimation via synthetic data. We assume that the scattering dominates the intrinsic size at these wavelengths (as is supported by the $\lambda^2$ scaling), so we use these fits to estimate the three parameters that characterize the long-wavelength scattering behavior: $\theta_{\rm maj,0}$, $\theta_{\rm min,0}$, $\phi_{\rm PA}$. 
\item Keeping $\theta_{\rm min,0}$ and $\phi_{\rm PA}$ fixed at the values obtained in step 1, we fit the $15-30\,{\rm cm}$ VLA data to obtain a tighter constraint on $\theta_{\rm maj,0}$. 
\item Having determined the three parameters of the long-wavelength ensemble-average image in steps 1 and 2, we use four additional pieces of evidence to constrain $r_{\rm in}$ and $\alpha$:
\begin{enumerate}
\item The nearly perfect scaling of image size as $\lambda^2$ down to 1.3\,cm and across the observing bandwidth at 1.3\,cm, combined with constancy of position angle and image anisotropy over this frequency range. These properties suggest that scattering must dominate over intrinsic structure at all wavelengths longer than 1.3\,cm and that the inner scale must exceed the diffractive scale at 1.3\,cm. 
\item The Gaussian scaling of visibility amplitude with baseline length at 1.3\,cm and tentative non-Gaussian scaling of visibility amplitude with baseline length at 7\,mm.
\item The magnitude of refractive visibility noise using long-baseline measurements at 1.3\,cm and 3.6\,cm, where the Gaussian image contribution is negligible.
\item An upper limit of 3\% on image size fluctuations at 7\,mm, as determined by historical data. 
\end{enumerate}
\item We then repeat steps 1-3 using the full scattering model ($\theta_{\rm maj,0}$, $\theta_{\rm min,0}$, $\phi_{\rm PA}$, $\alpha$, $r_{\rm in}$) estimated in the previous pass to estimate a new set of scattering parameters. 
\end{enumerate}

% \subsection{Limitations in the Fitting Procedure}

\section{Observations and Gaussian Fits}
\label{sec::Observations}

We now provide details on the specific observations that we use to constrain our scattering model. While previous scattering studies have generally relied on compiling large sets of observations and then averaging across multiple epochs to reduce parameter uncertainties, we instead consider a small number of high-quality observations and analyze each with a full scattering error budget.

\subsection{VLA Observations at 20cm}

\begin{figure}[t]
\centering
\includegraphics[width=\columnwidth]{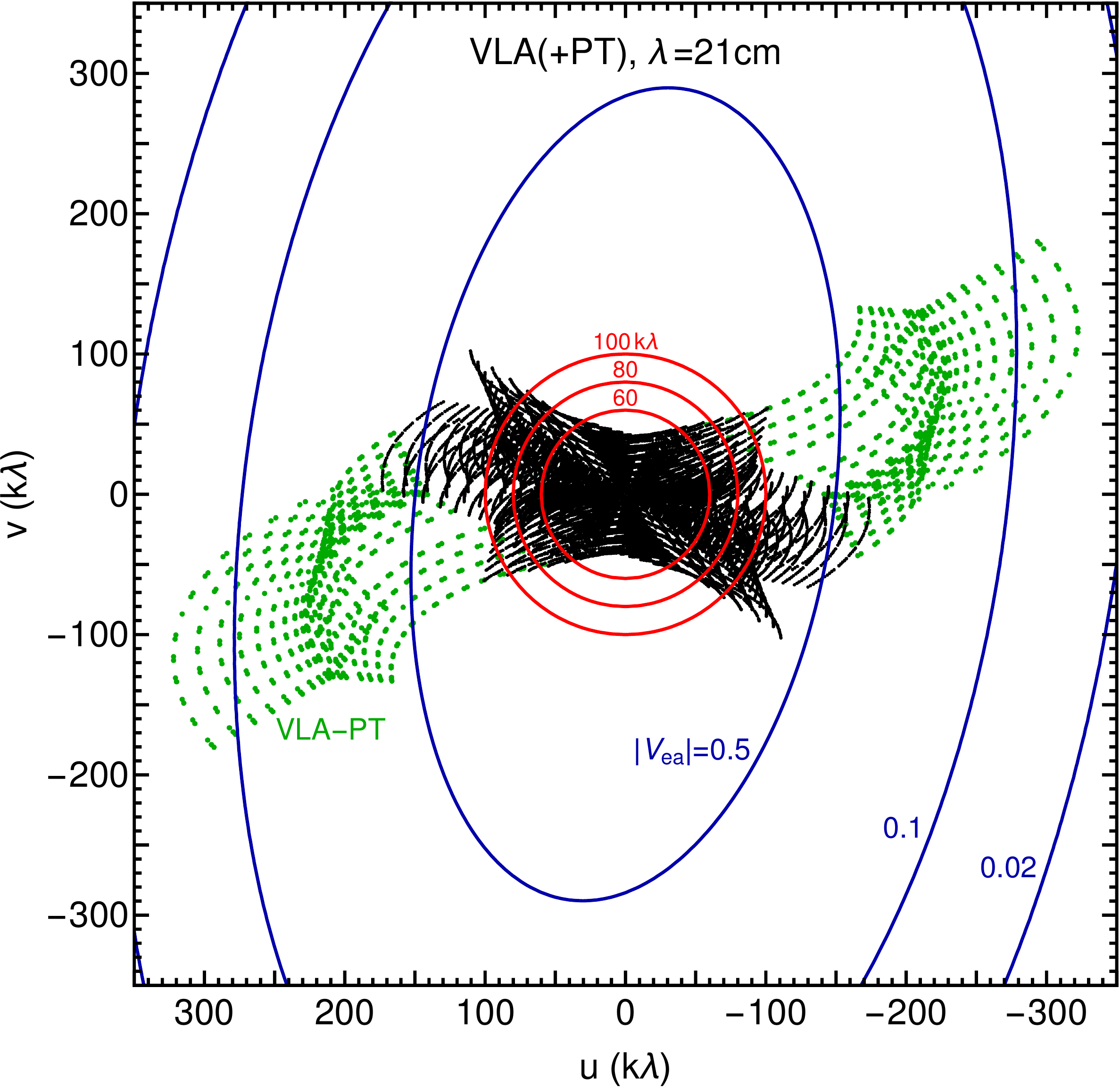}
\caption
{ 
Baseline coverage of the VLA at the representative wavelength $\lambda=21.06\,{\rm cm}$, with baselines to the single VLBA antenna at Pie Town in green. Red circles show our three cuts in $|\mathbf{u}|$ to eliminate flux density from diffuse emission near \sgra. Blue ellipses show contours at which the estimated ensemble-average Gaussian visibility function falls to $0.5$, $0.1$, and $0.02$ of the zero-baseline value.     
}
\label{fig::VLA_u-v_Coverage}
\end{figure}

The longest wavelength observations we examine are with the Karl G. Jansky Very Large Array (VLA). These observations span wavelengths from $15-29$\,cm. The recorded bandwidth was divided into 16 spectral windows, each 64\,MHz. Of the 16 original windows, 4 were flagged by the VLA calibration pipeline in CASA. We analyzed each spectral window independently. For each, we averaged in frequency and in 1-minute intervals.  Figure~\ref{fig::VLA_u-v_Coverage} shows representative baseline coverage for one spectral window. 

The long-wavelength data are subject to a challenge for Gaussian model fitting that does not affect our other observations. Namely, short baselines measure significant flux density that is not associated with \sgra; it is diffuse emission from the local Galactic Center environment. To eliminate contributions from this emission, we imposed a minimum baseline length $u_{\rm min}$ for visibilities used in the Gaussian fits. On baselines longer than ${\sim}50\,{\rm k}\lambda$, we do not see any indications of contaminating emission (e.g., via non-zero closure phases). Thus, we repeated Gaussian fits using $u_{\rm min} = 60\,{\rm k}\lambda$, $80\,{\rm k}\lambda$, and $100\,{\rm k}\lambda$, and we then used the scatter of these solutions as our estimate for the measurement uncertainty.

Because the VLA baselines only modestly resolve \sgra, we did not find stable results among the different spectral windows when fitting all Gaussian parameters separately. In addition, the fitted position angle was highly degenerate with the major axis size; smaller position angles produce a smaller major axis size because of the anisotropic baseline coverage (see Figure~\ref{fig::VLA_u-v_Coverage}). To mitigate this problem, we instead fit the Gaussian parameters holding the minor axis and position angle fixed to the values determined by 3.6 and 1.3\,cm observations (i.e., we only fit for the total flux density and major axis size in each spectral window). With this reduction, we obtained self-consistent estimates of the major axis among all spectral windows (see Table~\ref{tab::All_Fits}).

Fitting a single $\lambda^2$ scattering law to our measured major axis sizes yields $\theta_{\rm maj,0} = 1.3799 \pm 0.0067\,{\rm mas}$.  This uncertainty does not account for refractive scattering effects, and it underestimates thermal uncertainty because the measurements with varying $u_{\rm min}$ have identical thermal noise. 
The uncertainty in the assumed position angle, $\sigma_{\rm PA} \sim 0.2^\circ$, gives an additional uncertainty of ${\approx}\, (0.004\,{\rm mas}) \times \left(\sigma_{\rm PA}/1^{\circ} \right)$, which is negligible. The uncertainty in the assumed minor axis is likewise negligible.

To estimate the total uncertainty, we repeated the Gaussian fits using multi-frequency sets of synthetic data generated from 10 simulated realizations of the scattering. For each realization, we estimated a scattering law from the fitted Gaussian parameters. Note that while these data do not include diffuse emission, we used the same procedure with cutoffs of $u_{\rm min} = 60\,{\rm k}\lambda$, $80\,{\rm k}\lambda$, and $100\,{\rm k}\lambda$ for these data. These estimates of $\theta_{\rm maj,0}$ had a scatter of $0.011\,{\rm mas}$ relative to the ensemble-average value for the simulations. This scatter accounts for refractive noise, thermal noise, and systematic noise from gain errors, but it does not account for systematic uncertainty from diffuse structure. Adding our two estimated uncertainties at quadrature yields our final estimate: $\theta_{\rm maj,0} = 1.380 \pm 0.013\,{\rm mas}$. 

We also reanalyzed the VLA observations reported in \citet{Bower_2006} using the same procedure as for the VLA observations. These observations included the single VLBA antenna at Pie Town (PT) in addition to the VLA, thereby extending the longest baselines by a factor of ${\approx}\,2$. However, they had the disadvantage of a radio transient located only $2.7''$ south of \sgra, with a flux density that was ${\sim}5\%$ of \sgra\ \citep{Bower_Transient}. For our analysis, we adopted an image-domain approach to remove the transient. Namely, we performed maximum entropy imaging independently for each sub-band. For each image, we then computed the interferometric visibilities for the transient by windowing the image on a region of radius $1.8''$ centered on the transient. We subtracted these from the measured visibilities (because the Fourier relationship is linear) and use the remainder for Gaussian model fitting to \sgra. With this procedure, we found $\theta_{\rm maj,0} = 1.4082 \pm 0.0075$. This value is at modest tension ($1.9\sigma$) with the VLA-only results, especially because refractive effects are likely correlated between the two epochs (at these wavelengths, the refractive timescale is ${\sim}\,100\,{\rm years}$), so each of the two measurements would be similarly biased.  Because uncorrected contamination from the transient may bias the measured Gaussian values for \sgra, we adopt the measurement and uncertainty of the VLA-only results.

{
\begin{deluxetable*}{l|ccc|cccc}
\tablecaption{Summary of elliptical Gaussian fits to \sgra.}
\tablehead{
\colhead{$\lambda$}  & Instrument & Expt{.} & Obs.\ Date & \colhead{$\theta_{\rm maj}$} & \colhead{$\theta_{\rm min}$} & \colhead{P.A.}\\
\colhead{cm}         &            &            &         & \colhead{$\mu{\rm as}$}      & \colhead{$\mu{\rm as}$}      & \colhead{deg}
}
\startdata
28.84 & VLA & 15A-310 & 20 August, 2015 & $1147000 \pm 31000$ & --- & --- \\
27.17 & VLA & 15A-310 &                  & $1017000 \pm 11000$ & --- & --- \\ 
23.22 & VLA+PT & AB1134 & 1 \& 4 October, 2004 & $664000 \pm 121000$ & --- & --- \\ 
22.05 & VLA & 15A-310 &                  & $672000 \pm 28000$ & --- & --- \\ 
21.96 & VLA+PT & AB1134 &                  & $682000 \pm 14000$ & --- & --- \\ 
21.06 & VLA & 15A-310 &                  & $609000 \pm 16000$ & --- & --- \\ 
20.89 & VLA+PT & AB1134 &                  & $624000 \pm 4000$ & --- & --- \\ 
20.15 & VLA & 15A-310 &                  & $546000 \pm 12000$ & --- & --- \\ 
19.78 & VLA+PT & AB1134 &                  & $544000 \pm 6000$ & --- & --- \\ 
18.56 & VLA & 15A-310 &                  & $489000 \pm 6000$ & --- & --- \\ 
18.00 & VLA+PT & AB1134 &                  & $464000 \pm 12000$ & --- & --- \\ 
17.85 & VLA & 15A-310 &                  & $435000 \pm 3000$ & --- & --- \\ 
17.47 & VLA+PT & AB1134 &                  & $423000 \pm 5000$ & --- & --- \\ 
17.19 & VLA & 15A-310 &                  & $395000 \pm 13000$ & --- & --- \\ 
16.59 & VLA & 15A-310 &                  & $371000 \pm 14000$ & --- & --- \\ 
15.49 & VLA & 15A-310 &                  & $329000 \pm 5000$ & --- & --- \\ 
14.99 & VLA & 15A-310 &                  & $308000 \pm 7000$ & --- & --- \\ 
3.598 & VLBA(+GBT) & BG221B  & 09 April, 2014    & $18290 \pm 310$ & $9110 \pm 170$  & $82.2 \pm 0.8$  \\
1.261 & VLBA+GBT & BG221A  & 07 March, 2014    & $2255 \pm 61$  & $1243 \pm 39$   & $81.9 \pm 0.2$ \\
0.698 & KaVA     & r14308a & 04 November, 2014 & $741  \pm 19$   & $434 \pm 8$     & $81.2 \pm 0.6$ \\
0.348 & VLBA+LMT & BD183C  & 27 April, 2015    & $215 \pm 4$     & $139 \pm 4$     & $80.9 \pm 3.0$\\
0.131 & EHT      & 2013 Campaign & 21-27 March, 2013 & $59 \pm 6$      & $60 \pm 30$             & --- 
\enddata
\label{tab::All_Fits}
\tablecomments{Because of our Monte Carlo error estimation procedure, the size uncertainties are stated relative to the ensemble-average image. They account for thermal noise, systematic noise, limitations of our fitting procedure, and refractive variations of the image size. Note that the errors for each epoch are highly correlated (from all effects apart from thermal noise). }
\end{deluxetable*} 
}

\subsection{VLBA Observations at 3.6cm}
\label{sec::Obs_Xband}

We analyzed observations at $\lambda=3.6\,{\rm cm}$ taken with the VLBA in 2014. These observations also included the GBT, but we did not detect fringes between GBT and the inner VLBA, so we only use the inner six VLBA stations (Brewster: BR, Fort Davis: FD, Kitt Peak: KP, Los Alamos: LA, Owens Valley: OV, Pie Town: PT) for our analysis. These observations recorded four contiguous $128\,{\rm MHz}$ channels and spanned approximately 3.5\,hours. They used NRAO~530 as a calibration source. After a global fringe search in AIPS \citep{Greisen_2003}, we averaged the data in frequency and in 30-second intervals before Gaussian fitting. 

\begin{figure}[t]
\centering
\includegraphics[width=0.48\textwidth]{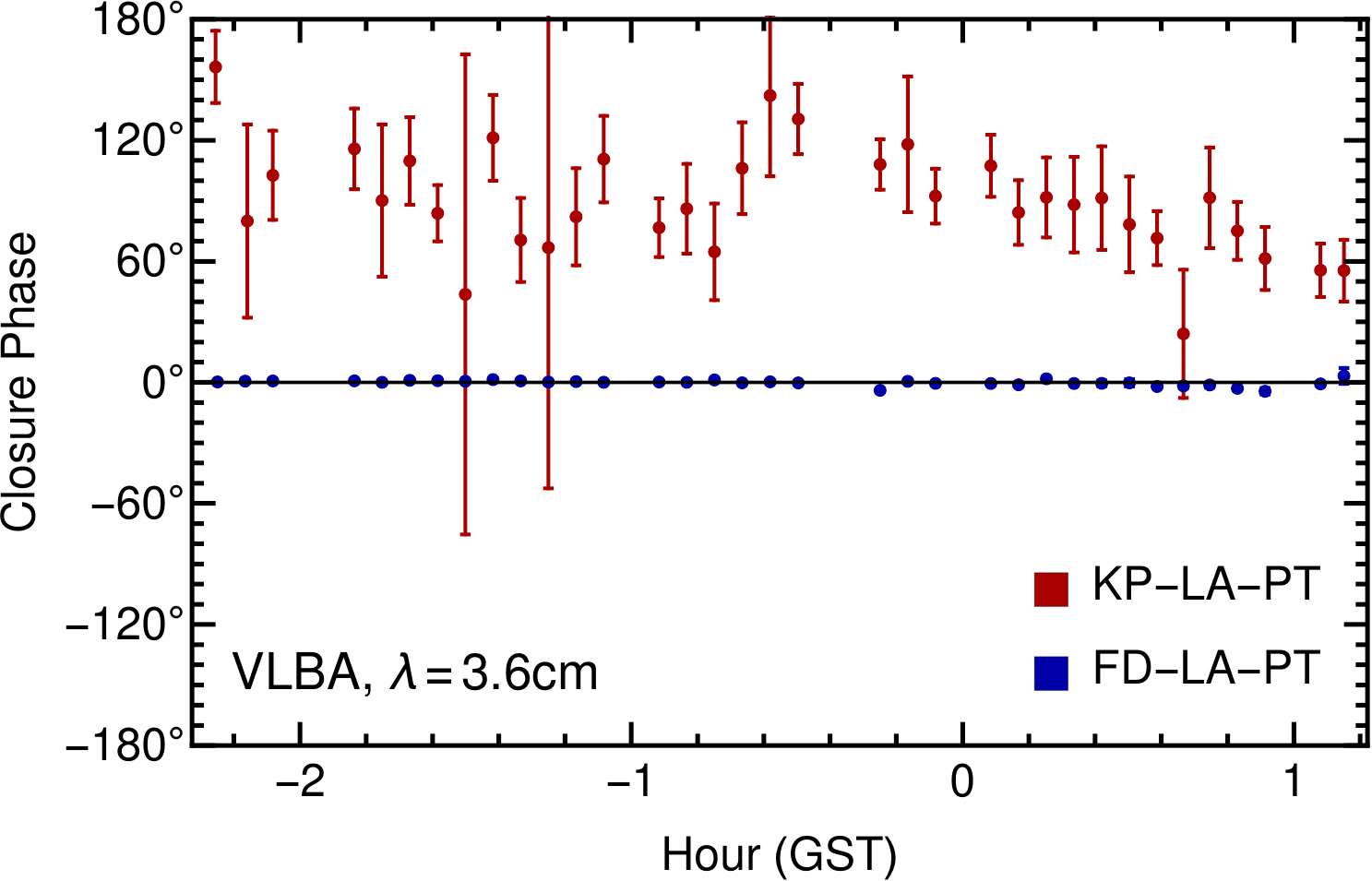}
\caption
{ 
Example closure phases from VLBA observations of \sgra\ at $\lambda=3.6\,{\rm cm}$. We first computed closure phases on 30-second intervals, then (vector) averaged them in 10-minute blocks. We estimated uncertainties via bootstrap resampling within each block. Closure phases on the triangle FD-LA-PT are close to zero, as is expected because these baselines are all dominated by the symmetric Gaussian image rather than substructure (see Figure~\ref{fig::Xband_Data_Summary}). In contrast, closure phases on the triangle KP-LA-PT are markedly non-zero, demonstrating the clear imprint of refractive substructure breaking symmetry of the smooth ensemble-average image. 
}
\label{fig::Xband_CPhase}
\end{figure}

Even without detailed analysis, the effects of refractive substructure are evident in the closure phases of these data. On triangles that resolve the source, the closure phases are markedly non-zero, demonstrating that the underlying image is inconsistent with \emph{any} smooth, scatter-broadened structure (see Figure~\ref{fig::Xband_CPhase}). Our Gaussian fitting procedure gives the major and minor axis sizes to within an estimated uncertainty of less than 2\%, even when including refractive effects in the error budget (see Table~\ref{tab::All_Fits}). 

\begin{figure*}[t]
\centering
\includegraphics[width=\textwidth]{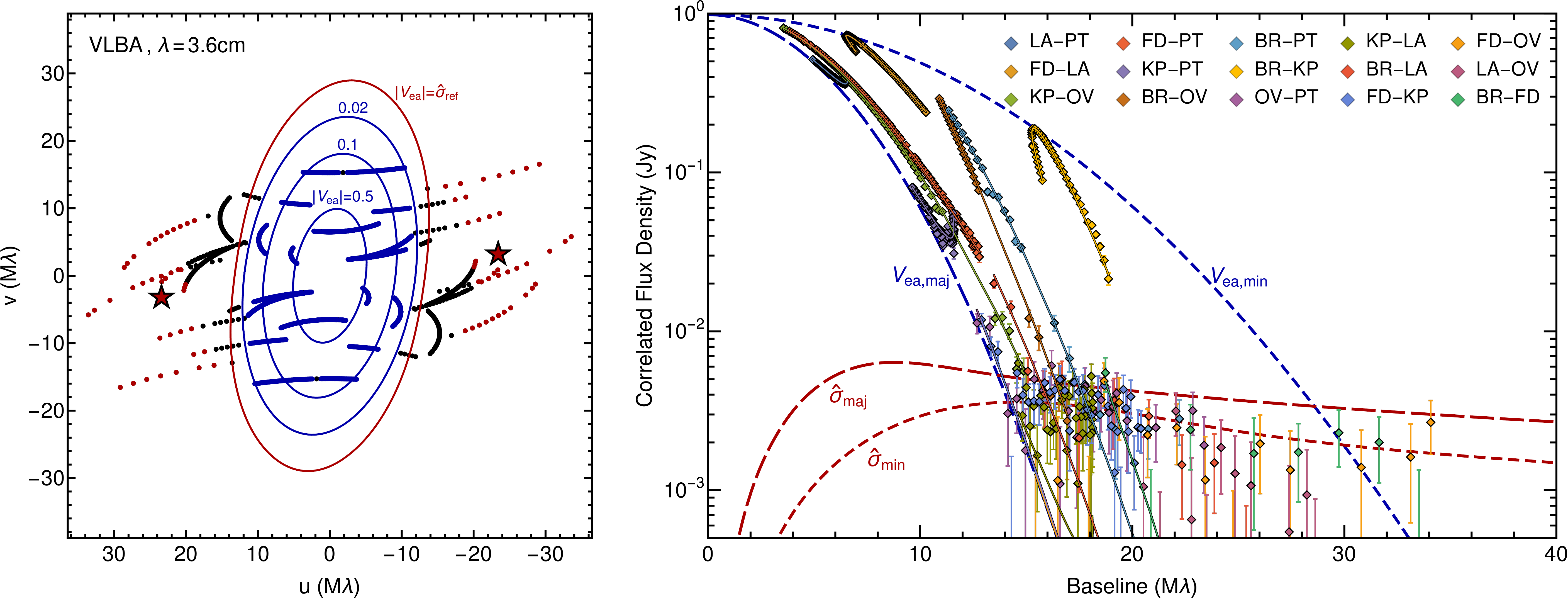}
\caption
{ 
Self-calibrated VLBA data at $\lambda=3.6\,{\rm cm}$ (see \S\ref{sec::Obs_Xband}). (left) Final $u$-$v$ coverage after cuts described in \S\ref{sec::Obs_Xband}. Blue points indicate those used for the Gaussian model fits; red points indicate those used to estimate the long-baseline refractive noise, with the red stars denoting their baseline vector average. Blue ellipses show 0.5, 0.1, and 0.02 contour levels of the fitted Gaussian; the red contour shows where the Gaussian visibility is equal to the rms renormalized refractive noise. (right) Correlated flux density as a function of baseline length. The long/short dashed blue curve shows the best-fit Gaussian visibility function along the major/minor axis. The red curves show the corresponding renormalized refractive noise. Points (with $\pm 1\sigma$ uncertainties) and Gaussian model curves are colored by baseline; baseline labels are ordered by median baseline length. Because of refractive noise, we expect systematic departures at the level of $\hat{\sigma}$ from the Gaussian model curves for each visibility; for a single baseline, these visibilities will be highly correlated over time (see, e.g., the BR-KP and the KP-PT visibilities). 
}
\label{fig::Xband_Data_Summary}
\end{figure*}

After the Gaussian fits, we self-calibrated the full data set to the Gaussian model. However, this procedure must be done with care because the longest baselines are dominated by refractive noise and are inconsistent with the pure Gaussian model. While our approximate prescription for model fitting with substructure (i.e., simply inflating the thermal noise with the rms renormalized refractive noise; see \S\ref{sec::Gaussian_Model_Fitting}) gives reliable results for Gaussian model fitting, we found that it could downward bias long-baseline visibilities. To perform the self-calibration without biasing long-baseline visibility amplitudes, we first derived time-dependent gain solutions using only ``short'' baselines, for which the Gaussian model visibility was four times the renormalized refractive noise. We then applied these self-calibration solutions to all baselines. We then dropped any visibilities that did not have simultaneous self-calibration solutions for both stations. In this way, long baselines that are dominated by refractive noise still obtain reliable self-calibration to the Gaussian model. Figure~\ref{fig::Xband_Data_Summary} shows our 3.6\,cm data after self calibration in this way. However, these data only have eight baselines with consistently strong detections (and there are six stations to self-calibrate), so we cannot reliably synthesize an image. 

% To compare this, we estimate the noise as a function of rin and alpha. Need to fit the ensemble-average size for each case. 
% 
% Position angle: $82.6 \pm 0.4$
% 
% Major Axis: $18.2 \pm 0.7~{\rm mas}$. Normalization is $1.40 \pm 0.05$. 
% 
% Minor Axis: $10.9 \pm 0.4~{\rm mas}$. Normalization is $0.84 \pm 0.03$.

\begin{figure*}[t]
\centering
\includegraphics[width=\textwidth]{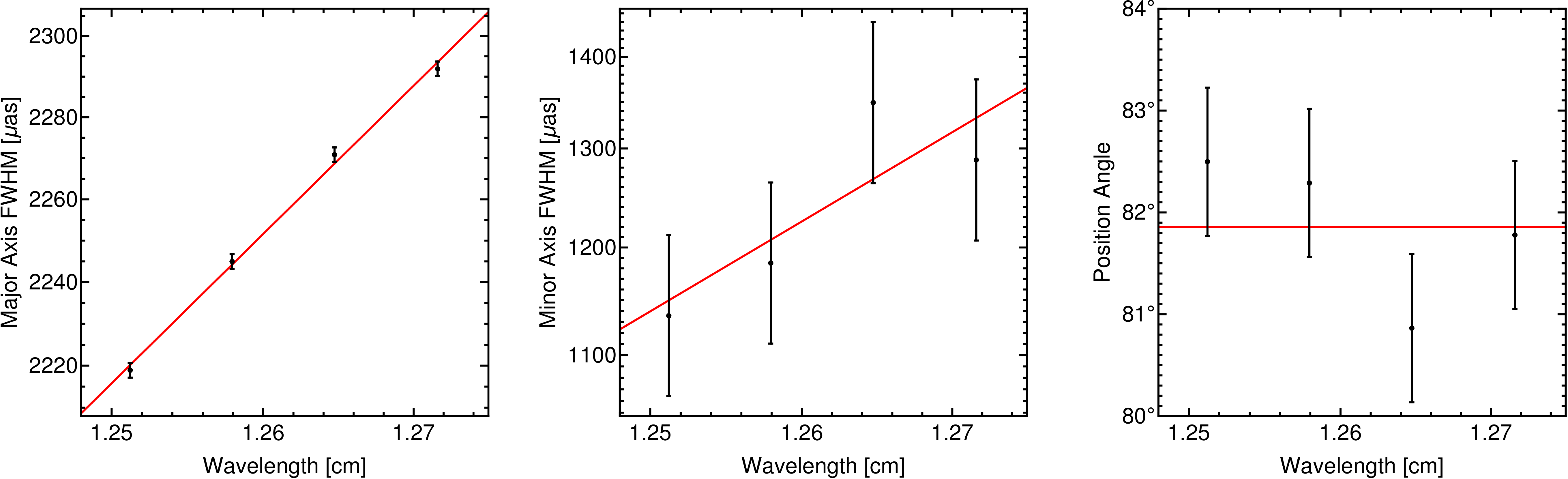}\\
\caption
{ 
Gaussian parameters for the four sub-bands of the $\lambda=1.26\,{\rm cm}$ observations fitted independently. The scatter among sub-bands when fitting the major axis scattering law is approximately $2\,\mu{\rm as}$, or roughly $0.08\%$ of the image size. However, the uncertainty from refractive fluctuations of the image size (which will give nearly identical bias for each sub-band) is approximately ${\approx}\, 20\,\mu{\rm as}$, or $1\%$ of the image size. Thus, the estimated major axis uncertainty relative to the ensemble average size is dominated by refractive image distortion. The close agreement with a $\lambda^2$ scaling law (shown in red) strongly suggests that intrinsic structure is heavily subdominant to scatter broadening at this wavelength and also that the inner scale must be larger than the diffractive scale, $ r_{\rm in} \gsim 300\,{\rm km}$, because otherwise the wavelength dependence of the scattering kernel steepens, $\theta \propto \lambda^{1+\frac{2}{\alpha}}$.
}
\label{fig::Kband_MultiIF}
\end{figure*}

\begin{figure}[t]
\centering
\includegraphics[width=\columnwidth]{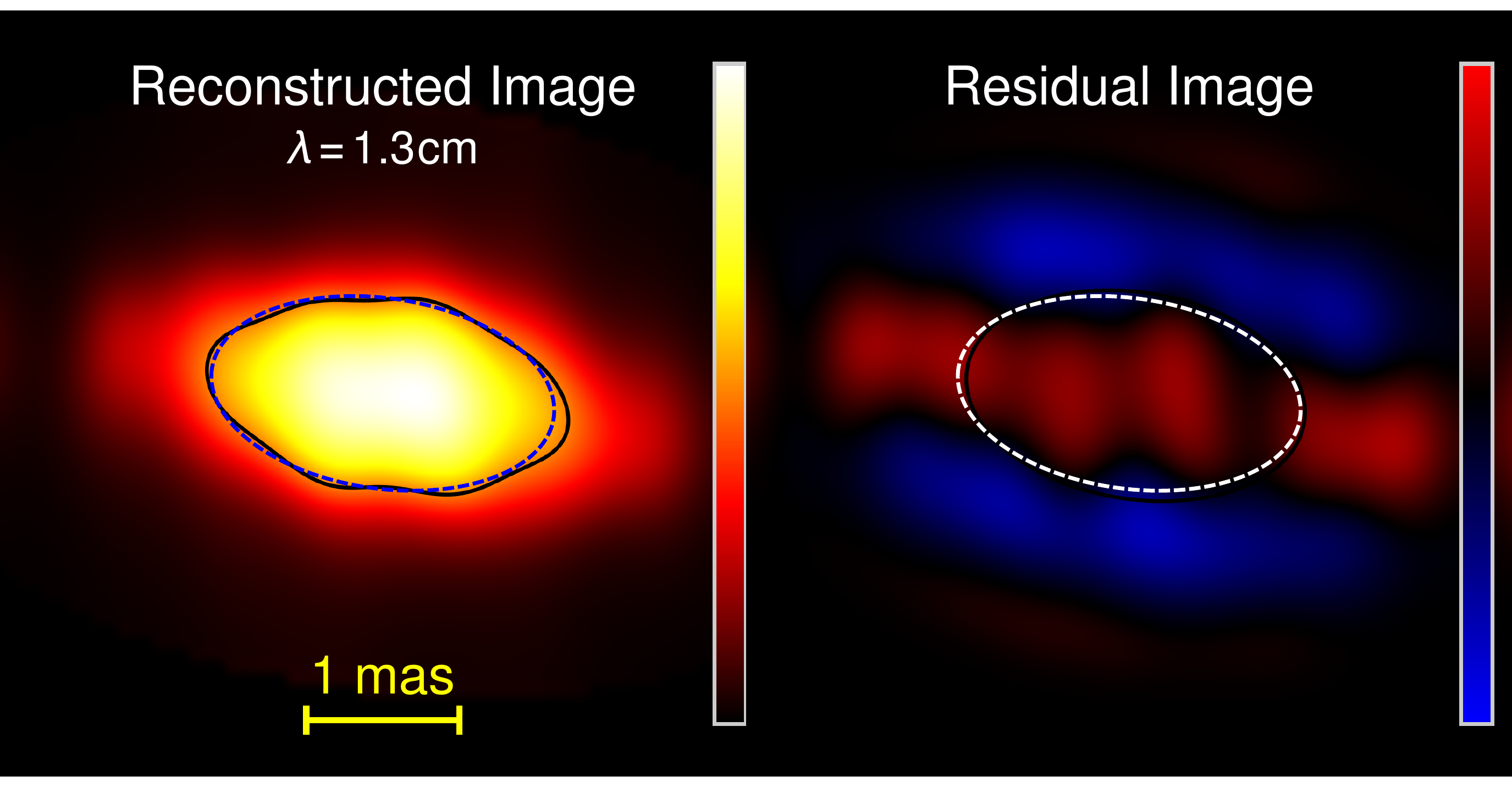}\\
\caption
{ 
(left) Reconstructed image at $\lambda=1.3\,{\rm cm}$. The color scale is linear and ranges from $0 - 0.33\,{\rm Jy}/{\rm mas}^2$. The dashed blue ellipse shows the Gaussian half-maximum contour from model fitting; the solid black line shows the half-maximum contour of the reconstructed image. Substructure is apparent through the subtle distortions from a smooth Gaussian image. (right) Residual image after subtracting the best-fit Gaussian image. The color scale is linear, and the range extends over $\pm 0.033\,{\rm Jy}/{\rm mas}^2$.
}
\label{fig::KBand_Image}
\end{figure}

\begin{figure*}[t]
\centering
\includegraphics[width=0.98\textwidth]{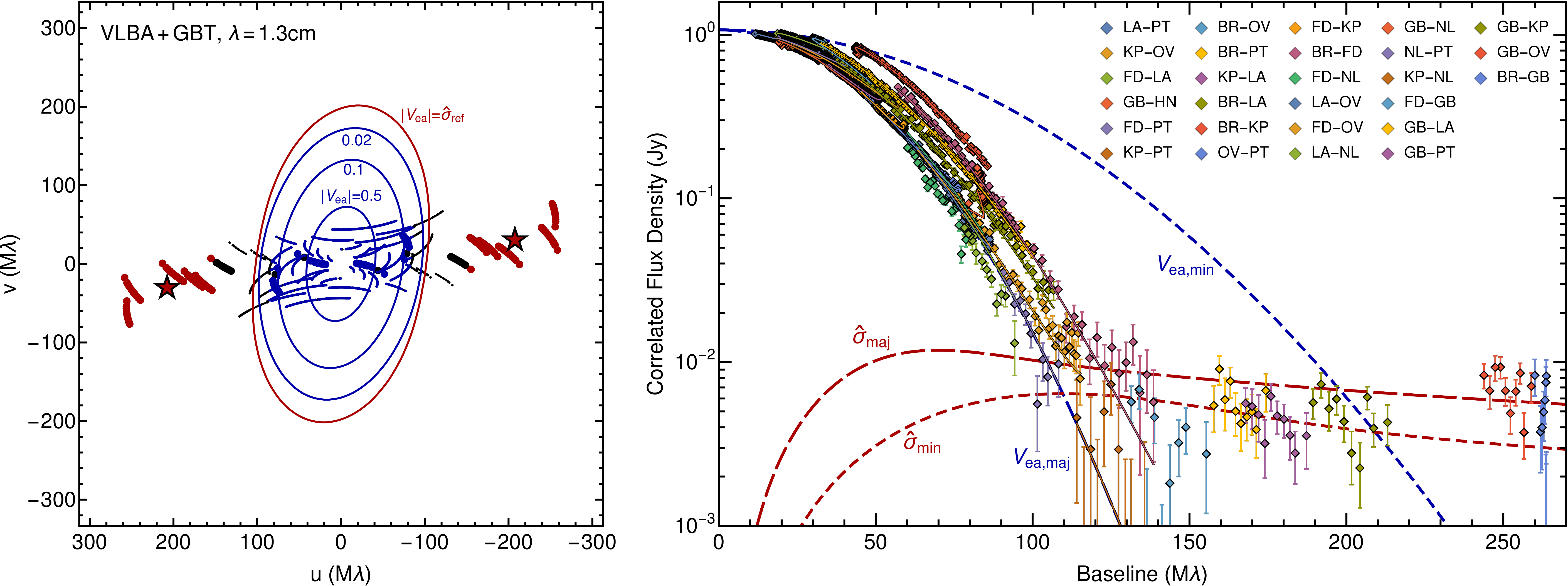}
\caption
{ 
VLBA+GBT observations at $\lambda=1.3\,{\rm cm}$ (see \S\ref{sec::Obs_Kband}). Panels are as described for Figure~\ref{fig::Xband_Data_Summary}. Larger points in the left panel denote baselines to the GBT. For clarity, we only show long baselines to GBT on this plot (omitting long baselines to NL and HN, which sample similar $(u,v)$ coordinates but with less sensitivity). 
}
\label{fig::Self-cal_wRefractive}
\end{figure*}

\subsection{VLBA Observations at 1.3cm}
\label{sec::Obs_Kband}

We analyzed observations at $\lambda=1.3\,{\rm cm}$ taken with the VLBA+GBT in 2014. These observations were analyzed in \citet{Gwinn_2014}, which reported the initial discovery of refractive substructure in \sgra. As with the 3.6\,cm data, these observations recorded four contiguous $128\,{\rm MHz}$ channels, spanned approximately 3.5\,hours, and used NRAO~530 as a calibration source. They include strong detections to the VLBA antennas at North Liberty (NL) and Hancock (HN) in addition to the sites noted in \S\ref{sec::Obs_Xband}. After a global fringe search in AIPS \citep{Greisen_2003}, we averaged the data in frequency and in 30-second intervals before Gaussian fitting. However, we averaged the data to four 128\,MHz sub-bands and separately analyzed each. 

For these data, the overall baseline coverage is well matched to the scattered image, and we can tightly constrain the Gaussian parameters separately within each sub-band. For each, the thermal uncertainty on the major axis size is less than $0.1\%$, and we clearly identify the $\lambda^2$ scaling of image size across the four sub-bands (see Figure~\ref{fig::Kband_MultiIF}). However, the uncertainty from refractive distortion is an order of magnitude larger, so we can only constrain the ensemble-average FWHM to within approximately 1\%. 

For these data, there is sufficient baseline coverage to reliably synthesize an image. Figure~\ref{fig::KBand_Image} shows a maximum entropy image reconstruction using the \texttt{eht-imaging} library \citep{Chael_2016}. The effects of refractive substructure are evident in substructure of the image. However, the effects of substructure are most striking in the visibility domain, where long baselines from GBT to the inner VLBA give strong detections on baselines for which the Gaussian visibility contribution is negligible. Figure~\ref{fig::Self-cal_wRefractive} shows the final self-calibrated visibilities, following the procedure described in \S\ref{sec::Obs_Xband}.

% Combining the measured Gaussian parameters at 1.3~cm and 3.6~cm, we obtain: 
% \begin{align}
% \theta_{\rm maj, 0} &= 1.37 \pm 0.02,\\
% \theta_{\rm min, 0} &= 0.82 \pm 0.02,\\
% \phi_{\rm PA} &= 82.4 \pm 0.3
% \end{align}

% PA: $82.4 \pm 0.3$
% 
% Minor Normalization: $0.82 \pm 0.02$. 

\subsection{KaVA Observations at 7mm}
\label{sec::KaVA_Observations}

\begin{figure*}[t]
\centering
\includegraphics[width=0.98\textwidth]{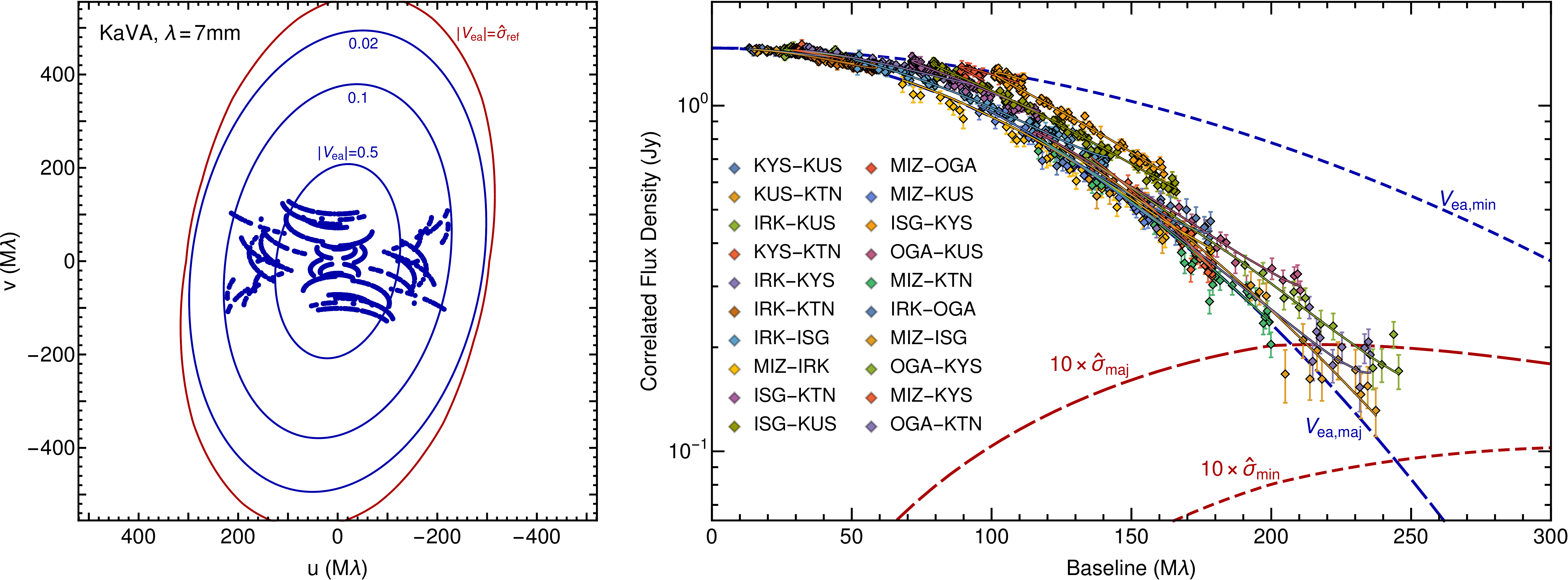}
\caption
{ 
KaVA observations at $\lambda=0.7\,{\rm cm}$ (see \S\ref{sec::KaVA_Observations}). Panels are as described for Figure~\ref{fig::Xband_Data_Summary}. 
}
\label{fig::KaVA_Data}
\end{figure*}

The KaVA array has been conducting regular monthly monitoring of \sgra\ at 7\,mm since September 2014 as part of the KaVA AGN large program \citep{Kino_2015,Zhao_2017}. The KaVA baselines range from 300 to 2300\,km and provide excellent $(u,v)$ coverage for \sgra\ observations (Figure~\ref{fig::KaVA_Data}; see also~\citet{Akiyama_2014}). In particular, the KaVA coverage along the North-South direction is significantly better than VLBA coverage at this frequency, so the KaVA data are better suited to estimate the minor axis size. 

%GY: the duration should 5.5~hours. Corrected
We analyzed data from the experiment r14308a, which were obtained in November 2014. The data were recorded with 256\,MHz total bandwidth, spanned 5.5~hours, and had an on-source time for \sgra\ of 220~minutes. NRAO~530 and two nearby SiO masers (OH 0.55-0.06, VX~Sgr) were observed as calibrators \citep{Cho_2017}. The correlated data were analyzed with AIPS in a standard pipeline. Most stations had good fringe detections in this experiment. After a global fringe search, the data were averaged in 30-second intervals and across the entire bandwidth for Gaussian model-fitting. See \citet{Zhao_2018} for more details of the monitoring and data analysis.

Figure~\ref{fig::KaVA_Data} shows the data from this observation, after our Gaussian model fitting and self-calibration. For these observations, the contribution of renormalized refractive noise is insignificant and is only comparable to the thermal noise on the longest baselines. 

\subsection{VLBA+LMT Observations at 3.5mm}

For $\lambda=3.5\,{\rm mm}$, we analyzed data from the first VLBI observations using the LMT in concert with the VLBA. These observations recorded 480\,MHz of bandwidth and spanned 7.5~hours (with approximately 3.4~hours on \sgra). We averaged the data in 10-second intervals and across the full bandwidth before Gaussian model fitting. For additional details about these observations, see \citet{Ortiz_2016}, who originally reported and analyzed them. 

Using our Gaussian fitting procedure, we found values and uncertainties very close to those reported in \citet{Ortiz_2016} using self-calibration. This agreement is expected because the only significant adaptation in our current approach is to include refractive noise, and the renormalized refractive noise is less than the thermal noise for all points. More recent data, reported by  \citet{Brinkerink_2016}, also includes the GBT and shows marked non-Gaussianity in the closure phases. For these data, which achieve significantly better sensitivity, including refractive noise in the error budget for model fitting may be significant.

\subsection{EHT Observations at 1.3mm}
\label{sec::EHT}

Since 2007, the Event Horizon Telescope (EHT) has observed \sgra\ using a 1.3\,mm VLBI array with stations in California, Arizona, and Hawaii. Recently, \citet{Lu_2018} reported observations that included a fourth station (in Chile). With only 3-4 stations, there is insufficient baseline coverage to create an image. In addition, the measured visibilities are markedly non-Gaussian \citep{Johnson_2015,Fish_2016,Lu_2018}, as expected because of complex intrinsic structure from optically thin emission near the black hole. Nevertheless, even under these circumstances, the image FWHM is still a meaningful quantity that is reliably constrained by sparse coverage because it represents universal behavior of the interferometric visibility function on short baselines (see \S\ref{sec::Gaussian_Model_Assumption}). Thus, we will now estimate this characteristic FWHM and its uncertainty at 1.3\,mm using previously published EHT data.
% Because is useful for model assessment at 1.3\,mm and for comparison with the FWHM at longer wavelengths, 

Because the EHT baseline joining CARMA-SMT (California-Arizona) does not significantly resolve \sgra\ at 1.3\,mm, current EHT measurements primarily constrain the source size in the direction of the California-Hawaii and Arizona-Hawaii baselines, close to East-West (i.e., roughly along the major axis of the scattering kernel). Early detections were consistent with a Gaussian image having a FWHM of approximately $40\,\mu{\rm as}$ \citep{Doeleman_2008,Fish_2011}. However, more recent measurements with improved sensitivity and calibration find visibility amplitudes on the shortest Hawaii baselines (California-Hawaii) that are strongly inconsistent with the Gaussian model \citep{Johnson_2015,Lu_2018}. Thus, the appropriate FWHM is not that of the Gaussian fits, which are incompatible with the data, but can instead be estimated by computing the characteristic FWHM of models that do fit the short- and intermediate baseline visibility amplitudes. 

One such model is an annulus. The fitted annulus in \citet{Johnson_2015} gives a characteristic FWHM of $58.5\,\mu{\rm as}$ for the intrinsic source (as defined in Eq.~\ref{eq::fwhm_maj}). For comparison, the annulus model from \citet{Doeleman_2008} gave $51.5\,\mu{\rm as}$. Two-Gaussian model fits that also include closure phase measurements and baselines to APEX give FWHMs of $55.2\,\mu{\rm as}$ and $60.4\,\mu{\rm as}$ along the East-West direction or $62.5\,\mu{\rm as}$ and $60.5\,\mu{\rm as}$ along the major axis of the scattering \citep[for Models A and B of][]{Lu_2018}. Because the East-West scatter-broadening is ${\lsim}\,20\,\mu{\rm as}$ at this frequency, our revisions to the scattering model and remaining uncertainties have little effect on the estimated intrinsic FWHM. The uncertainties are instead dominated by the sparse baseline coverage, and we estimate a plausible range of $51-63\,\mu{\rm as}$ for the FWHM of the intrinsic source along the major axis of the scattering based on the span of these fitted models. Note that this range extends beyond the expected diameter of the black hole shadow ($51 \pm 3\,\mu{\rm as}$), so it does not necessitate that the accretion flow is viewed at large inclination. Accounting for both source and scattering uncertainties, we adopt a plausible range of $53-66\,\mu{\rm as}$ for the FWHM of the scattered image of \sgra\ at 1.3\,mm along the major axis of the scattering kernel.

The North-South FWHM at 1.3\,mm is comparatively poorly constrained. \citet{Krichbaum_1998} reported detections of \sgra\ at $\lambda=1.4\,{\rm mm}$ on the baseline joining Pico Veleta and an antenna of the IRAM interferometer at Plateau de Bure. These observations had a baseline length $|\mathbf{u}| \approx 0.7 \times 10^9$, but the baseline was aligned close to East-West (position angle approximately $70^\circ$ East of North). \citet{Lu_2018} have recently reported detections at $\lambda=1.3\,{\rm mm}$ on baselines from APEX to California and Arizona, which are oriented close to North-South, but these heavily resolve the source. Thus, they are unreliable for estimating a FWHM using Eq.~\ref{eq::fwhm_maj} or for computing the second moment of a fitted model. Instead, we estimate a maximum size of the source along the scattering minor axis by requiring that the SMT-CARMA baseline amplitude be at least 80\% of the zero-baseline flux density over the GST range from $-0.5 - 4.0$~hours, as is supported by both a priori calibration \citep{Lu_2018} and polarization arguments \citep{Johnson_2015}. For a major axis FWHM of ${\sim}60\,\mu{\rm as}$, this requirement gives an upper limit to the minor axis FWHM of approximately $90\,\mu{\rm as}$. To obtain a corresponding lower limit, we require that the correlated flux density on the SMT-APEX baselines for the scattered image never exceeds 10\% of the zero-baseline flux density over the GST range from $0.0-2.5$~hours \citep[otherwise it would exceed measurements on this baseline;][]{Lu_2018}. This constraint only requires that the scattered source have a minor axis FWHM that exceeds $25\,\mu{\rm as}$. Combining these limits, we obtain a plausible range for the FWHM along the scattering minor axis direction of $25-90\,\mu{\rm as}$ (of course, the scattering position angle need not correspond to that of the scattered or unscattered image at 1.3\,mm). 

Finally, we note that the EHT has detected persistent non-zero closure phases of \sgra\ on the California-Arizona-Hawaii triangle, demonstrating that the scattered image structure is not point symmetric \citep{Fish_2016}. However, these results do not imply that the intrinsic or scattered FWHM is asymmetric because the non-zero closure phases may be produced by image substructure. For instance, Model~B in \citet{Lu_2018} fits both the visibility amplitudes and closure phases but has little asymmetry in the FWHM, with major and minor axes FWHMs of $60.5\,\mu{\rm as}$ and $60.3\,\mu{\rm as}$, respectively. 
% Thus, current EHT measurements are fully consistent with an intrinsic source that is symmetric overall, as quantified by the second moment or FWHM of the image. 

\section{Composite Constraints on the Scattering and Intrinsic Structure of Sgr A*}
\label{sec::Composite_Constraints}

We now use our Gaussian model fits and self-calibrated data to constrain the five parameters of our scattering model and estimate the intrinsic structure of \sgra. We derive constraints in two stages. First, in \S\ref{sec::Gauss_Constraints}, we constrain the three asymptotic Gaussian parameters using our fits to the long-wavelength observations (${\geq}\,1.3\,{\rm cm}$), for which scatter-broadening is dominant over intrinsic structure. Next, in \S\ref{sec::alpha_rin_constraints}, we jointly constrain $\alpha$ and $r_{\rm in}$ using the observed refractive scattering signatures and limits from the scattering kernel shape and wavelength dependence. With the scattering constraints in place, we show the estimated scattering properties and estimate the intrinsic size of \sgra\ in \S\ref{sec::Scattering_Intrinsic}.

\subsection{Constraining the Asymptotic Gaussian Parameters}
\label{sec::Gauss_Constraints}

The three asymptotic parameters of our scattering model can be estimated directly from the Gaussian fits to long-wavelength data. These parameters can also be directly compared with the results of previous studies.

For the major axis normalization, our analysis of the VLA data from $15-30\,{\rm cm}$ gives $\theta_{\rm maj,0} = 1.380 \pm 0.013~{\rm mas}$. For comparison, our fits to the $3.6\,{\rm cm}$ VLBA observation gives $\theta_{\rm maj,0} = 1.412 \pm 0.024~{\rm mas}$. Thus, the two estimates are consistent to within their stated uncertainties. We will adopt the VLA estimate and uncertainty for our constraint on $\theta_{\rm maj,0}$.

Because we could not reliably fit the minor axis and position angle using the VLA or VLA$+$PT data, we use VLBI measurements at shorter wavelengths to estimate these parameters. The minor axis of the scattering is small enough at 1.3\,cm that intrinsic structure may be significant. Taking only the 3.6\,cm measurement and full uncertainty gives $\theta_{\rm min, 0} = 0.703 \pm 0.013~{\rm mas}$. This estimate represents an upper limit to the scattering size because we have not included a contribution from intrinsic structure. However, our representative intrinsic source size derived below using the full set of shorter-wavelength data (see \S\ref{sec::Scattering_Intrinsic}) would bias this upward by only ${\lsim}\,0.01~{\rm mas}$, which is within our measurement uncertainty. 

Despite the relatively complete baseline coverage at 3.6\,cm (see Figure~\ref{fig::Xband_Data_Summary}), the position angle is rather poorly constrained at this wavelength. The reason for the poor constraint is that there are only eight baselines that are dominated by the Gaussian structure, and these baselines must constrain the (time-dependent) self-calibration solutions for the six participating stations. For comparison, among those same six stations, the 1.3\,cm data have fifteen baselines that are dominated by the Gaussian structure. Thus, the self-calibration at 1.3\,cm is heavily over-constrained, and the measured position angle has small uncertainties despite the more limited baseline coverage. Because we find a position angle that is consistent with a constant value over wavelengths from 3.5\,mm to 3.6\,cm, it is unlikely that intrinsic structure changes the position angle appreciably at wavelengths of 1.3 or 3.6\,cm. In addition, for the scattering model of \citet{Psaltis_2018}, the position angle of the scattering kernel is independent of wavelength. Thus, we estimate the scattering position angle by combining the measured position angles at 1.3~cm and 3.6~cm, giving $\phi_{\rm PA} = 81.9 \pm 0.2$.

% Thus, because the we infer a source that is largely isotropic and also find a position angle that is consistent with a constant value over wavelengths from 3.5\,mm to 3.6\,cm, 
Table~\ref{tab::GaussianParameters} compares our newly derived Gaussian parameters with previously reported estimates. Note that the three observations used to derive our parameter estimates (2015 VLA observations, and VLBA observations at 3.6 and 1.3\,cm) were not used by any of these previous studies. Relative to past work, the major and minor axes are consistent with the values found by \citet{Shen_2005}, but our major axis normalization is $4.7\sigma$ larger than the estimate of \citet{Bower_2006} and $3\sigma$ larger than \citet{Bower_2015} (both relied on the same VLA+PT image-domain analysis at long wavelengths). Our major axis uncertainty is similar to these previous results, largely because of the increased error budget to accommodate refractive fluctuations, while our minor axis uncertainty is significantly smaller than all past work. While our position angle is somewhat larger than most previous studies, it is close to the value and uncertainty estimated by  \citet{Bower_2015}.

% One benefit of our parametrization for the scattering model is that these Gaussian parameters are decoupled from the parameters that characterize the turbulence on larger scales. Thus, as estimates of the Gaussian parameters at long wavelengths continue to be refined, the scattering model can immediately adopt updated values without simultaneously re-fitting for $\alpha$ and $r_{\rm in}$. 

{
\begin{deluxetable}{lccc}
% \tablewidth{0.5\columnwidth}
\tablecaption{Estimated Asymptotic Gaussian Scattering Parameters.}
\tablehead{
\colhead{Reference}  & \colhead{$\theta_{\rm maj, 0}$~(mas)} & \colhead{$\theta_{\rm min, 0}$~(mas)} & \colhead{P.A.~(deg)}
}
\startdata
\citet{Lo_1998}      & $1.430 \pm 0.020$ & $0.760 \pm 0.050$ & $80 \pm 3$\\
\citet{Shen_2005}    & $1.390 \pm 0.020$ & $0.690 \pm 0.060$ & $80$\\
\citet{Bower_2006}   & $1.309 \pm 0.015$ & $0.640^{+0.040}_{-0.050}$ & $78^{+0.8}_{-1.0}$\\
\citet{Lu_2011}      & $1.335 \pm 0.014$ & $0.817 \pm 0.042$ & ---\\
\citet{Psaltis_2015}\tablenotemark{a}\hspace{-0.25cm} & $1.320 \pm 0.040$ & $0.820 \pm 0.210$ & $77.8 \pm 9.7$ \\
\citet{Bower_2015}   & $1.320 \pm 0.020$ & $0.670 \pm 0.020$ & $81.8 \pm 0.2$ \\ 
This Work            & $1.380 \pm 0.013$ & $0.703 \pm 0.013$ & $81.9 \pm 0.2$
\enddata
\tablecomments{These parameters give the scattering kernel at the reference wavelength $\lambda_0 \equiv 1\,{\rm cm}$. }
% \tablenotetext{a}{\citet{Lu_2011} fitted a sample of historical measurements of \sgra\ in addition to their new measurements and .}
\tablenotetext{a}{Unlike the other entries in this table, \citet{Psaltis_2015} reanalyzed a sample of published Gaussian parameter fits rather than analyzing new or archival observations directly.}
\label{tab::GaussianParameters}
\end{deluxetable} 
}

\begin{figure*}[t]
\centering
\includegraphics[width=0.9\textwidth]{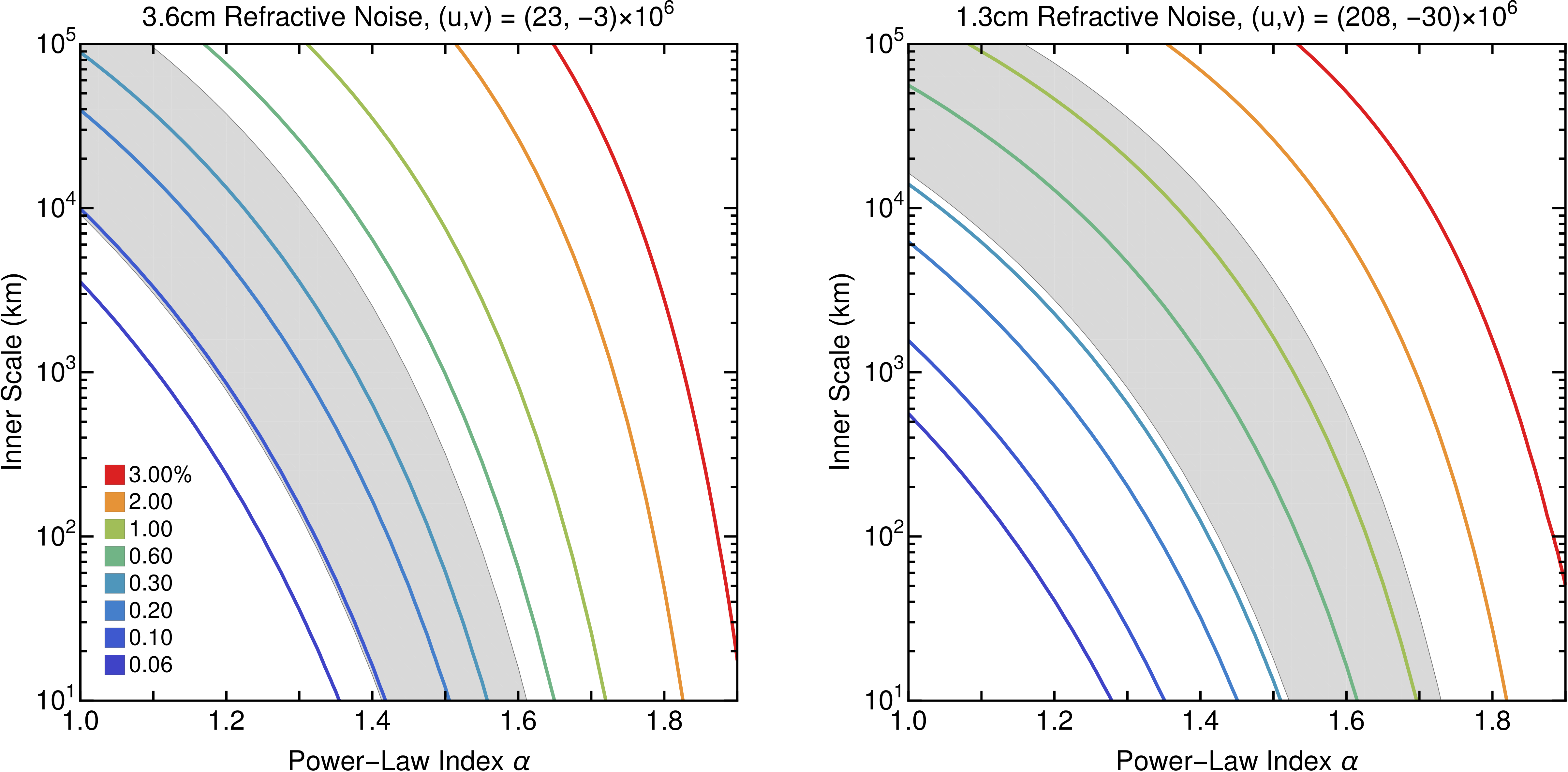}
\caption
{ 
(left) Expected amplitude of the renormalized refractive noise $\hat{\sigma}_{\rm ref}$ at $\lambda = 3.6\,{\rm cm}$ on the baseline $(u,v) = (23.4, -3.2) \times 10^6$ as a function of $\alpha$ and $r_{\rm in}$. Colored contours show predicted values of $\hat{\sigma}_{\rm ref}$; the gray shaded region shows the $95\%$ confidence range determined by the measured refractive noise (see \S\ref{sec::refractive_noise_constraint}). For the model values, we approximate the ensemble-average image size at each wavelength using our measured size. Thus, the assumed intrinsic size depends on $\alpha$ and $r_{\rm in}$ because the scattering kernel depends on them. (right) Amplitude of the refractive noise at $\lambda = 1.3\,{\rm cm}$ on the baseline $(u,v) = (207.6, -30.4) \times 10^6$ as a function of $\alpha$ and $r_{\rm in}$. The gray shaded region shows the $95\%$ confidence range determined by the measured refractive noise.
}
\label{fig::XKNoise}
\end{figure*}

\subsection{Constraining $\alpha$ and $r_{\rm in}$}
\label{sec::alpha_rin_constraints}

The remaining two parameters of our scattering model, $\alpha$ and $r_{\rm in}$, can be constrained in two ways: through a change in the scatter-broadening law from its asymptotic behavior at long wavelengths and through stochastic signatures of refractive scattering. For both types of constraints, the effects of $\alpha$ and $r_{\rm in}$ must be considered jointly; $\alpha$ will determine the asymptotic behavior at short wavelengths, but $r_{\rm in}$ determines the scale on which the scattering transitions between the two asymptotic regimes. Many previous efforts have constrained $\alpha$ by fitting the wavelength-dependence of scatter broadening to a power-law $\lambda^\beta$ or by quantifying Gaussianity of the scattered image \citep[e.g.,][]{Lo_1998,Bower_2004,Lu_2011}. However, these studies have implicitly assumed the limit $r_{\rm in} \rightarrow 0$, effectively fitting centimeter data to the properties of the scattering expected for the asymptotic regime $\lambda \rightarrow 0$. As we will demonstrate, jointly fitting the two parameters is imperative to derive meaningful parameter constraints for $\alpha$ and $r_{\rm in}$.

We will now derive a series of constraints $\alpha$ and $r_{\rm in}$. In \S\ref{sec::refractive_noise_constraint}, we derive constraints from the refractive noise on long baselines at 3.6 and 1.3\,cm. In \S\ref{sec::Image_Size_Fluctuations}, we determine constraints from the stringent limits on refractive fluctuations of the image size at 7\,mm. In \S\ref{sec::lambda_squared_constraint}, we derive constraints based on the $\lambda^2$ scaling of scatter broadening at centimeter wavelengths. In \S\ref{sec::gaussian_image_constraint}, we establish constraints based on Gaussianity of the scattered image at 1.3\,cm.  
While any of these individual constraints can only weakly constrain the parameter pair $(\alpha,r_{\rm in})$, the cumulative constraints are quite restrictive, summarized in Figure~\ref{fig::alpha_rin_limits}. We discuss these constraints and give our recommended characteristic values in \S\ref{sec::alpha_rin_characteristic}.

\subsubsection{Constraints from Refractive Noise at 3.6 and 1.3\,cm}
\label{sec::refractive_noise_constraint}

For a single long-baseline visibility measurement, the refractive noise is drawn from a circular Gaussian distribution. On baselines that heavily resolve the ensemble-average image, the visibility amplitude is then drawn from a Rayleigh distribution. However, the mean of this distribution is poorly constrained by a single measurement. Moreover, refractive noise among nearby baselines will be correlated, with a correlation length that is comparable to the length of baselines that begin to resolve the source \citep{Johnson_Narayan_2016}. Consequently, our long baselines at 3.6 and 1.3\,cm only sample a few independent realizations of the refractive noise. 

To combine measurements from multiple baselines, we adopted a simple procedure. First, we only examined visibilities that were reliably dominated by refractive noise, with negligible contribution from the ensemble-average structure. At 3.6\,cm, we used the cut $20 \times 10^6 < \left| \mathbf{u} \right| < 40 \times 10^6$, while at 1.3\,cm we used $150 \times 10^6 < \left| \mathbf{u} \right| < 300 \times 10^6$. Next, we performed an unweighted scalar average of the noise-debiased visibilities on these baselines. We use this average as an approximation to the mean renormalized refractive noise on the vector average of the baselines, for which we expect $\left \langle \left| V(\mathbf{b}) \right| \right \rangle = \frac{\sqrt{\pi}}{2} \hat{\sigma}_{\rm ref}(\mathbf{b}) \approx 0.89 \hat{\sigma}_{\rm ref}(\mathbf{b})$. The average baseline was $(u,v) = (23.4, -3.2) \times 10^6$ at 3.6\,cm and $(u,v) = (207.6, -30.4) \times 10^6$ at 1.3\,cm. 

In this way, we obtained an estimate of $\hat{\sigma}_{\rm ref}$ on a single baseline at each wavelength. This simplification facilitates direct comparisons with predictions for $\hat{\sigma}_{\rm ref}$ from a scattering model. To validate this reduction and determine a confidence interval for our refractive noise estimates, we generated 1000 simulated images of the scattering at both wavelengths. For each image, we calculated the visibilities on all the long baselines for the 3.6 and 1.3\,cm observations and computed the scalar average of the visibility amplitudes. At 3.6\,cm, the mean amplitude of the sampled refractive noise (averaged over all the long baselines and the multiple simulations) was within 10\% of the mean amplitude for refractive noise of the average baseline. For averaged visibility amplitudes for individual image realizations, 95\% of values fell between $0.40$ and $1.90$ times the expected mean value for the fixed baseline. For draws of a Rayleigh distributed random variable, the middle 95\% of samples will extend to $0.18$ and $2.2$ times the mean. Thus, our simple averaging scheme significantly tightens the bounds on $\hat{\sigma}_{\rm ref}$ by combining multiple correlated measurements. At 1.3\,cm, the mean amplitude of the refractive noise averaged over long baselines was within 0.1\% of the mean value on the average baseline. The middle 95\% of samples fell within the range of $0.45$ to $1.70$ times the expected mean noise amplitude on the fixed baseline. 

With this approach, we thereby estimate 95\% confidence intervals for $\sigma_{\rm ref}$ of $[0.096\%, 0.45\%]$ at $(u,v) = (23.4, -3.2) \times 10^6$ for 3.6\,cm, and $[0.32\%, 1.2\%]$ at $(u,v) = (207.6, -30.4) \times 10^6$ for 1.3\,cm (both are expressed as a fraction of the total flux density). Figure~\ref{fig::XKNoise} shows the expected values for $\sigma_{\rm ref}$ at both wavelengths as a function of $\alpha$ and $r_{\rm in}$; the gray shaded regions show the 95\% confidence intervals for $\alpha$ and $r_{\rm in}$ based on the refractive noise measurements at both wavelengths. 

% Notably, these measurements do not significantly constrain either $\alpha$ or $r_{\rm in}$, but they do narrowly constrain the pair.  

\subsubsection{Constraints from Refractive Fluctuations of the Image Size}
\label{sec::Image_Size_Fluctuations}

Refractive scattering causes variations in the observed angular size of a source \citep{Blandford_Narayan_1985}. For observations that span many refractive timescales, the observed level of variability can then be used to constrain the scattering model. Because the intrinsic source may also be time-variable, measurements of the image size variability can only give an upper limit for the variations attributable to scattering. As with other refractive effects, fluctuations of image size will increase with increasing $\alpha$ and $r_{\rm in}$. 

For \sgra, the most stringent constraints on image size fluctuations come from observations at $\lambda=7\,{\rm mm}$. Even without accounting for refractive noise, observations over the past ${\sim}$20 years consistently find a major axis size in the range of $680 - 750~\mu{\rm as}$ \citep[e.g.,][]{Krichbaum_1993,Backer_1993,Lo_1998,Bower_2004,Lu_2011,Akiyama_2014,Bower_2014b,Bower_2015,Zhao_2017}. A uniform distribution over the entire range $680 - 750~\mu{\rm as}$ has a standard deviation of $20.2\,\mu{\rm as}$, or fractional variations of $2.8\%$. Note that this range is inflated by measurement uncertainties in the reported sizes (in addition to scattering and intrinsic variability). Thus, we estimate that the fractional scatter of the major axis size at $\lambda=7\,{\rm mm}$ from refractive distortion is certainly less than $3\%$. 

We can compare this limit to the expected refractive fluctuations, which can be computed semi-analytically via the framework for renormalized refractive noise developed in the Appendix. 
Namely, on short baselines, the renormalized refractive noise is dominated by refractive fluctuations in image size. For a short baseline $\mathbf{u}$, the renormalized visibility (i.e., the visibility after normalizing the total flux density and centering the image) is given by
\begin{align}
\hat{V}(\mathbf{u}) = 1 - \frac{\pi^2}{4 \ln2} \theta_{\parallel}^2 u^2 + \mathcal{O}\left( \theta_{\parallel}^3 u^3 \right),
\end{align}
where $\theta_{\parallel}$ is the source size projected along the baseline direction (see Eq.~\ref{eq::VCZ} and \ref{eq::fwhm_maj}). Because of scattering, the instantaneous source size will not match the ensemble-average value, $\left\langle \theta_{\parallel} \right\rangle$; this discrepancy is what produces renormalized refractive noise $\hat{\sigma}_{\rm ref}(\mathbf{u})$ on short baselines. Explicitly,
\begin{align}
\nonumber \hat{\sigma}_{\rm ref}(\mathbf{u}) &= \sqrt{\left \langle \Delta \hat{V}(\mathbf{u})^2 \right \rangle}\\
\nonumber &\approx \frac{\pi^2}{4 \ln2} u^2 \sqrt{\left( \theta_{\parallel}^2 - \left \langle \theta_{\parallel} \right \rangle^2   \right)^2} \\
\nonumber &\approx \frac{\pi^2}{2 \ln2} u^2 \left \langle \theta_{\parallel} \right \rangle \sqrt{ \left \langle \Delta \theta_{\parallel}^2 \right \rangle} \\
\Rightarrow \dfrac{\sqrt{\langle \Delta \theta_{\parallel}^2 \rangle}}{\left \langle \theta_{\parallel} \right \rangle} &\approx \frac{2\ln 2}{\pi^2} \frac{\hat{\sigma}_{\rm ref}(\mathbf{u})}{u^2  \left \langle \theta_{\parallel} \right \rangle^2}.
\end{align}
% 
% 
% Thus, for a short baseline $\mathbf{u}$, the rms fluctuation of source size projected along the baseline direction ($\theta_{\parallel}$) is given by (N.B., compare with Eq.~\ref{eq::fwhm_maj})
% \begin{align}
% \sqrt{\left \langle \Delta \theta_{\parallel}^2 \right \rangle} &\approx \frac{2\ln 2}{\pi^2} \frac{\hat{\sigma}(\mathbf{u})}{u^2 \theta_{\parallel}^2} \left \langle \theta_{\parallel} \right \rangle.
% \end{align}
The red lines in the right panel of Figure~\ref{fig::alpha_rin_limits} show contours for the values of $\alpha$ and $r_{\rm in}$ that would produce $1\%$, $3\%$, and $5\%$ fractional fluctuations of the major axis size at $\lambda=7\,{\rm mm}$. For these calculations, we hold the ensemble-average image size fixed (approximating it by our measured size), so the intrinsic size is also a function of these parameters because the scattering kernel depends on them. The requirement that the fluctuations are smaller than 3\% then gives an $\alpha$-dependent upper-limit on $r_{\rm in}$. 

Observe that the shapes of the image fluctuation contours are very similar to those of the refractive noise at 1.3\,cm on the fixed baseline $\mathbf{u} = (207.6, -30.4) \times 10^6$. This similarity is expected because both effects are dominated by scattering modes on the same angular scale. Specifically, at 7\,mm, the dominant modes for image distortion are on the scale of the image size, $\theta_{\rm maj} \approx 0.7\,{\rm mas}$. At 1.3\,cm, the dominant modes for refractive noise are those matched to the angular resolution of the long baselines, $1/|\mathbf{u}| \sim 1\,{\rm mas}$.

\subsubsection{Constraints from the $\lambda^2$ Scaling of Scattered Size}
\label{sec::lambda_squared_constraint}

The constant scaling of image size as $\theta \propto \lambda^2$, stable image anisotropy, and constant position angle at wavelengths $\lambda \gsim 1.3\,{\rm cm}$ strongly argues against a departure of the angular broadening from the asymptotic $\lambda^2$ law in this interval, as that would require intrinsic structure to fortuitously offset the change in angular broadening. Likewise, these properties argue against intrinsic structure being significant at these wavelengths. This plausibility argument gives a lower bound on the inner scale because the angular broadening asymptotes to $\theta \propto \lambda^{1 + \frac{2}{\alpha}}$ as $\lambda \rightarrow 0$, with the transition when the diffractive scale becomes larger than the inner scale. 

The diffractive scale is larger at shorter wavelengths, so our most stringent constraints on $r_{\rm in}$ come from the shortest wavelengths that exhibit the $\lambda^2$ law. We have found close agreement with the $\lambda^2$ law across the observing bandwidth at 1.3\,cm (see Figure~\ref{fig::Kband_MultiIF}), so the inner scale must exceed the diffractive scale at 1.3\,cm: $r_{\rm in} \gsim 300~{\rm km}$. The limit is slightly higher for lower values of $\alpha$ because they asymptotically give a stronger departure from $\lambda^2$ scaling. The limit is slightly higher for the minor axis than for the major axis because the former has a larger diffractive scale. 

Blue lines in the right panel of Figure~\ref{fig::alpha_rin_limits} show $\alpha$-dependent lower limits on $r_{\rm in}$ using the simple condition that the $\lambda=1.3\,{\rm cm}$ angular broadening cannot be more than 5\% smaller than the value extrapolated from $\lambda \rightarrow \infty$ with a pure $\lambda^2$ scaling (i.e., the limit as $r_{\rm in} \rightarrow \infty$). Requiring that the scaling across the full $\lambda=1.3\,{\rm cm}$ bandwidth match a $\lambda^2$ law to within the uncertainties shown in Figure~\ref{fig::Kband_MultiIF} gives a similar constraint.

\subsubsection{Constraints from the Image Gaussianity}
\label{sec::gaussian_image_constraint}

We can also constrain $\alpha$ and $r_{\rm in}$ from the shape of the scatter-broadened image at a fixed wavelength. At long wavelengths, the scatter-broadening is Gaussian and the visibility function falls as $e^{-u^2}$, while at short wavelengths the visibility function falls as $e^{-u^\alpha}$. As in \S\ref{sec::lambda_squared_constraint}, this constraint is really a plausibility argument; the intrinsic source could fortuitously offset any change in the angular broadening function to produce a Gaussian image despite non-Gaussian scattering, and non-Gaussian source structure could mimic the behavior of a non-Gaussian scattering kernel. Thus, we focus these tests on our 1.3\,cm and 7\,mm observations, where the baseline coverage is excellent and source structure is subdominant to scatter broadening. 

Once again, the transition between the two scaling regimes depends on the inner scale. Specifically, the scattering kernel will depart from a Gaussian for baselines with physical lengths $b \gsim (1+M) r_{\rm in}$, where $M \approx 0.53$ for \sgra\ (see \S\ref{sec::Scattering_Geometry}). At 1.3\,cm, the longest baselines that are not dominated by refractive noise are ${\sim}100\,{\rm M}\lambda \approx 1300\,{\rm km}$, so these observations can, in principle, constrain $r_{\rm in}$ to be greater than ${\sim}800~{\rm km}$. The lower limit is expected to increase with decreasing $\alpha$ because of a sharper deviation from the Gaussian kernel with decreasing $\alpha$. 

\begin{figure}[t]
\centering
\includegraphics[width=\columnwidth]{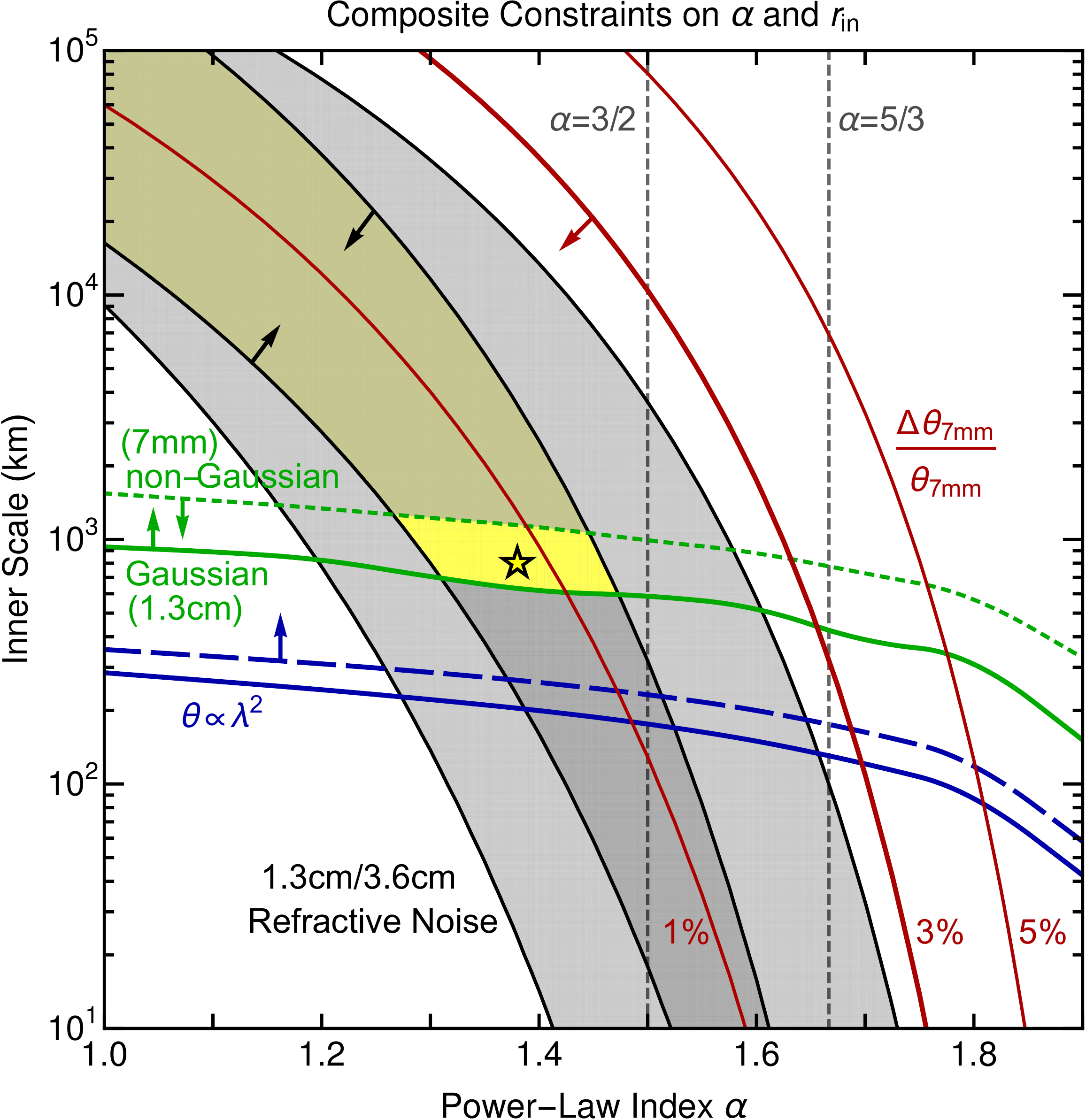}
\caption
{ 
Composite constraints on $\alpha$ and $r_{\rm in}$. Blue lines show $\alpha$-dependent \emph{lower} limits on $r_{\rm in}$ from the $\lambda^2$ scaling of image size (major axis: solid, minor axis: dashed; see \S\ref{sec::lambda_squared_constraint}). The solid green line shows the \emph{lower} limit on $r_{\rm in}$ from the measured Gaussian image shape at 1.3\,cm, while the dotted green line shows the (more tentative) \emph{upper} limit on $r_{\rm in}$ from the measured non-Gaussian image shape at 7\,mm (see \S\ref{sec::gaussian_image_constraint}). Red lines show contours of 1\%, 3\%, and 5\% for rms fluctuations of the major axis size at $\lambda=7\,{\rm mm}$; these fluctuations are constrained to be less than 3\%, giving an $\alpha$-dependent \emph{upper} limit on $r_{\rm in}$ (see \S\ref{sec::Image_Size_Fluctuations}). Combining these constraints, the yellow shaded region shows the plausible range of $\alpha$ and $r_{\rm in}$; the darker yellow region shows the range without including the non-Gaussian measurement at 7mm. Overall, we find that $\alpha \lsim 1.47$ and $r_{\rm in} \gsim 600~{\rm km}$. The star marks our recommended characteristic values: $\alpha=1.38$ and $r_{\rm in}=800\,{\rm km}$ (see \S\ref{sec::alpha_rin_characteristic}).
}
\label{fig::alpha_rin_limits}
\end{figure}

To derive constraints on $\alpha$ and $r_{\rm in}$ using image Gaussianity tests, we fit our 1.3\,cm and 7\,mm observations with the full, non-Gaussian kernel of our scattering model. For each case, we included refractive noise in the error budget as we did for Gaussian fits. For the 1.3\,cm fits, we used a point-source model for the intrinsic structure. This procedure then quantifies the baseline length at which visibilities become inconsistent with a Gaussian curve; this break is insensitive to the distinction between intrinsic structure and scattering because of the convolution action of scattering in the visibility domain. We found the best fits to the 1.3\,cm data were those with $r_{\rm in} \rightarrow \infty$ (giving a perfectly Gaussian image). Thus, the fits give an $\alpha$-dependent \emph{lower} limit for $r_{\rm in}$. The solid green curve shown in Figure~\ref{fig::alpha_rin_limits} corresponds to the values with an increase of 4 in the total chi-squared of the fitted model, corresponding to a $2\sigma$ confidence contour. These limits range from $r_{\rm in} \gsim 520\,{\rm km}$ for $\alpha=1.6$ to $r_{\rm in} \gsim 930\,{\rm km}$ for $\alpha=1.0$, in line with expectations from the simple calculation in the previous paragraph.  

For the 7\,mm data, intrinsic structure is non-negligible, so we fixed the Gaussian scattering parameters to the estimates from \S\ref{sec::Gauss_Constraints} and then fit for the three parameters of an anisotropic intrinsic Gaussian source along with $\alpha$ and $r_{\rm in}$. These fits showed strong indication of non-Gaussian structure, with an increase in total chi-squared for a purely Gaussian model of 19.2 relative to the best fitting models with a finite inner scale (i.e., a ${\gsim}4\sigma$ preference for a non-Gaussian image). The fits then provide an $\alpha$-dependent \emph{upper} limit for $r_{\rm in}$. This test must be interpreted with caution, as the intrinsic source structure is non-negligible at this wavelength and may be non-Gaussian, although it is expected to be Gaussian on baselines that do not significantly resolve the intrinsic source (see \S\ref{sec::Gaussian_Model_Assumption}). Because the KaVA baselines only modestly resolve the scattered source, this assumption is likely acceptable. Nevertheless, we still use a slightly higher threshold for these results relative to those at 1.3\,cm because the plausibility argument is weaker. Thus, the dotted green curve shown in Figure~\ref{fig::alpha_rin_limits} corresponds to the values with an increase of 9 in the total chi-squared of the fitted model relative to the best-fit model, corresponding to a $3\sigma$ confidence contour. However, when the residual gain priors on the a priori calibration are unconstrained, the significance of the finite inner scale is only ${\approx}\,1\sigma$. Thus, we regard this detection of visibility amplitude non-Gaussianity and the corresponding upper limit on $r_{\rm in}$ as tentative.

\begin{figure*}[ht]
\centering
\includegraphics[width=1.0\textwidth]{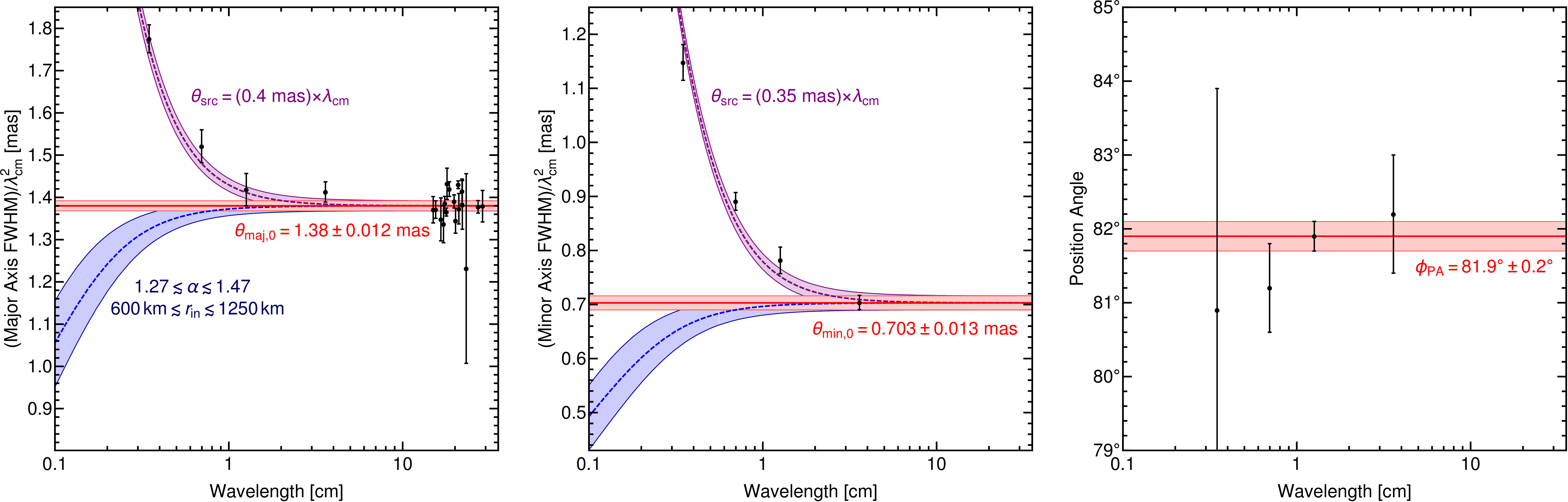}
\caption
{ 
Gaussian parameters of the scattered image of \sgra\ as a function of wavelength. Major and minor axes sizes are normalized by $\lambda^{-2}$. Black points show our measurements with $\pm1\sigma$ uncertainties (Table~\ref{tab::All_Fits}). The red region shows our fitted asymptotic parameters $\theta_{\rm maj,0}$, $\theta_{\rm min,0}$, and $\phi_{\rm PA}$ with their respective uncertainties. The blue region shows the plausible range of values for the scattering kernel based on our constraints on all scattering parameters, including $\alpha$ and $r_{\rm in}$ (see Figure~\ref{fig::alpha_rin_limits}). The purple region shows the size corresponding to a simple source model, with $\theta_{\rm src} \propto \lambda$, added at quadrature to the scattering law, with the associated scattering model uncertainty. The blue and purple dashed curves show the results corresponding to our recommended characteristic values for $\alpha$ and $r_{\rm in}$. 
}
\label{fig::Multifreq_Constraints}
\end{figure*}

\begin{figure*}[ht]
\centering
\includegraphics[width=0.95\textwidth]{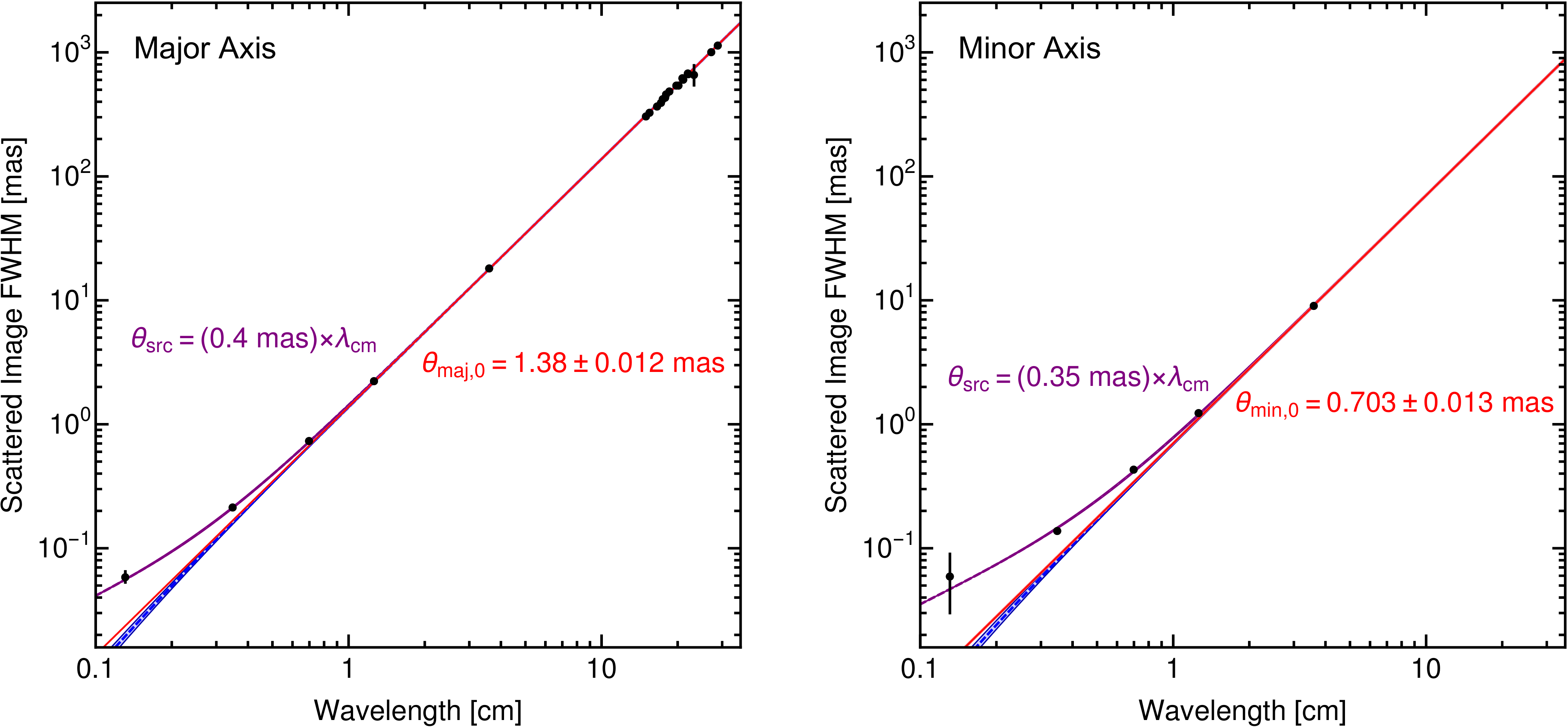}
\caption
{ 
Same as Figure~\ref{fig::Multifreq_Constraints}, but without normalizing sizes by $\lambda^{-2}$. 
}
\label{fig::Multifreq_Constraints_2}
\end{figure*}

\begin{figure*}[th]
\centering
\includegraphics[width=1.0\textwidth]{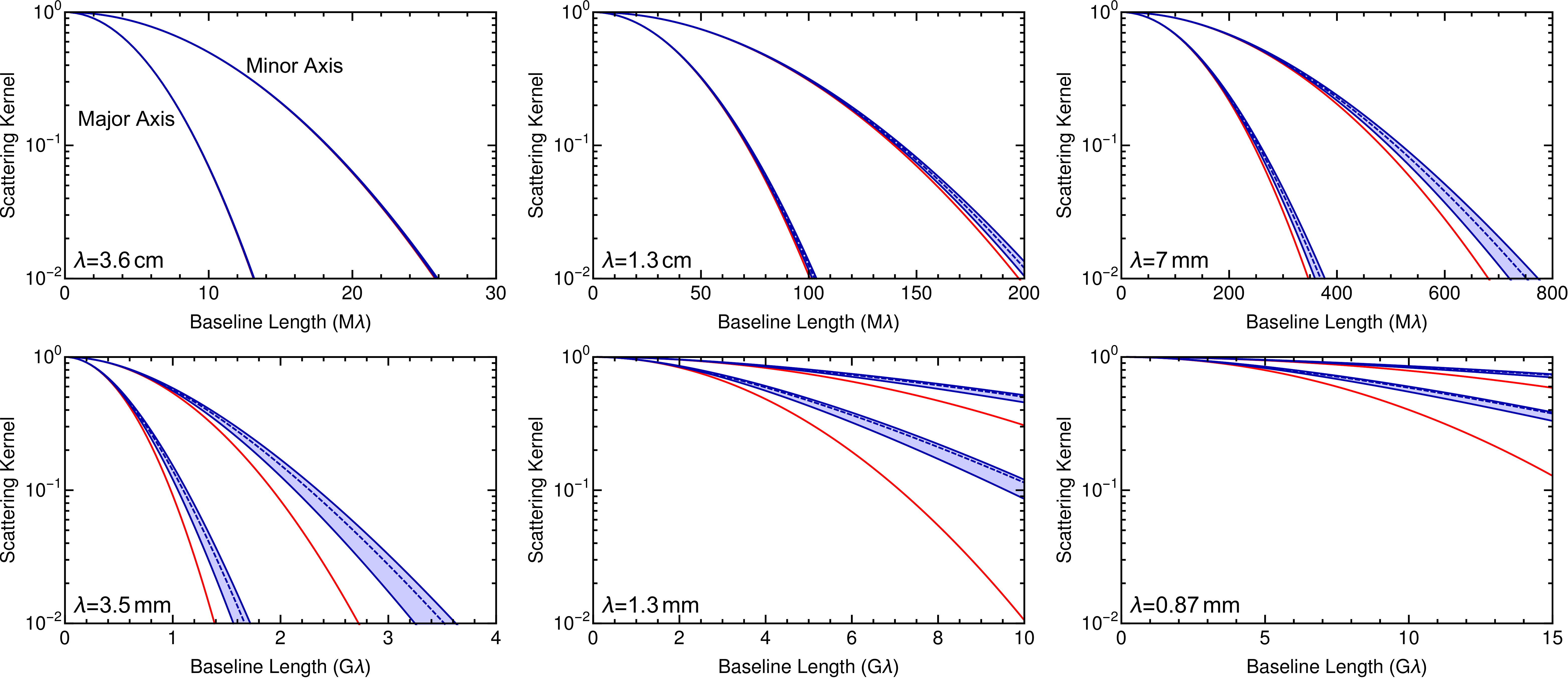}\\
\caption
{ 
The scattering kernel of \sgra\ as a function of baseline length along the major and minor axes at representative wavelengths. For each panel, the red curves correspond to a Gaussian kernel with our measured scattering parameters, the blue shaded region shows the plausible range of values based on our constraints on the scattering model, and the dashed blue line shows the kernel corresponding to our recommended characteristic values for the scattering parameters. For baselines much longer than the inner scale, the kernel is non-Gaussian, falling as $e^{-b^\alpha}$ rather than $e^{-b^2}$. For the EHT, which has baselines extending to ${\sim}6\,{\rm G}\lambda$ along the major axis and to ${\sim}9\,{\rm G}\lambda$ along the minor axis at 1.3\,mm, the expected kernel differs significantly from the Gaussian prediction, and we expect that the effects of scattering are substantially weaker than have been assumed. Because of its intrinsic structure, \sgra\ is heavily resolved for baselines of $3-6\,{\rm G}\lambda$ \citep[e.g.,][]{Lu_2018}, the remaining kernel uncertainties may have minimal effects for imaging at 1.3 and 0.87\,mm (see, e.g., Figure~\ref{fig::Multifreq_Constraints_2}). 
}
\label{fig::Scattering_Kernel}
\end{figure*}

\subsubsection{Recommended Characteristic Values for $\alpha$ and $r_{\rm in}$}
\label{sec::alpha_rin_characteristic}

% The constraints in \S\ref{sec::lambda_squared_constraint}-\ref{sec::Image_Size_Fluctuations} significantly narrow the plausible parameter space for $\alpha$ and $r_{\rm in}$. We now give some additional considerations to motivate specific characteristic values and uncertainties for each.

For our recommended characteristic value for the inner scale, we adopt $r_{\rm in} = 800\,{\pm}\,200\,{\rm km}$, based on the combined image Gaussianity tests at 1.3\,cm and 7\,mm and the refractive noise constraints for 3.6 and 1.3\,cm. However, while the lower limit at ${\sim}600\,{\rm km}$ is rather firm, we regard the upper limit as somewhat tentative without confirmation from additional 7\,mm observations and from tests at other wavelengths. 

To obtain a characteristic value for $\alpha$, we then use the joint likelihood function of the 1.3 and 3.6\,cm refractive noise with $r_{\rm in}$ as estimated above. As in \S\ref{sec::refractive_noise_constraint}, we used 1000 scattering realizations at both wavelengths. For each realization, we sampled and then averaged visibilities on the long baselines, following our procedure for the data. This sample then provides an estimate for the likelihood function for $\hat{\sigma}$ on the characteristic baseline at each wavelength. The joint likelihood function then gives $\alpha = 1.38^{+0.08}_{-0.04}$, where $\alpha=1.38$ is the maximum likelihood estimate.

{
\begin{deluxetable}{lc}
\tablecaption{Estimated Scattering Model for \sgra.}
 \tablehead{
\colhead{Parameter} & \colhead{Estimate}
 }
\startdata
\sidehead{\bf Geometrical Parameters} 
Scattering Screen Magnification & $M = D/R = 0.53 \pm 0.08$\vspace{0.1cm}\\
Earth-Scattering Distance & $D = 2.7 \pm 0.3\,{\rm kpc}$\vspace{0.1cm}\\
\sgra-Scattering Distance & $R = 5.4 \pm 0.3\,{\rm kpc}$\\
\sidehead{\bf Scattering Parameters} 
Reference Wavelength     & $\lambda_0 \equiv 1\,{\rm cm}$\vspace{0.1cm}\\
Gaussian Major Axis FWHM & $\theta_{\rm maj, 0} = 1.380 \pm 0.013\ {\rm mas}$\vspace{0.1cm}\\
Gaussian Minor Axis FWHM & $\theta_{\rm min, 0} = 0.703 \pm 0.013\ {\rm mas}$\vspace{0.1cm}\\
Gaussian Position Angle  & $\phi_{\rm PA}       = 81.9^\circ \pm 0.2^\circ$\vspace{0.1cm}\\
Power-Law Index of $D_{\phi}(\mathbf{r})$          & $\alpha = 1.38^{+0.08}_{-0.04}$\vspace{0.1cm}\\
Power-Law Index of $Q(\mathbf{q})$ and $P_{n_{\rm e}}(\mathbf{q})$     & $\beta = \alpha+2 = 3.38^{+0.08}_{-0.04}$\vspace{0.1cm}\\
Inner Scale              & $r_{\rm in} = 800 \pm 200\,{\rm km}$\\ 
\sidehead{\bf Scattering Transitions}
Gaussian-Inertial Kernel Transition   & $\lambda=5\,{\rm mm}$\vspace{0.1cm}\\
Weak-Strong Scattering Transition   & $\lambda=0.2\,{\rm mm}$
\enddata
\tablecomments{Values and uncertainties for $\alpha$, $\beta$, and $r_{\rm in}$ use our tentative upper-limit from non-Gaussianity at 7\,mm (see Figure~\ref{fig::alpha_rin_limits}). We define the Gaussian-inertial kernel transition as the wavelength for which the diffractive scale and inner scale are equal. At significantly longer wavelengths, the scattering kernel will be Gaussian (determined by the 3 asymptotic Gaussian parameters); at significantly shorter wavelengths, the shape will be determined by $\alpha$. The weak-strong transition is the wavelength for which the diffractive scale and refractive scale are equal. For Kolmogorov turbulence, $\alpha = 5/3$ and $\beta = 11/3$.}
\label{tab::ScatteringParameters}
\end{deluxetable} 
}

\vspace{0.2cm}

\subsection{The Scattering Kernel and Intrinsic Structure of Sgr A*}
\label{sec::Scattering_Intrinsic}

Using the scattering model determined in \S\ref{sec::Gauss_Constraints}-\ref{sec::alpha_rin_constraints}, we now explore the expected scattering properties for \sgra\ and estimate its wavelength-dependent intrinsic size. Table~\ref{tab::ScatteringParameters} summarizes our estimates for the parameters of this scattering model and provide additional derived quantities. 

Figures~\ref{fig::Multifreq_Constraints} and \ref{fig::Multifreq_Constraints_2} show our estimated major axis FWHM, minor axis FWHM, and position angle as a function of wavelength (these values are given in Table~\ref{tab::All_Fits}). 
After normalizing by $\lambda^{-2}$, these sizes show a significant increase with decreasing wavelength for $\lambda \lsim 1.3\,{\rm cm}$. Because $\alpha<2$, the scattering law can only become steeper than $\lambda^2$ at short wavelengths. Thus, this increase in normalized size robustly indicates intrinsic structure at millimeter wavelengths. 

Figures~\ref{fig::Multifreq_Constraints} and \ref{fig::Multifreq_Constraints_2} also show the estimated FWHM of the scattering kernel, including the uncertainty spanned by the plausible range of $\alpha$ and $r_{\rm in}$ (see Figure~\ref{fig::alpha_rin_limits}). 
For these estimates, we do not define the FWHM using the second derivative of the visibility amplitude on a zero-baseline (as in \S\ref{sec::Gaussian_Model_Assumption}). We instead identify the baseline length at which the scattering kernel falls to half, $\exp\left[ -\frac{1}{2}D_\phi\left(\mathbf{u}_{1/2}/(1+M)\right) \right] \equiv 1/2$, and then derive a representative image FWHM based on the relationship for a Gaussian image: $\theta_{\rm FWHM} = \frac{2\ln 2}{\pi u_{1/2}}$. 
%We choose this alternate (but still observationally motivated) definition because the second derivative of the kernel at zero baseline is independent of the inner scale even though the shape of the scattering kernel strongly depends on it (see Figure~\ref{fig::Scattering_Kernel}). 
The kernel uncertainties become larger at shorter wavelengths because of our limited constraints on $\alpha$ and $r_{\rm in}$. At 1.3\,mm, the uncertainty is ${\sim}20\%$ on both the major and minor axis FWHM.  

Figure~\ref{fig::Scattering_Kernel} shows our estimates of the visibility domain scattering kernel and its uncertainties at six representative wavelengths. This kernel is required to ``deblur'' measurements of \sgra, so it is fundamental to scattering mitigation strategies \citep{Fish_2014,Stochastic_Optics}. 
At millimeter wavelengths, kernel differs significantly from the prediction of the simple Gaussian/$\lambda^2$ model, and the remaining uncertainty in the scattering kernel on long baselines is significant, primarily because of uncertainties on $\alpha$ and $r_{\rm in}$. Thus, we expect that the blurring effects of scattering are significantly weaker than have been assumed for \sgra. 

\begin{figure}[t]
\centering
\includegraphics[width=\columnwidth]{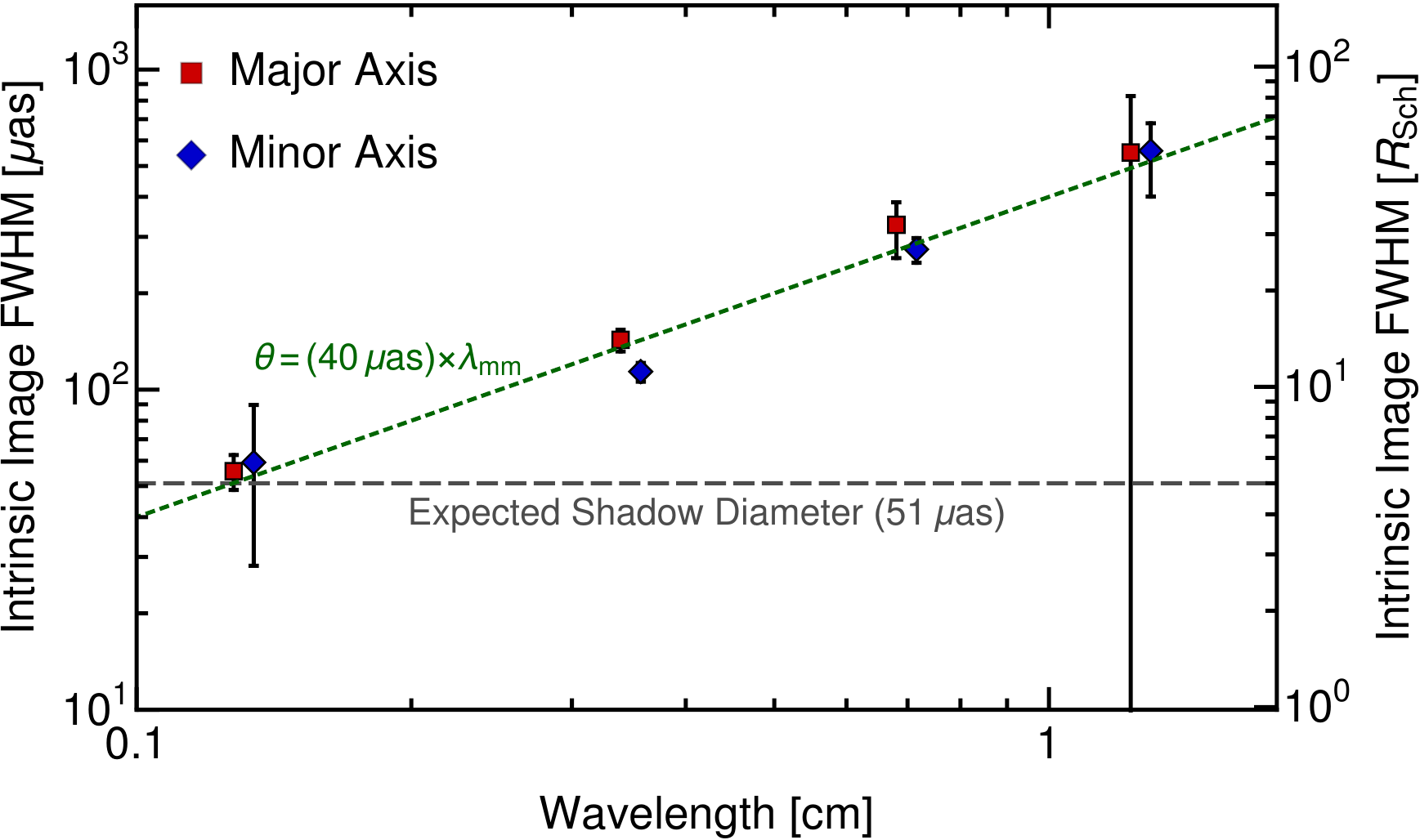}\\
\caption
{ 
Estimated intrinsic size of \sgra\ along the directions of the major and minor axes of the scattering kernel. The error bars denote the range of values accounting for uncertainties in the full scattering model and for $\pm 1\sigma$ measurement uncertainty (see Table~\ref{tab::All_Fits}). Diamonds show central values for measured parameters and they use our recommended characteristic values for $\alpha$ and $r_{\rm in}$. The major/minor axis markers are offset slightly in wavelength, for visual clarity. We also show the expected diameter of the black hole ``shadow'' and plot a simple isotropic source model with size directly proportional to wavelength. We do not find evidence for significant intrinsic anisotropy at any wavelength or for a steep scaling of intrinsic size with wavelength. 
}
\label{fig::Intrinsic_Size}
\end{figure}

With the estimated size of the scattering kernel in place, we can deconvolve the scattering from the observed Gaussian size to estimate an intrinsic FWHM at each wavelength.  Table~\ref{tab::Intrinsic_Size} and Figure~\ref{fig::Intrinsic_Size} show the estimated intrinsic size along the major and minor axes of the scattering kernel as a function of wavelength. The plotted uncertainties account for uncertainties in our estimates of the scattered size (from thermal noise, systematic effects, and refractive distortion) and in the estimated scattering kernel. The inferred intrinsic size is nearly isotropic and scales approximately as $\theta \propto \lambda$. The largest intrinsic anisotropies are at 3.5\,mm (up to ${\sim}1.3{:}1$) and 1.3\,mm (up to ${\sim}2{:}1$, but poorly constrained). 
Because the total flux density $I_0$ of \sgra\ rises with frequency over this range \citep[see, e.g.,][]{Lu_2011,Bower_2015_spectrum}, the brightness temperature $T_{\rm b} \propto \lambda^2 \theta_{\rm src}^{-2} I_0$ also rises. For instance, using $\theta_{\rm src} \sim (40\,\mu{\rm as}) \times \lambda_{\rm mm}$, we estimate $T_{\rm b} \sim 1.1 \times 10^{10}\,{\rm K}$ at 1.3\,cm and $T_{\rm b} \sim 3.1 \times 10^{10}\,{\rm K}$ at 1.3\,mm.

{
\begin{deluxetable}{lcccc}
\tablecaption{Estimated Intrinsic Size of \sgra.}
\tablehead{
\colhead{$\lambda$ (cm)}  & \multicolumn{2}{c}{$\theta_{\rm maj}$ ($\mu{\rm as}$)} & \multicolumn{2}{c}{$\theta_{\rm min}$ ($\mu{\rm as}$)} \\
 & \colhead{Char.} & \colhead{Plausible} & \colhead{Char.} & \colhead{Plausible}
}
\startdata
3.598 & $4000^{+1200}_{-1900}$ & $4000^{+1800}_{-4000}$ & --- & ---\vspace{0.1cm}\\
1.261 & $550^{+210}_{-370}$ & $550^{+280}_{-550}$ & $560^{+80}_{-90}$ & $560^{+120}_{-160}$\vspace{0.1cm}\\
0.698 & $327^{+41}_{-46}$ & $327^{+58}_{-69}$ & $274^{+12}_{-13}$ & $274^{+23}_{-25}$\vspace{0.1cm}\\
0.348 & $143^{+6}_{-6}$ & $143^{+11}_{-12}$ & $114^{+5}_{-5}$ & $114^{+7}_{-8}$\vspace{0.1cm}\\
0.131 & $56^{+6}_{-6}$ & $56^{+7}_{-7}$ & $59^{+30}_{-31}$ & $59^{+30}_{-31}$
\enddata
\label{tab::Intrinsic_Size}
\tablecomments{These estimates of intrinsic FWHM correspond to directions along the major and minor axes of the scattering kernel. We give uncertainties for both the assumed characteristic scattering parameters (which only account for $\pm1\sigma$ measurement uncertainties) and for the full range of plausible scattering parameters (which account for remaining uncertainties in the scattering kernel). We omit the estimated intrinsic minor axis at 3.6\,cm because the minor axis scattering was estimated from this measurement (see \S\ref{sec::Gauss_Constraints}). }
\end{deluxetable} 
}

\ \\ \ \\

\section{Discussion}
\label{sec::Discussion}

\subsection{The Intrinsic Structure of Sgr A*}

Our estimates for the approximately linear wavelength dependence of intrinsic size are typical for stratified emission in synchrotron self-absorbed systems \citep[e.g.,][]{Blandford_Konigl_1979,Falcke_Markoff_2000,Davelaar_2018}, and they are plausible for both disk- and jet-dominated models for the radio emission of \sgra. However, the lack of asymmetry in the inferred intrinsic size and the stable position angle of the scattered image both argue against intrinsic structure that is highly asymmetric for $3.5\,{\rm mm} \lsim \lambda \lsim 1.3\,{\rm cm}$. For instance, the model of \citet{Falcke_Markoff_2000} predicts an image asymmetry of roughly $4{:}1$, and recent GRMHD simulations of jets show asymmetry of ${\sim}2{:}1$ at 7\,mm \citep{Davelaar_2018}. The intrinsic size of \sgra\ we find is qualitatively consistent with RIAF models \citep[e.g.,][]{Ozel_2000,Yuan_2003,Yuan_Narayan_2014}, which are more plausible for producing a nearly isotropic image at wavelengths as long as 1.3\,cm. Recent GRRMHD simulations show good agreement with our estimated size at 1.3\,mm and also show a similar size trend, but they generally underpredict the size at 7\,mm and 1.3\,cm \citep{Chael_2018}, perhaps highlighting the contribution from a non-thermal population of electrons.  

Note that there has not been consistency in how image FWHM from simulations is defined. For comparison with Gaussian image sizes reported here and elsewhere, simulations should compute the image FWHM from the second moment along principal axes of the image brightness distribution (see \S\ref{sec::Gaussian_Model_Assumption}). While some simulation papers have adopted this convention for comparisons \citep[e.g.,][]{Moscibrodzka_2009,Moscibrodzka_2012,Davelaar_2018,Chael_2018}, others have developed ad hoc definitions for the reported image size \citep[e.g.,][]{Ozel_2000,Falcke_Markoff_2000,Psaltis_2015b,Chan_2015} or do not state their procedure for estimating the size. An alternative is to fit or compare simulations directly to measured interferometric visibilities \citep[e.g.,][]{Broderick_2009,Dexter_2010,Pu_2016,Kim_2016,Broderick_2016,Gold_2016}.

\subsection{Implications for Interstellar Scattering}

We now reevaluate our assumptions for the scattering of \sgra, and we discuss implications of our findings.

\subsubsection{The Outer Scale of Turbulence}
\label{sec::Outer_Scale}

All of our calculations and model fits have assumed that the outer scale of turbulence is effectively infinite. We now evaluate this assumption a posteriori. In particular, \citet{Goldreich_2006} estimated that $r_{\rm out} \lsim 10^{11}\,{\rm cm} \times \left( \frac{R}{130\,{\rm pc}} \right)^{5/2} \times \left(\frac{T}{10^4\,{\rm K}} \right)^{3/4}$. For the previously assumed value of $R = 130\,{\rm pc}$ \citep{Lazio_Cordes_1998}, they noted that this scale is unacceptably small, as it produces too much heating and it does not correspond to a reasonable astronomical scale for nonlinear density fluctuations. In addition to these objections, our measurements of refractive noise give a lower bound for the outer scale because refractive noise will be suppressed on angular scales larger than ${\sim}\,r_{\rm out}/D$. Thus, our measurements of refractive noise on baselines with $|\mathbf{u}| \sim 10^7$ at 3.6\,cm show that $r_{\rm out} \gsim (2\pi)^{-1} D/10^7 \sim 10^{14}\,{\rm cm}$.  With the modified distance to the scattering (see \S\ref{sec::Scattering_Geometry} and \citet{Bower_Magnetar_2014}), the problems identified by \citet{Goldreich_2006} are mitigated, as we now discuss in detail. 

Specifically, an upper limit on the outer scale can be estimated as the scale on which the scattering power spectrum requires density fluctuations of order unity. Suppose that the scattering material is statistically homogeneous over a region of length $z$ along the line of sight. Electron density fluctuations $\delta n_{\rm e}(\ell)$ on a scale $\ell$ then introduce corresponding screen phase fluctuations of $\delta \phi(\ell) \sim r_{\rm e} \lambda \sqrt{\ell z} \delta n_{\rm e}(\ell)$ (because of the random walk through $z/\ell$ regions; see \S\ref{sec::Background}). Taking $\delta n_{\rm e}(\ell)/n_{\rm e} \sim (\ell/r_{\rm out})^{(\alpha-1)/2}$, we obtain,
\begin{align}
 \delta \phi(\ell) \sim n_{\rm e} \lambda r_{\rm e} \sqrt{z \ell} \left(\frac{\ell}{r_{\rm out}} \right)^{(\alpha-1)/2},\\
\nonumber \Rightarrow r_{\rm out} \lsim \left(n_{\rm e} \lambda r_{\rm e} \sqrt{z r_{\rm diff}} \right)^{2/(\alpha-1)} r_{\rm diff},
\end{align}
where $r_{\rm diff} \approx \lambda/((1+M) \theta_{\rm scatt})$ is the diffractive scale (i.e., $\delta \phi(r_{\rm diff}) \sim 1$). For \sgra, $r_{\rm diff} \approx 10^8\,{\rm cm}$ at $\lambda = 1\,{\rm cm}$. With our characteristic value $\alpha = 1.38$, we then find
\begin{align}
\label{eq::rout_max}
r_{\rm out} \lsim (10\,{\rm pc}) \times \left( \frac{n_{\rm e}}{10\,{\rm cm}^{-3}} \right)^{5.26} \left( \frac{z}{10\,{\rm pc}} \right)^{2.63}\!.
\end{align}
For comparison, \citet{Armstrong_1995} estimate $r_{\rm out} \gsim 30\,{\rm pc}$ for the scattering material within 1\,kpc. 
For Eq.~\ref{eq::rout_max} to violate our measured lower limit for the 3.6\,cm refractive noise would require much lower electron densities $n_{\rm e} \lsim 0.1$ or larger values of $\alpha$ (approximately $\alpha \gsim 5/3$, for the characteristic values of $z$ and $n_{\rm e}$ given in Eq.~\ref{eq::rout_max}), although these results are highly sensitive to the screen thickness, $z$. From the dispersion measure of the Galactic Center magnetar \citep[e.g.,][]{Eatough_2013,Kravchenko_2016}, we can only estimate an upper bound on the plasma density, $n_{\rm e} < (180\,{\rm cm}^{-3}) / (z / 10\,{\rm pc})$. Regardless, the outer scale required by our scattering model is not implausible.

\citet{Goldreich_2006} have also provided an alternative model for interstellar scattering from folded magnetic field structures. Their proposed model reproduces the $\lambda^2$ scaling and Gaussian scatter-broadening for \sgra, but it also predicts significantly suppressed refractive scintillation. Furthermore, intrinsic structure would be blurred out on small angular scales from the scattering in this model. Thus, our measurements of image substructure at 1.3\,cm conclusively reject the folded field model for the scattering of \sgra\ in its simplest form. However, we will demonstrate later that this model is compatible with our measurements if the inner scale (corresponding to the thickness of current sheets in this model) is significantly larger than expected: $r_{\rm in} \sim 2\times 10^6\,{\rm km}$. In this case, the \citet{Goldreich_2006} spectrum would instead produce significantly \emph{enhanced} refractive effects at millimeter wavelengths.

\subsubsection{The Inner Scale of Turbulence}
\label{sec::Inner_Scale}

Our measurements constrain the inner scale of turbulence, both through plausibility arguments related to the $\lambda^2$ dependence of the angular broadening and image Gaussianity and by relating the scattering power on large scales (refractive noise) to that on small scales (the diffractive blurring). Ultimately, our most stringent lower limit on the inner scale comes from the image Gaussianity at 1.3\,cm, giving $r_{\rm in} \gsim 600\,{\rm km}$. Likewise, the 7\,mm data show a statistically significant departure from a Gaussian image, with a preference for $r_{\rm in} \lsim 1000\,{\rm km}$, although we regard this upper limit as tentative (see \S\ref{sec::gaussian_image_constraint}). Thus, we have adopted a recommended characteristic value of $r_{\rm in} = 800\,{\rm km}$. While the scattering of \sgra\ is anomalously strong, the dissipation mechanism for turbulence in the ISM may be universal. Thus, we now compare our estimate for $r_{\rm in}$ with previous theoretical and observational estimates. 

Using VLBI measurements of the angular broadening for several heavily scattered objects, \citet{Spangler_Gwinn_1990} estimated an inner scale of $50-200\,{\rm km}$. 
Based on weak scintillation measurements at centimeter wavelengths, \citet{Armstrong_1995} constrained the inner scale for the nearby ISM (within 1\,kpc) to be less than ${\sim}5\,{\times}\,10^4\,{\rm km}$. \citet{Rickett_2009} estimated $r_{\rm in} = 70\,{-}\,100\,{\rm km}$ from the pulse broadening of PSR~J1644-4559. \citet{Smirnova_2010} estimated $r_{\rm in} = 350\pm 150\, {\rm km}$ from the pulse broadening of PSR~B2111+46. Each of these studies has its own limitations. For instance, the pulsar analyses assumed isotropic scattering, and \citet{Rickett_2009} noted that a (finely-tuned) anisotropy would allow an arbitrarily large inner scale. Perhaps the most significant difficulty in our study of \sgra\ is that intrinsic source structure becomes significant for the baselines and wavelengths that are sensitive to a direct estimate of the inner scale for \sgra. Nevertheless, our lower limit on the inner scale is quite robust. 

\citet{Goldreich_Sridhar_1995} suggest that the inner scale in the ISM may approach the ion Larmor radius, and \citet{Spangler_Gwinn_1990} proposed that the inner scale corresponds to the larger of the ion inertial length and the ion Larmor radius in the scattering medium. The ion inertial length is $\ell_{\rm i} = V_{\rm A}/\Omega_{\rm i} \approx 230/\sqrt{n_{\rm e}/{\rm cm}^{-3}}\,{\rm km}$, where $V_{\rm A} = B/\sqrt{4\pi n_{\rm e} m_{\rm i}}$ is the Alfv\'en speed and $\Omega_{\rm i} = e B/(m_{\rm i} c)$ is the ion cyclotron frequency. The ion Larmor radius is $r_{\rm i} = v_{\rm th}/\Omega_{\rm i} \approx 930\,{\rm km} \times \left( \frac{B}{1\,\mu{\rm G}} \right)^{-1} \left( \frac{T}{10^4\,{\rm K}} \right)^{1/2}$, where $v_{\rm th} = \sqrt{k T/m_{\rm i}}$ is the ion thermal speed. Given the strong scattering of \sgra, it is likely that the ion Larmor radius will then determine the inner scale in this model. The required $B \sim 1\,\mu{\rm G}$ is somewhat lower than expected for magnetic fields in the ISM at the galactocentric distance $R \sim 5.5\,{\rm kpc}$ \citep[e.g.,][]{Han_2006}, and it may suggest that the inner scale is a few times larger than $r_{\rm i}$. 

In terms of a specific model for the scattering of \sgra, \citet{Sicheneder_2017} have proposed that the scattering may arise in a single \ion{H}{2} region along the line of sight, with density $n_{\rm e} \sim 200\,{\rm cm}^{-3}$ and radius ${\sim} 3\,{\rm pc}$. They note that this region can also produce the observed rotation measure of the Galactic center magnetar SGR~J1745-2900, if the field strengths in the scattering material are $15-70\,\mu{\rm G}$. In this model, the ion inertial length is only $\ell_{\rm i} \sim 10-20\,{\rm km}$ and the ion Larmor radius is $r_{\rm i} \lsim 60\,{\rm km}$. Thus, our estimates of an inner scale that is significantly higher than either of these values support the scenario in which the large RM of the magnetar arises from a local contribution near the Galactic Center \citep{Eatough_2013,Desvignes_2018}. For smaller magnetic fields, $B \sim 1\,\mu{\rm G}$, the parameters identified by \citet{Sicheneder_2017} remain plausible for the scattering.

\subsubsection{The Power-Law Index of Turbulence}
\label{sec::Power-Law}

\begin{figure}[t]
\centering
\includegraphics[width=\columnwidth]{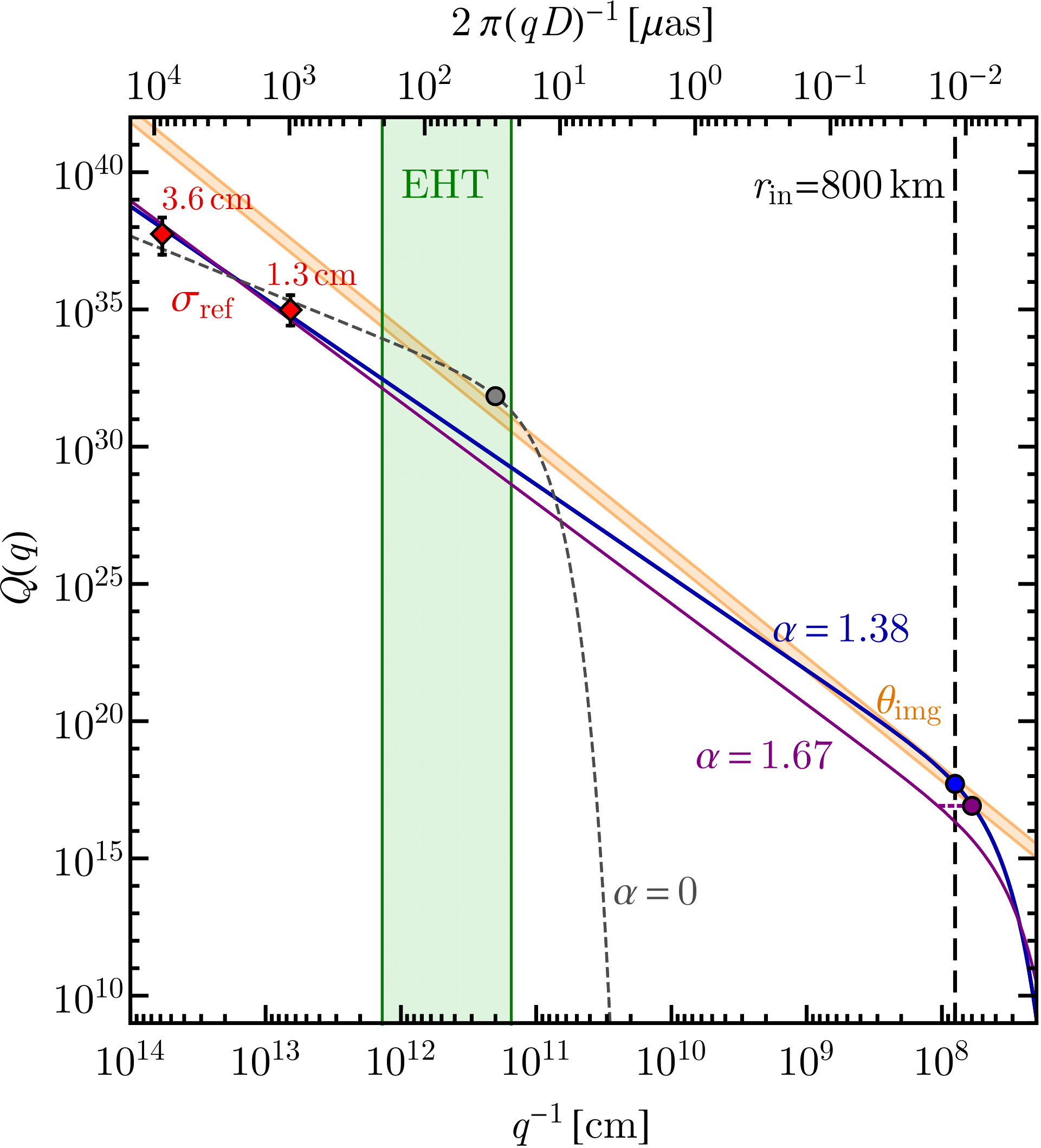}\\
\caption
{ 
Summary of constraints on the power spectrum of phase fluctuations, $Q(\mathbf{q})$. Refractive noise on long baselines at 3.6 and 1.3\,cm constrains the power in wavenumbers $q^{-1}\,{\sim}\,10^{13} - 10^{14}\,{\rm cm}$ (red diamonds). Asymptotic Gaussian angular broadening constrains the power in wavenumbers $q\,{\sim}\,r_{\rm in}^{-1}$; the corresponding constraint on $Q$ depends strongly on $r_{\rm in}$ and weakly on $\alpha$. The orange band shows the angular broadening constraint as a function of $q=r_{\rm in}^{-1}$ over the range $1 < \alpha < 5/3$. Three models are plotted: a Kolmogorov spectrum with our minimum allowed inner scale (purple; $\alpha=5/3$, $r_{\rm in}=600\,{\rm km}$), our recommended characteristic model (blue; $\alpha=1.38$, $r_{\rm in}=800\,{\rm km}$), and a \citet{Goldreich_2006} spectrum (dashed gray; $\alpha=0$, $r_{\rm in}=2 \times 10^6\,{\rm km}$). Corresponding colored circles show the constraint on $Q$ from angular broadening for each model. While a Kolmogorov spectrum is compatible with the refractive noise measurements taken alone, it would require spectral flattening or additional power near the inner scale to be compatible with the measured angular broadening, as shown by the horizontal purple dashed line. The green shaded region shows the range of modes that contribute refractive noise to EHT images of \sgra\ ($\sigma_{\rm ref} \propto \sqrt{Q}$). Note that refractive noise predictions from our model for the EHT are rather insensitive to possible generalizations that would allow $\alpha=5/3$. However, the \citet{Goldreich_2006} spectrum would increase refractive noise by a factor of ${\approx}10$ relative to our characteristic model.  
}
\label{fig::Q_Constraints}
\end{figure}

Figure~\ref{fig::Q_Constraints} shows our constraints on the power spectrum of phase fluctuations, $Q(\mathbf{q})$, along the direction of the scattering major axis. Refractive noise on a long baseline $\mathbf{u}$ is dominated by refractive modes with $\mathbf{q} \sim 2\pi \mathbf{u}/D$. Thus, our measurements of refractive noise at 3.6 and 1.3\,cm constrain the power in wavenumbers $q^{-1}\,{\sim}\,10^{13} - 10^{14}\,{\rm cm}$. In addition, our measurements of the asymptotic Gaussian angular broadening constrain the power in wavenumbers $q^{-1}\,{\sim}\,r_{\rm in}$, with the exact constraint also weakly dependent on $\alpha$: $Q(r_{\rm in}^{-1}) \propto r_{\rm in}^4/\Gamma(1-\alpha/2)$. 

As is evident from Figure~\ref{fig::Q_Constraints}, larger values of the inner scale require a flatter power spectrum for the measured refractive noise to be compatible with the measured angular broadening (see also Figure~\ref{fig::alpha_rin_limits}). Allowing arbitrarily small inner scales, we find $\alpha \lsim 1.6$, while including our derived constraints on the inner scale, we obtain $\alpha < 1.47$. Thus, a Kolmogorov spectrum ($\alpha=5/3$) is incompatible with our measurements, as is an $\alpha=3/2$ spectrum \citep[see, e.g.,][]{Iroshnikov_1964,Kraichnan_1965,Sridhar_1994,Goldreich_Sridhar_1995}. Our results are at tension with measurements for the local ISM \citep[e.g.,][]{Armstrong_1995}, the wavelength dependence of pulsar temporal broadening \citep[e.g.,][]{Lohmer_2001,Bhat_2004,Lewandowski_2013}, and VLBI of heavily scattered sources \citep{Spangler_Gwinn_1990}, all of which tend to infer somewhat larger values of $\alpha$. 
We now outline possible generalizations to our scattering model that might render higher values of $\alpha$, including a Kolmogorov spectrum, feasible. 

The first possibility is an outer scale of turbulence that is similar to the scales probed at 3.6\,cm, thereby reducing the 3.6\,cm refractive noise but perhaps not the 1.3\,cm refractive noise (which probes smaller scales). The 3.6\,cm refractive noise corresponds to scattering modes with a transverse scale of ${\sim}4\,{\rm AU}$ on the scattering screen, so the required outer scale is $r_{\rm out} \sim 1\,{\rm AU}$. This value is somewhat smaller than the lower limit estimated by \citet{Armstrong_1995}. Moreover, this is a finely-tuned constraint, requiring the outer scale to be precisely matched to our observing parameters --- a smaller outer scale would be inconsistent with the observed 1.3\,cm refractive noise, and a larger outer scale would not affect the 3.6\,cm noise.

A second possibility is that there is extra power located near the dissipation scale. Spectral flattening near the inner scale has been seen in the solar wind \citep[e.g.,][]{Neugebauer_1975,Celnikier_1983,Coles_1991} and possibly also in the ISM \citep{Smirnova_2010}. A pile-up in power by a factor of ${\sim}15$ would reconcile our measurements with a Kolmogorov spectrum (see Figure~\ref{fig::Q_Constraints}); a factor of ${\approx}\,2$ would be needed for an $\alpha = 3/2$ spectrum. 

A more radical possibility, which is not excluded by our data, is that the spectrum is extremely shallow and the inner scale is correspondingly large. In particular, the model of \citet{Goldreich_2006} produces a power spectrum with $\alpha=0$, which would be consistent with all our measurements if $r_{\rm in} \sim 2\times 10^6\,{\rm km}$. This inner scale is a factor of 20 larger than the characteristic value used by \citet{Goldreich_2006} and requires an outer scale that is 400 times larger than expected, or a few kpc. Nevertheless, our measurements are insufficient to rule out this type of power spectrum, and it would produce refractive signatures for the EHT that are approximately 10 times stronger than those predicted by our recommended characteristic model. Thus, we expect continued studies at 1.3\,mm (and possibly 3.5\,mm) will be able to conclusively confirm or reject this scattering model for \sgra.

% Our current constraints on $\alpha$ are rather indirect. By combining many epochs at 1.3\,cm, the baseline dependence of the refractive noise could be estimated, giving $\alpha$ directly.  Alternatively, measurements of the scattering kernel shape at baselines longer than the inner scale could also estimate $\alpha$; this may be possible with expanded arrays at 7 or 3.5\,mm, depending on how complex the intrinsic structure is at those wavelengths. 

\subsection{Sensitivity to the Assumed Scattering Model}
\label{sec::Sensitivity_to_Model_Assumptions}

We have analyzed all our data in the context of a single scattering model. The anisotropy in this model is determined by the magnetic field wander along the line of sight relative to its preferred orientation (which determines the minor axis of the scattering ellipse). \citet{Psaltis_2018} give three representative models for the field wander:  ``von Mises,'' ``Dipole,'' and ``Boxcar.'' The von~Mises model represents the angular field wander using a generalized Gaussian distribution for circular quantities; the Dipole model uses a change of variables to rescale the principal axes of the power spectrum; and the Boxcar model has a power spectrum that is isotropic across a restricted range of angles and is zero elsewhere. 
Because of the efficient computational tools developed in Appendix~\ref{sec::Efficient_Computation}, all of our results have used the Dipole model. We now evaluate how sensitive our conclusions are to this choice.

\citet{Psaltis_2018} show that the shape of the scattering kernel is almost independent of the choice of scattering model. However, the refractive noise along the minor axis is sensitive to the scattering model. Because our measurements of refractive noise are predominantly along the major axis, our results are not strongly affected by the choice of scattering model.

For example, the mean refractive noise on our long baselines at 3.6\,cm changes by less than $\pm 2\%$ among the three scattering models (well within our uncertainty from sampling only a few elements of the refractive noise). Likewise, the 95\% confidence intervals are almost identical for the three models. For the average refractive noise on long baselines at 1.3\,cm, the von Mises and Dipole models agree to within 1\%, but the Boxcar model differs by 5\%. Again, these differences are negligible within our error budget. 

Thus, we conclude that our specific choice of scattering model is irrelevant for our results. Equivalently, our current measurements provide no firm guidance for discriminating among these models for the magnetic field wander. Future measurements of refractive noise on long baselines along the minor axis could immediately rule out the Boxcar model and would be sensitive to differences between the von Mises and Dipole models \citep[see, e.g., Figures~9 and 13 in][]{Psaltis_2018}.

\subsection{Implications for Continued Studies of Sgr A*}

For the scattering parameters that we have identified, a Gaussian scattering kernel is likely a good approximation for \sgra\ for centimeter wavelengths, although the full non-Gaussian kernel shape should be used for continued studies at millimeter wavelengths. The full kernel shape is especially important for scattering mitigation in imaging with EHT data \citep{Fish_2014,Stochastic_Optics}.

\begin{figure}[t]
\centering
\includegraphics[width=\columnwidth]{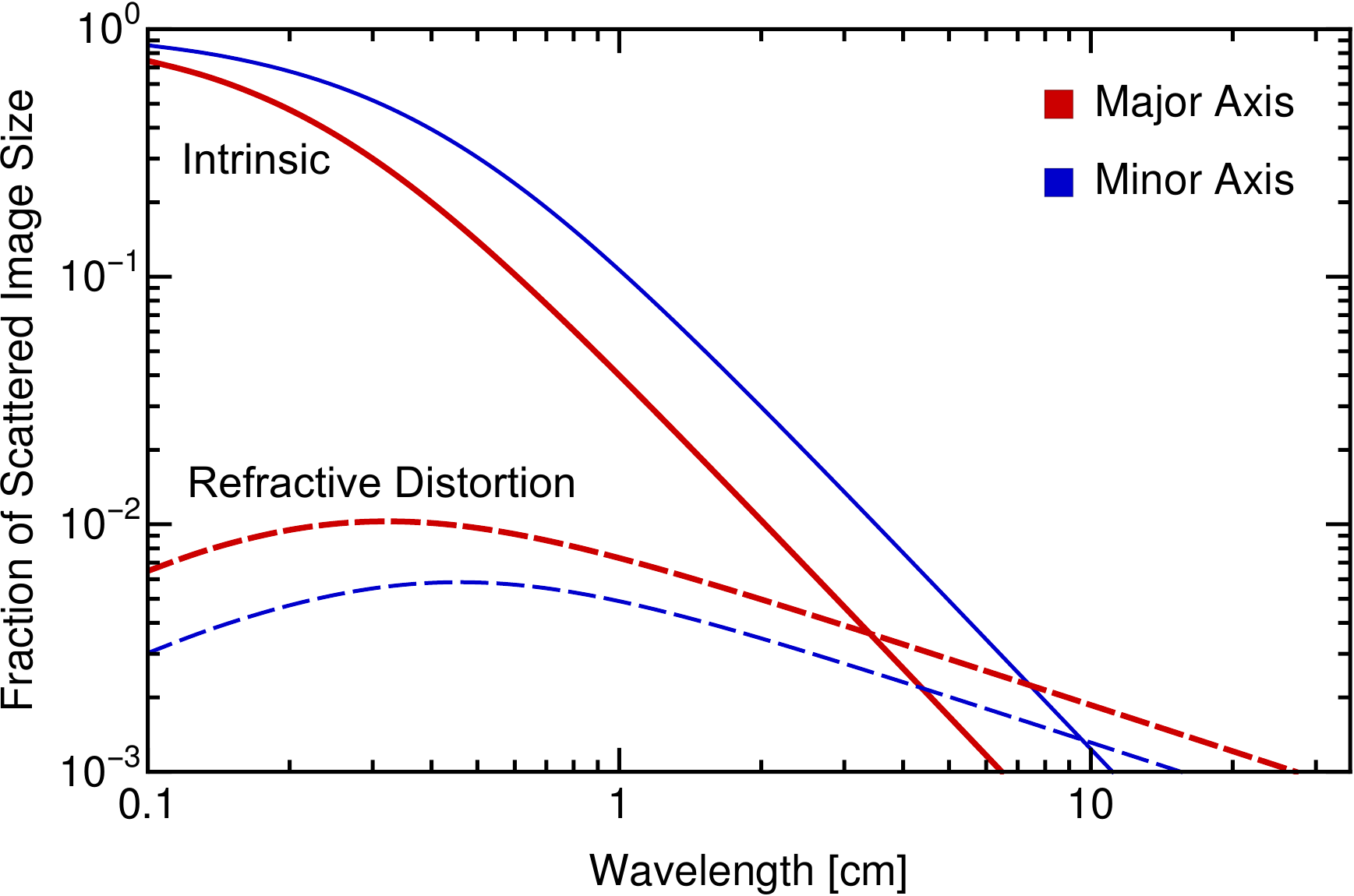}\\
\caption
{ 
Comparison of the wavelength-dependent fractional effects from intrinsic image structure and from refractive image distortion. The intrinsic curves show the fractional increase in the scattered image size because of intrinsic structure: $1 - \theta_{\rm scatt}/\theta_{\rm ea}$, where $\theta_{\rm scatt}$ is the size of a scattered point source and $\theta_{\rm ea}$ is the angular size of an extended source using our estimated wavelength-dependent size of \sgra. The refractive distortion curves show the expected fractional fluctuations of image size among different observing epochs because of refractive scattering (see \S\ref{sec::Image_Size_Fluctuations}). Intrinsic structure can only be reliably estimated when refractive jitter is significantly smaller than the intrinsic contribution, irrespective of the observing sensitivity or baseline coverage. Requiring that the fractional increase from intrinsic size must be at least three times the rms distortion, we estimate that intrinsic properties for the major axis can only be reliably constrained for observations with $\lambda \lsim 1.3\,{\rm cm}$, while intrinsic properties for the minor axis can only be constrained for observations with $\lambda \lsim 3.6\,{\rm cm}$.
}
\label{fig::Fractional_Size_Effects}
\end{figure}

We have shown that refractive noise is a critical component of the error budget when model fitting to observations of \sgra. In addition, we caution that inferences of intrinsic size should account for refractive image distortion. At long wavelengths, stochastic changes from refractive distortion can exceed the contribution of intrinsic structure (see Figure~\ref{fig::Fractional_Size_Effects}). Our results suggest that intrinsic structure can only be securely decoupled from refractive distortion for $\lambda \lsim 3.6\,{\rm cm}$ (minor axis) or $\lambda \lsim 1.3\,{\rm cm}$ (major axis). Note that these are fundamental limitations; they would apply even if the sensitivity and baseline coverage of the observations were perfect.

\begin{figure}[t]
\centering
\includegraphics[width=\columnwidth]{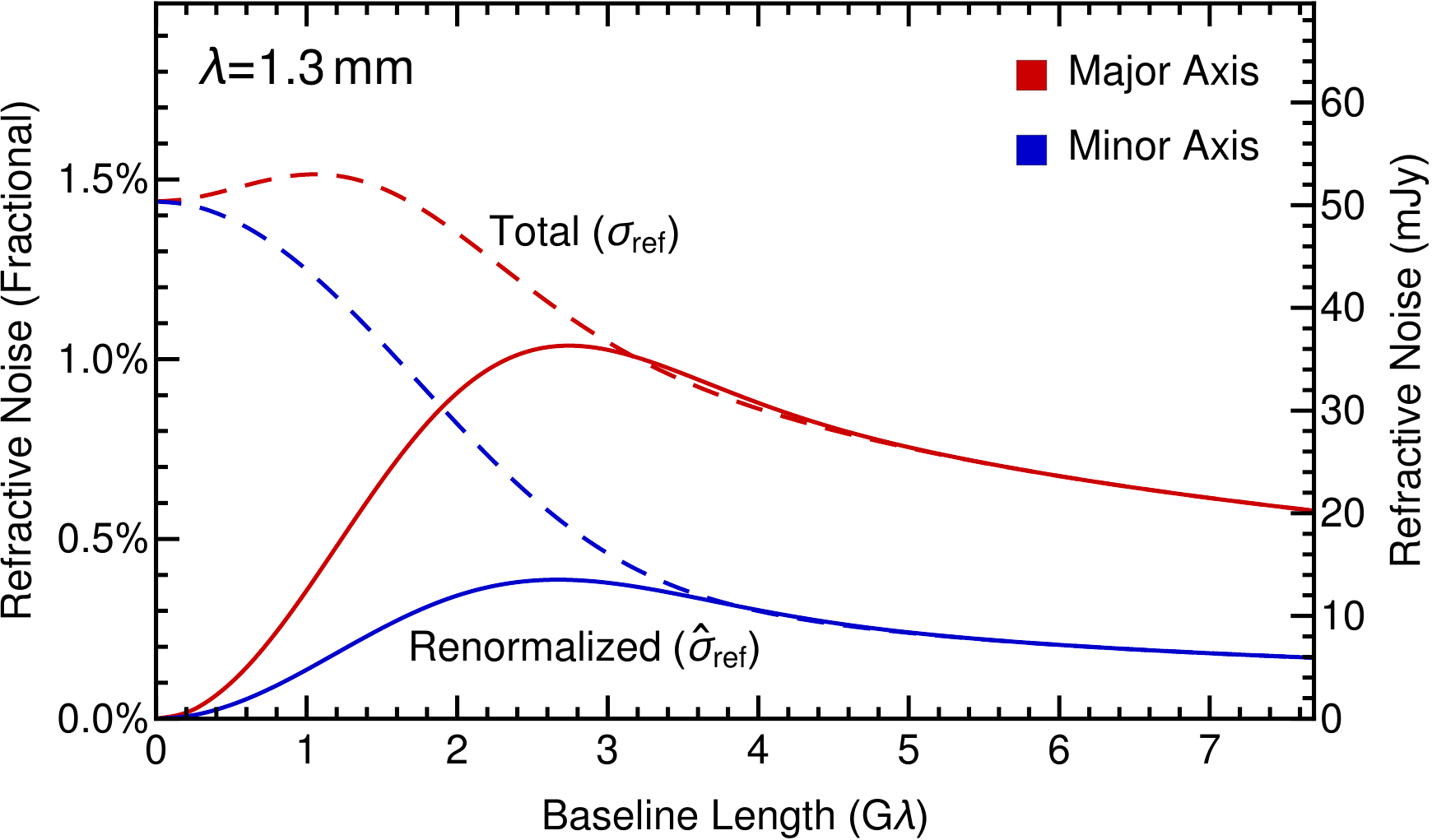}\\
\caption
{ 
Expected rms of refractive noise (dashed) and renormalized refractive noise (solid) for \sgra\ at 1.3\,mm (i.e., for EHT observations). The curves correspond to our recommended characteristic scattering model and an isotropic intrinsic Gaussian source with FWHM $\theta_{\rm src} = 52\,\mu{\rm as}$. To express the refractive noise in units of flux density, we assume a total flux density of 3.5\,Jy for \sgra. 
}
\label{fig::EHT_Refractive_Noise}
\end{figure}

Our work has two significant implications for imaging \sgra\ with the EHT. First, we have shown that the scattering kernel may be much smaller than has been estimated, so the blurring effects of scattering may be less severe than have been assumed (see Figure~\ref{fig::Scattering_Kernel}). Second, we find $\alpha \lsim 1.47$, which produces significantly less refractive noise than the standard Kolmogorov picture; it predicts that the renormalized refractive noise is at most ${\sim}1\%$ of the zero-baseline flux density (see Figure~\ref{fig::EHT_Refractive_Noise}). This estimate reinforces the conclusions of \citet{Fish_2016} and \citet{Lu_2018} that refractive noise is unlikely to be a significant component of the error budget for past EHT observations. Both these implications improve the prospects for horizon-scale imaging at 1.3\,mm. However, continued observations must also account for the possibility of strong refractive effects from shallow spectra, such as from the model with $\alpha=0$ and $r_{\rm in} = 2\times 10^6\,{\rm km}$. While this model remains speculative and lacks support from other lines of sight, it would produce striking differences from our characteristic model for EHT observations (an increase in refractive noise by a factor of 10) and should be tested further with long baseline measurements at 1.3\,mm and 3.5\,mm.

% 
% 
% . 
% 
% . The refractive noise is significantly smaller than what was expected for the case of a $40\,\mu{\rm as}$ source with Kolmogorov scattering \citep{Johnson_Gwinn_2015}. 

% Primary remaining uncertainties: extrapolation of the scattering size, and the shape of the scattering kernel.
% 
% To extrapolate refractive noise estimates from 1.3\,cm to the EHT... Key point: The refractive noise extrapolation to EHT wavelengths is fairly secure, even if the scattering size at 1.3mm is poorly constrained. The key point is to measure the refractive noise at comparable wavenumbers. The longest baselines so far are K-band, which extend to 250~M$\lambda$ at 1.3cm, which is 2.5~G$\lambda$ at 1.3mm. This is comparable to the EHT. Scales with baseline as $b^{-\alpha/2}$. Only a 20\% uncertainty after a change of baseline length by a factor of 2 over the range from $\alpha \in [1.3, 1.9]$. 

\ \\

\section{Summary}
\label{sec::Summary}

We have analyzed observations of \sgra\ at wavelengths from 1.3\,mm to 30\,cm using a physically motivated model for its scattering \citep[developed in][]{Psaltis_2018}. At long wavelengths, the angular broadening from scattering is an anisotropic Gaussian, and its size scales as $\theta_{\rm scatt} \propto \lambda^2$. At shorter wavelengths, the shape and wavelength dependence of the scattering depend on the inner scale of turbulence, $r_{\rm in}$, and on the power-law index of the scattering, $\alpha$. Using a new prescription to perform model fitting with refractive noise in the error budget, we are able to estimate the asymptotic Gaussian scattering parameters to excellent accuracy (see Table~\ref{tab::GaussianParameters}). In addition, we show that $\alpha \lsim 1.47$ and $r_{\rm in} \gsim 600\,{\rm km}$ (our recommended characteristic values for these parameters are $\alpha = 1.38$ and $r_{\rm in} = 800\,{\rm km}$). Our recommended scattering parameters are summarized in Table~\ref{tab::ScatteringParameters}.

After deconvolving the effects of scattering from our estimated sizes of the scatter broadened images, we find that the intrinsic image of \sgra\ is nearly isotropic, with FWHM $\theta_{\rm src} \sim (40\,\mu{\rm as}) \times \lambda_{\rm mm}$ from 1.3\,mm to 1.3\,cm. While this linear wavelength dependence for the emission size is natural for both disk- and jet-dominated models, the nearly isotropic image shape strongly favors disk models. At 1.3\,mm, where the emission region is expected to be largely optically thin, our estimated image size is consistent with predictions from recent GRRMHD simulations \citep[e.g.,][]{Chael_2018}. 

For ISM scattering, our most surprising conclusion is that a Kolmogorov spectrum ($\alpha=5/3$) in the inertial range is incompatible with our observations. However, our constraints on $\alpha$ are somewhat indirect, as they relate the refractive power in large scattering modes to the diffractive power from small scattering modes (see Figure~\ref{fig::Q_Constraints}). For a generalized scattering model, larger values of $\alpha$ are possible but would require continuous injection of energy on ${\lsim}{\rm AU}$-scales, a pile-up of energy near the inner scale (${\sim}10^3\,{\rm km}$), or a small outer scale for the turbulence (${\sim}1\,{\rm AU}$). 

We have also shown that the inner scale cannot be smaller than $600\,{\rm km}$ and is likely $r_{\rm in} \approx 800\,{\rm km}$. This scale is comparable to the ion Larmor radius for regions of the ISM with weak magnetic fields $B \sim 1\,\mu{\rm G}$, thereby supporting identification of the ion Larmor radius (or a few times this radius) with the dissipation scale of ISM turbulence. This estimate also suggests that the rotation measure associated with the scattering material is modest, and hence that the rotation measure of the Galactic Center magnetar is dominated by local contributions \citep{Eatough_2013} rather than from the scattering material \citep{Sicheneder_2017}. 
Our estimated $r_{\rm in}$ is also comparable to the ion inertial length for $n_{\rm e} \sim 0.1\,{\rm cm}^{-3}$ and requires $n_{\rm e} \gsim 0.1\,{\rm cm}^{-3}$ if the inner scale is determined by the larger of these characteristic two plasma length scales \citep{Spangler_Gwinn_1990}. However, we cannot conclusively rule out much shallower spectra with correspondingly larger inner scales. 

While our primary objective has been to constrain the parameters of our specific scattering model, our observations also constrain alternative theories for the scattering of \sgra. For example, \citet{Goldreich_2006} have proposed a model in which the scattering is caused by an ensemble of folded current sheets in the ISM. While this model naturally reproduces the $\lambda^2$ scaling of angular broadening and the Gaussian image at long radio wavelengths, it predicts an absence of refractive scattering effects. This model would not produce scattering substructure in images, and any intrinsic substructure would be blurred out by small-scale scattering modes. Hence, the pronounced long-baseline refractive noise at 1.3\,cm enables us to firmly reject this alternative scattering model for \sgra\ in its simplest form \citep[see also][]{Gwinn_2014}. However, the model is compatible with our measurements if the thickness of the current sheets is significantly larger than expected, corresponding to $r_{\rm in} \sim 2\times 10^6\,{\rm km}$, in which case it would instead produce strongly \emph{enhanced} refractive effects at millimeter wavelengths. Thus, we expect that continued observations with the GMVA and EHT will be sufficient to firmly support or reject this model.

Our results highlight the importance of including refractive noise when fitting models to radio observations of \sgra. Refractive uncertainties can plausibly explain many of the discrepancies in past measurements of the size of \sgra, such as those identified by \citet{Psaltis_2015}. In addition, we have shown that refractive effects likely prohibit a meaningful study of intrinsic structure at wavelengths longer than $1.3\,{\rm cm}$ (or $3.6\,{\rm cm}$ for the minor axis; see Figure~\ref{fig::Fractional_Size_Effects}). Nevertheless, our results also show that both the blurring and substructure from scattering may be significantly smaller at 1.3\,mm than expected. Thus, the prospects for deeper study of \sgra\ at millimeter wavelengths, including imaging with the EHT, are excellent.

\acknowledgements{We thank the MPIfR internal referee, Thomas Krichbaum, and the anonymous ApJ referee for detailed comments that significantly improved the manuscript. We thank the National Science Foundation (AST-1440254, AST-1716536, AST-1312651) and the Gordon and Betty Moore Foundation (GBMF-5278) for financial support of this work. This work was supported in part by the Black Hole Initiative at Harvard University, which is supported by a grant from the John Templeton Foundation.
This work was supported in part by the Fundamental Research Program No.~28 of the Presidium of the Russian Academy of Sciences and the government of the Russian Federation (agreement 05.Y09.21.0018).
%The VLA is operated by the National Radio Astronomy Observatory, the VLBA is operated by the Long Baseline Observatory,  the GBT is operated by the Green Bank Observatory. 
The National Radio Astronomy Observatory, the Long Baseline Observatory and the Green Bank Observatory are facilities of the National Science Foundation operated under cooperative agreement by Associated Universities, Inc.
This work made use of the Swinburne University of Technology software correlator \citep{Deller_2011}, developed as part of the Australian Major National Research Facilities Programme and operated under licence.
}

Facilities: \facility{VLA}, \facility{EVLA}, \facility{VLBA}, \facility{VERA}, \facility{KVN}, \facility{EHT}, \facility{LMT}, \facility{GBT}.

\begin{appendix}

\section{Calculating Renormalized Refractive Noise}
\label{sec::CalculatingRenormalizedRefractiveNoise}

As described in \S\ref{sec::Refractive_Noise}, refractive noise includes contributions that may not be appropriate for the relevant error budget. For instance, the refractive noise on a zero-baseline corresponds to refractive modulation of the total flux density, which may be absorbed into model parameters. Likewise, the variance in the imaginary part of visibilities on short baselines is produced by image wander, which is only relevant for observations with absolute phase referencing. We now derive ``renormalized'' refractive noise expressions, which remove the contributions of flux density modulation and/or image wander.  We now derive expressions to efficiently compute properties of renormalized refractive noise semi-analytically, following the methodology and notation of \citet{Johnson_Narayan_2016} \citep[see also][]{Blandford_Narayan_1985}.  

To proceed, we will first define the renormalized average visibility $\hat{V}_{\rm a}(\mathbf{b})$ as the visibility corresponding to a scattered image that has been normalized to have unit total flux density (e.g., $\hat{V}_{\rm a}(\mathbf{0})=1$) and that has been shifted such that its brightness distribution is centered on the origin:
\begin{align}
\hat{V}_{\rm a}(\mathbf{b}) &\equiv \frac{V_{\rm a}(\mathbf{b})}{V_{\rm a}(\mathbf{0})}  e^{2\pi i \mathbf{b} \cdot \mathbf{x}_{0,\rm a}/(\lambda D)}\\
\nonumber &= \frac{V_{\rm ea}(\mathbf{b}) + \Delta V_{\rm a}(\mathbf{b})}{V_{\rm ea}(\mathbf{0}) + \Delta V_{\rm a}(\mathbf{0})}  e^{2\pi i \mathbf{b} \cdot \Delta\mathbf{x}_{0,\rm a}/(\lambda D)} e^{2\pi i \mathbf{b} \cdot \mathbf{x}_{0,\rm ea}/(\lambda D)}\\
\nonumber &\approx \frac{e^{2\pi i \mathbf{b} \cdot \mathbf{x}_{0,\rm ea}/(\lambda D)}}{V_{\rm ea}(\mathbf{0})} \left[V_{\rm ea}(\mathbf{b}) + \Delta V_{\rm a}(\mathbf{b}) - \frac{V_{\rm ea}(\mathbf{b})}{V_{\rm ea}(\mathbf{0})} \Delta V_{\rm a}(\mathbf{0}) + 2\pi i V_{\rm ea}(\mathbf{b}) \mathbf{b} \cdot \Delta\mathbf{x}_{0,\rm a}/(\lambda D) \right].
\end{align}
In these expressions and throughout the remainder of this paper, a subscript ``a'' denotes a quantity in the average image regime, while ``ea'' denotes a quantity in the ensemble-average image regime \citep{NarayanGoodman89,GoodmanNarayan89}. The original image centroid, $\mathbf{x}_{0,\rm a}$, is given as a transverse displacement on the scattering screen. Thus, in angular units, the centroid is at $\boldsymbol{\eta}_{0,\rm a} \equiv \mathbf{x}_{0,\rm a}/D$. We assume that the refractive effects are only a small perturbation of the ensemble-average image, so the final expression only includes refractive terms to linear order. 

The renormalized refractive noise can then be written
\begin{align}
\label{eq::NormalizedNoise}
\Delta \hat{V}_{\rm a}(\mathbf{b}) &\approx \frac{e^{2\pi i \mathbf{b} \cdot \mathbf{x}_{0,\rm ea}/(\lambda D)}}{V_{\rm ea}(\mathbf{0})} \left[ \Delta V_{\rm a}(\mathbf{b}) - \frac{V_{\rm ea}(\mathbf{b})}{V_{\rm ea}(\mathbf{0})} \Delta V_{\rm a}(\mathbf{0}) + 2\pi i V_{\rm ea}(\mathbf{b}) \mathbf{b} \cdot \Delta\mathbf{x}_{0,\rm a}/(\lambda D) \right].
% &\equiv \int d^2\mathbf{r}\, f_{\hat{\rm V}}(\mathbf{r}; \mathbf{b}, \lambda) \phi(\mathbf{r}). 
\end{align}
The prefactor in this expression only depends on the ensemble-average visibility. It normalizes the ensemble-average image to have unit flux density and to be centered on the origin. The first term inside the square brackets is the full refractive noise of the average image. The remaining two terms remove the contributions from flux modulation and from image wander, respectively. To simplify the remainder of our discussion, we will assume that the ensemble-average image is centered on the origin: $\mathbf{x}_{0,{\rm ea}} = \mathbf{0}$ and $\Delta \mathbf{x}_{0,{\rm a}} = \mathbf{x}_{0,{\rm a}}$.  

To estimate properties of the refractive noise, we must determine the function $f_{\hat{\rm V}}(\mathbf{r}; \mathbf{b}, \lambda)$ defined by $\Delta \hat{V}_{\rm a}(\mathbf{b}) \equiv \int d^2\mathbf{r}\, f_{\hat{\rm V}}(\mathbf{r}; \mathbf{b}, \lambda) \phi(\mathbf{r})$, where $\phi(\mathbf{r})$ is the refractive component of the scattering screen phase (i.e., consisting only of modes with wavelengths longer than the Fresnel scale). For example, the average visibility is approximated as \citep[see][Eq.~11-12]{Johnson_Narayan_2016}
\begin{align}
\label{eq::Vavg}
V_{\rm a}(\mathbf{b}) &\approx V_{\rm ea}(\mathbf{b}) + \int d^2\mathbf{r}\,  f_{\rm V}(\mathbf{r};\mathbf{b},\lambda) \phi(\mathbf{r}) \equiv V_{\rm ea}(\mathbf{b}) + \Delta V_{\rm a}(\mathbf{b}),\\
\nonumber f_{\rm V}(\mathbf{r};\mathbf{b},\lambda) &\equiv r_{\rm F}^2 e^{-i \mathbf{r} \cdot \mathbf{b}/(D \lambdabar)} \left[ \frac{i}{D\lambdabar} \mathbf{b} \cdot \nabla I_{\rm ea}(\mathbf{r}) - \nabla^2 I_{\rm ea}(\mathbf{r}) \right].
\end{align}
These equations arise from the approximate representation of scattering in the geometric optics regime, which gives the scattered image $I_{\rm a}(\mathbf{r})$ in terms of the ensemble-average image $I_{\rm ea}(\mathbf{r})$ and the refractive scattering screen phase gradients: 
\begin{align}
I_{\rm a}(\mathbf{r}) &\approx I_{\rm ea}(\mathbf{r} + r_{\rm F}^2 \nabla \phi_{\rm r}(\mathbf{r}))\\
\nonumber &\approx I_{\rm ea}(\mathbf{r}) + r_{\rm F}^2 \left[\nabla \phi_{\rm r}(\mathbf{r})\right] \cdot \left[ \nabla I_{\rm ea}(\mathbf{r}) \right].
\end{align}

To denote the integral correspondence in Eq.~\ref{eq::Vavg}, we will introduce the shorthand $\Delta V_{\rm a}(\mathbf{b})  \leftrightarrow f_{\rm V}(\mathbf{r};\mathbf{b},\lambda)$. Obviously, we then have $\Delta V_{\rm a}(\mathbf{0})  \leftrightarrow f_{\rm V}(\mathbf{r};\mathbf{0},\lambda)$. Finally, note that
\begin{align}
\left. \nabla_{\mathbf{b}} V_{\rm a}(\mathbf{b}) \right\rfloor_{\mathbf{b}=0} = - 2\pi i V_{\rm a}(\mathbf{0}) \mathbf{x}_{0,\rm a}/(\lambda D) = - 2\pi i V_{\rm a}(\mathbf{0}) \Delta \mathbf{x}_{0,\rm a}/(\lambda D), 
\end{align}
where $\nabla_{\mathbf{b}}$ denotes a gradient with respect to baseline (not the directional derivative $\mathbf{b} \cdot \nabla$). This general identity relates an image centroid to the corresponding visibility gradient at zero baseline. We will use the notation that $\left. \nabla_{\mathbf{b}} V_{\rm a}(\mathbf{b}) \right\rfloor_{\mathbf{b}=0} \equiv \nabla_{\mathbf{b}=\mathbf{0}} V_{\rm a}(\mathbf{b})$. Thus, $\Delta \mathbf{x}_{0,\rm a} = i \frac{\lambda D}{2\pi} \frac{1}{V_{\rm a}(\mathbf{0})}\nabla_{\mathbf{b}=\mathbf{0}} V_{\rm a}(\mathbf{b}) \approx i \frac{\lambda D}{2\pi} \frac{1}{V_{\rm ea}(\mathbf{0})}\nabla_{\mathbf{b}=\mathbf{0}} V_{\rm a}(\mathbf{b})$. Consequently, 
\begin{align}
 \Delta \mathbf{x}_{0,\rm a} \leftrightarrow i \frac{\lambda D}{2\pi} \frac{1}{V_{\rm ea}(\mathbf{0})} \nabla_{\mathbf{b}=\mathbf{0}} f_{\rm V}(\mathbf{r};\mathbf{b},\lambda). 
\end{align}
Putting everything together, we obtain
\begin{align}
\label{eq::fVhat}
 \Delta \hat{V}_{\rm a}(\mathbf{b}) \leftrightarrow f_{\hat{\rm V}}(\mathbf{r};\mathbf{b},\lambda) \equiv \frac{1}{V_{\rm ea}(\mathbf{0})} 
 \left[ f_{\rm V}(\mathbf{r};\mathbf{b},\lambda) - \frac{V_{\rm ea}(\mathbf{b})}{V_{\rm ea}(\mathbf{0})} f_{\rm V}(\mathbf{r};\mathbf{0},\lambda) - \frac{V_{\rm ea}(\mathbf{b})}{V_{\rm ea}(\mathbf{0})} \mathbf{b} \cdot \nabla_{\mathbf{b}=\mathbf{0}} f_{\rm V}(\mathbf{r};\mathbf{b},\lambda) \right].
\end{align}
For computational purposes, we require the Fourier-conjugate quantity, $\tilde{f}_{\hat{\rm V}}(\mathbf{q};\mathbf{b},\lambda) \equiv \int d^2\mathbf{r}\, f_{\hat{\rm V}}(\mathbf{r};\mathbf{b},\lambda) e^{-i \mathbf{q} \cdot \mathbf{r}}$, which follows trivially from Eq.~\ref{eq::fVhat}:
\begin{align}
\label{eq::fvhatt}
 \tilde{f}_{\hat{\rm V}}(\mathbf{q};\mathbf{b},\lambda) = \frac{1}{V_{\rm ea}(\mathbf{0})} 
 \left[ \tilde{f}_{\rm V}(\mathbf{q};\mathbf{b},\lambda) - \frac{V_{\rm ea}(\mathbf{b})}{V_{\rm ea}(\mathbf{0})} \tilde{f}_{\rm V}(\mathbf{q};\mathbf{0},\lambda) - \frac{V_{\rm ea}(\mathbf{b})}{V_{\rm ea}(\mathbf{0})} \mathbf{b} \cdot \nabla_{\mathbf{b}=\mathbf{0}} \tilde{f}_{\rm V}(\mathbf{q};\mathbf{b},\lambda) \right],
\end{align}
where $\tilde{f}_{\rm V}(\mathbf{q};\mathbf{b},\lambda)$ is \citep[][Eq.~15]{Johnson_Narayan_2016}
\begin{align}
  \tilde{f}_{\rm V}(\mathbf{q};\mathbf{b},\lambda)  &= r_{\rm F}^2 \mathbf{q} \cdot \left[ \mathbf{q} + (1+M)^{-1} r_{\rm F}^{-2} \mathbf{b} \right] V_{\rm ea}\left( (1+M)r_{\rm F}^2 \mathbf{q}  + \mathbf{b} \right).
\end{align}
As in Eq.~\ref{eq::NormalizedNoise}, the prefactor in Eq.~\ref{eq::fvhatt} normalizes the ensemble-average image to have unit total flux density. The first term in the brackets gives the usual contribution of refractive noise, the second term eliminates noise from refractive flux modulation, and the third term eliminates noise from refractive position wander. 

We can also express Eq.~\ref{eq::fvhatt} explicitly in terms of the ensemble-average visibility and its gradient:
\begin{align}
\nonumber \mathbf{b} \cdot \nabla_{\mathbf{b}=\mathbf{0}} \tilde{f}_{\rm V}(\mathbf{q};\mathbf{b},\lambda) &=  \frac{\mathbf{q} \cdot \mathbf{b}}{1+M} V_{\rm ea}\left( (1+M)r_{\rm F}^2 \mathbf{q} \right) + r_{\rm F}^2 \left| \mathbf{q} \right|^2 \mathbf{b} \cdot \nabla V_{\rm ea}\left( (1+M)r_{\rm F}^2 \mathbf{q} \right)\\
 \Rightarrow  \tilde{f}_{\hat{\rm V}}(\mathbf{q};\mathbf{b},\lambda) &= \frac{r_{\rm F}^2}{V_{\rm ea}(\mathbf{0})} \left\{ \mathbf{q} \cdot \left[ \mathbf{q} + (1+M)^{-1} r_{\rm F}^{-2} \mathbf{b} \right] \left[ V_{\rm ea}\left( (1+M)r_{\rm F}^2 \mathbf{q}  + \mathbf{b} \right) 
 - \frac{V_{\rm ea}(\mathbf{b})}{V_{\rm ea}(\mathbf{0})} V_{\rm ea}\left( (1+M)r_{\rm F}^2 \mathbf{q} \right)\right] \right. \\
\nonumber & \hspace{8.5cm} - \left. \left| \mathbf{q} \right|^2 \frac{V_{\rm ea}(\mathbf{b})}{V_{\rm ea}(\mathbf{0})}  \mathbf{b} \cdot \nabla V_{\rm ea}\left( (1+M)r_{\rm F}^2 \mathbf{q} \right) \right\}.
\end{align}
In this expression, the terms that eliminate flux modulation and image wander are mixed. 

A benefit of these representations is that we can easily calculate the corresponding functions for the real or imaginary components of the normalized refractive noise; e.g., $ \mathrm{Re}\left[ \Delta \hat{V}_{\rm a}(\mathbf{b}) \right] \leftrightarrow f_{\hat{\rm V},{\rm re}}(\mathbf{r};\mathbf{b},\lambda)$. These functions are necessary to estimate the full covariance matrix of the complex refractive noise among different interferometric baselines. Specifically, because $f_{\hat{\rm V}}(\mathbf{r};-\mathbf{b},\lambda) = f_{\hat{\rm V}}^\ast(\mathbf{r};\mathbf{b},\lambda)$ and the Fourier transform is linear, we obtain
\begin{align}
\tilde{f}_{\hat{\rm V},\rm re}(\mathbf{q};\mathbf{b},\lambda) &= \frac{1}{2} \left[ \tilde{f}_{\hat{\rm V}}(\mathbf{q};\mathbf{b},\lambda) + \tilde{f}_{\hat{\rm V}}(\mathbf{q};-\mathbf{b},\lambda) \right],\\
\nonumber \tilde{f}_{\hat{\rm V},\rm im}(\mathbf{q};\mathbf{b},\lambda) &= \frac{1}{2i} \left[ \tilde{f}_{\hat{\rm V}}(\mathbf{q};\mathbf{b},\lambda) - \tilde{f}_{\hat{\rm V}}(\mathbf{q};-\mathbf{b},\lambda) \right].
\end{align}   
   
Using these functions, we can compute statistical properties of the renormalized refractive noise using the expressions given in \citet{Johnson_Narayan_2016}, with $\tilde{f}_{{\rm V}}$ replaced by $\tilde{f}_{\hat{\rm V}}$. For example, the variance of the renormalized refractive noise on a baseline $\mathbf{b}$ is
\begin{align}
\label{eq::sigmarref_Q}
\left \langle \left| \Delta \hat{V}_{\rm a}(\mathbf{b}) \right|^2 \right \rangle &= \frac{\lambda^2}{(2\pi)^4} \int d^2\mathbf{q}\, \left| \tilde{f}_{\hat{\rm V}}(\mathbf{q};\mathbf{b},\lambda)  \right|^2 Q(\mathbf{q}).
\end{align}

% \begin{figure*}[t]
% \centering
% \includegraphics[width=1.0\textwidth]{Renormalized_Noise_Example.pdf}
% \caption
% { 
% Example showing renormalized refractive noise. Add $\sqrt{2}$ times imaginary rms? Need to update this figure!?!
% }
% \label{fig::Renormalized_Noise_Example}
% \end{figure*}

\section{Efficient Computation of Refractive Noise}
\label{sec::Efficient_Computation}

Even the simplified expressions for refractive noise (e.g., Eq.~\ref{eq::sigmarref_Q}) remain numerically expensive, and an efficient approximation is necessary for our fitting framework. We now derive a suitable approximation by making two key simplifications: we approximate the ensemble-average visibility as an elliptical Gaussian, and we approximate the power spectrum $Q(\mathbf{q}) \propto q^{-(\alpha+2)}$ by a sum of exponentials. With these approximations, the (renormalized) refractive noise integrals (e.g., Eq.~\ref{eq::sigmarref_Q}) become Gaussian integrals and can be computed analytically. The first of these approximations is likely excellent for \sgra; the second can achieve any desired accuracy based on a simple prescription that we now develop.

Specifically, we use the framework of \citet{Psaltis_2018} to define the power spectrum of phase fluctuations in the scattering screen. In this framework, the power spectrum arises from a wandering magnetic field direction throughout the scattering medium; anisotropic scattering arises if the field has a preferred direction, with the major axis of the scattering orthogonal to the preferred field direction. In this framework, the power spectrum takes the form,
\begin{align}
Q (\mathbf{q}) = \bar{Q} \, (q\rin)^{-(\alpha+2)} \exp(-q^2\rin^2)\, P(\phi_q-\phi_0),
\end{align}
where $P(\phi_q-\phi_0)$ describes the (normalized) angular distribution of scattering power, and the overall normalization $\bar{Q}$ is given by,
\begin{align}
\bar{Q} =
\frac{2}{\Gamma\left(1-\frac{\alpha}{2}\right)}\,\left[\frac{\rin^2
    (D+R)}{\frac{\sqrt{2 \ln 2}}{\pi} \lambdabar_0^2 R}\right]^2\,(\theta_{\rm maj,0}^2 +
\theta_{\rm min,0}^2)\,.
\end{align}
We will use the ``dipole'' model of \citet{Psaltis_2018}:
\begin{align}
  P(\phi_q-\phi_{0,\rm PA}; k_{\zeta}) = \frac{\left[1+k_{\rm \zeta}\sin^2(\phi_q-\phi_{0,\rm PA})\right]^{-(\alpha+2)/2}}{2\pi \;_2F_1\left(1/2,1+\alpha/2;1;-k_{\rm \zeta}\right)}\;,
    \label{eq:model2}
\end{align}
where $k_{\rm \zeta}$ is determined by the asymptotic ($\lambda \rightarrow \infty$) asymmetry of the scatter-broadening:
\begin{align}
\left( \frac{\theta_{\rm maj,0}}{\theta_{\rm min,0}} \right)^2 &= 
\frac{
  \,_2F_1\left(\frac{\alpha+2}{2},\frac{1}{2};2;-k_{\rm \zeta}\right)
}{
  \,_2F_1\left(\frac{\alpha+2}{2},\frac{3}{2};2;-k_{\rm \zeta}\right)\;
  }.
\end{align}
With this model, the power spectrum can be written
\begin{align}
 Q (\mathbf{q}) = \frac{\bar{Q} \rin^{-(\alpha+2)}}{2\pi \;_2F_1\left(1/2,1+\alpha/2;1;-k_{\rm \zeta}\right)} \, \left(q^2 \left[1+k_{\rm \zeta}\sin^2(\phi_q-\phi_{0,\rm PA})\right]   \right)^{-(\alpha+2)/2} \exp(-q^2\rin^2)\, .
\end{align}

\begin{figure*}[t]
\centering
\includegraphics[width=0.32\textwidth]{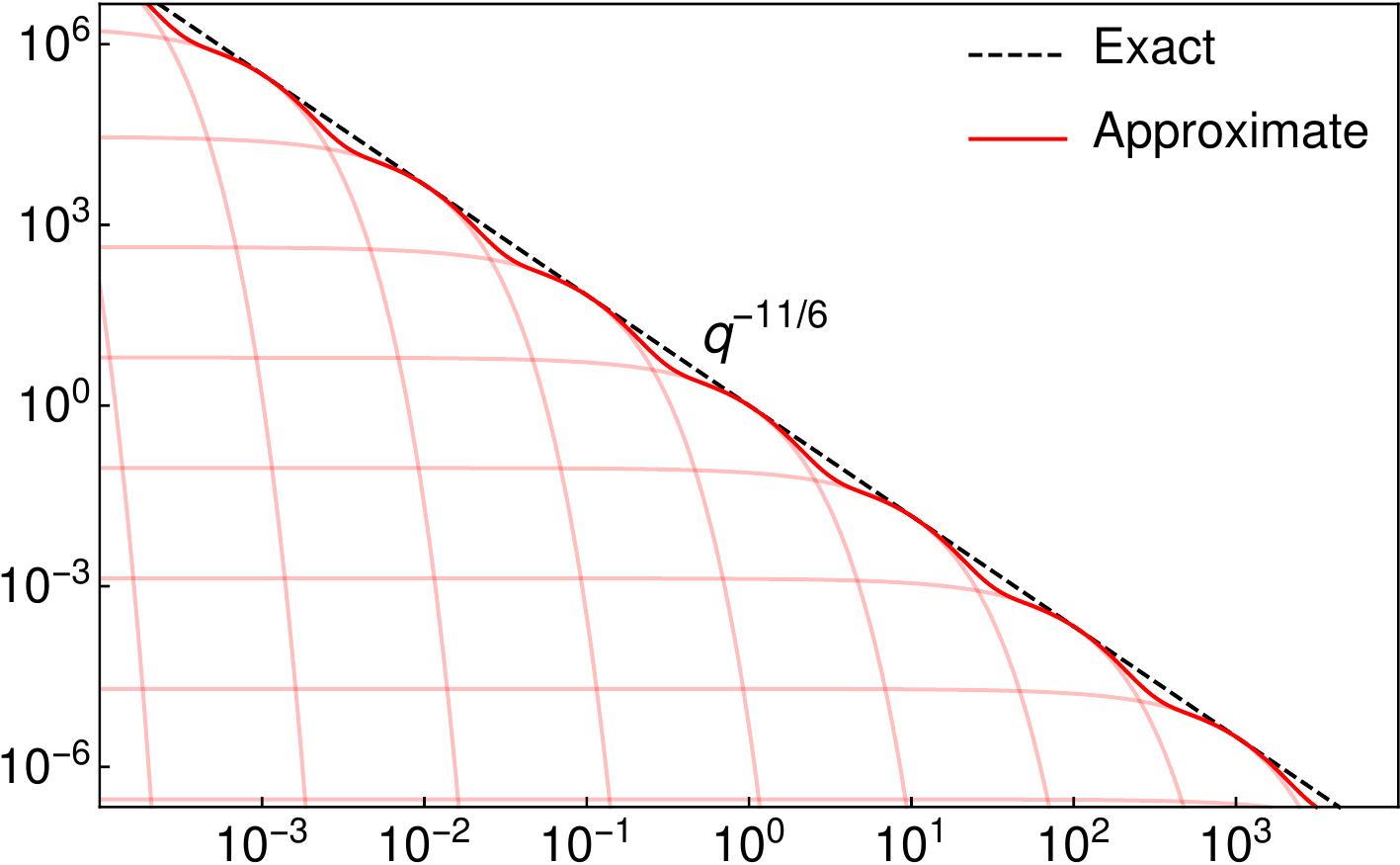}
\includegraphics[width=0.328\textwidth]{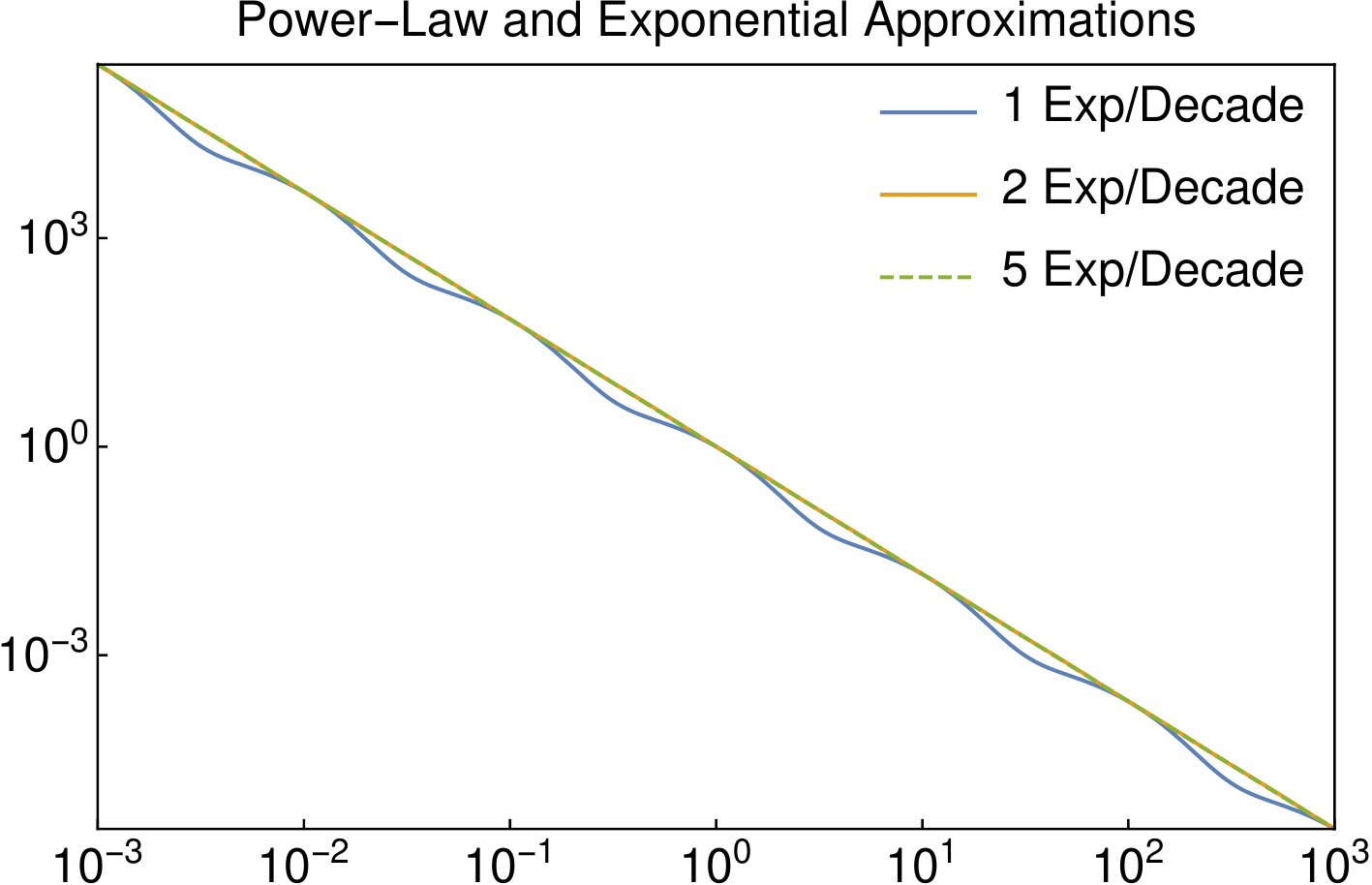}
\includegraphics[width=0.33\textwidth]{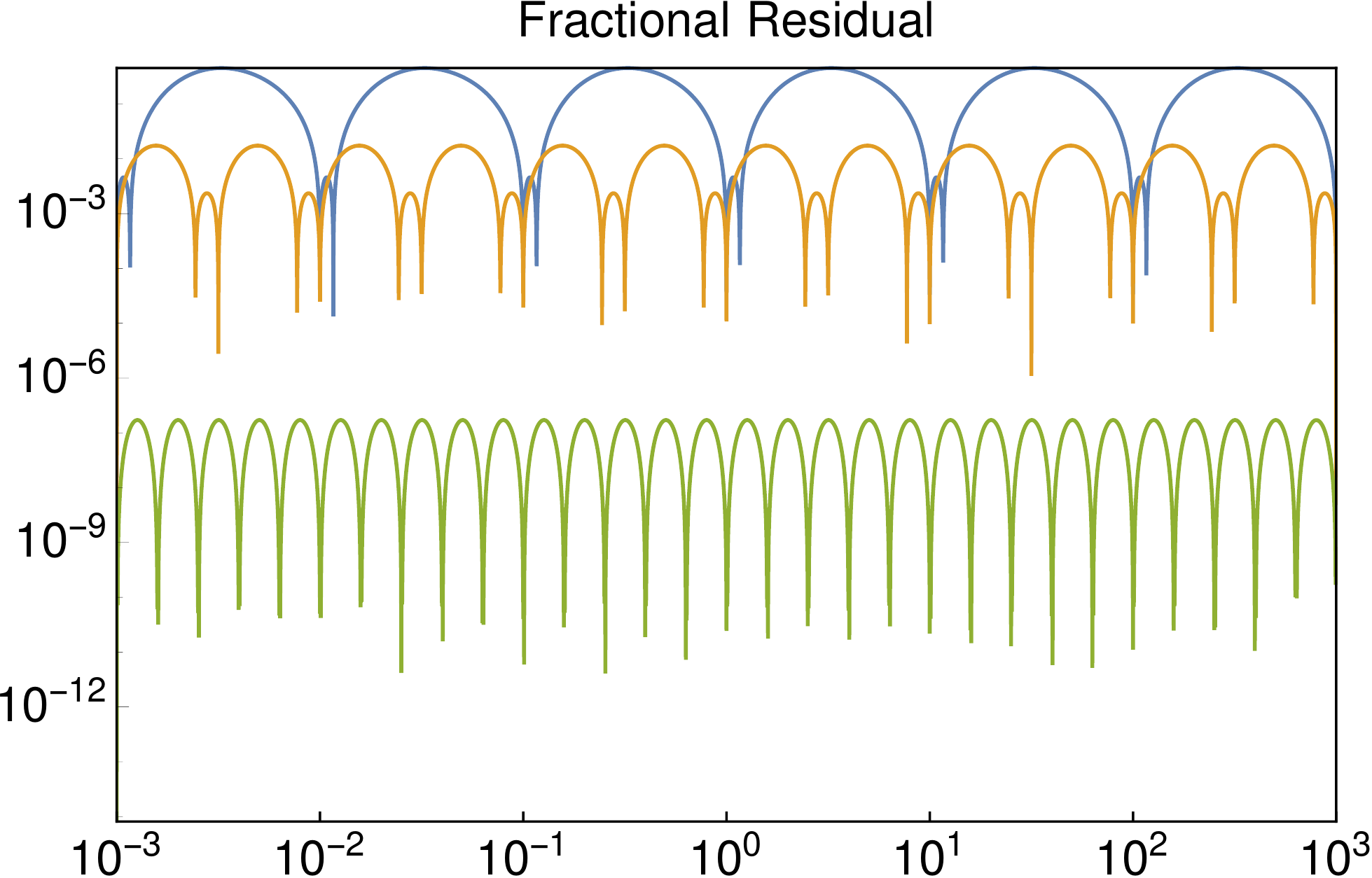}
\caption
{ 
Approximating a power-law using a sum of exponentials. In all panels, the exact power-law is $q^{-11/6}$. The left panel shows how our exponential approximation is constructed; in this example, one exponential per decade is summed. Each of these components matches the power law and its derivative at the reference point, $q_i$, and (in log-log space) each has the form of a soft step function. The exponentials are then summed and normalized at the chosen global reference point (here, at $q=1$). The center panel compares the approximations with 1, 2, and 5 exponentials per decade, and the right panel shows the fractional residual in each case. 
}
\label{fig::PowerLawApproximation}
\end{figure*}

We can now derive an approximation of $Q(\mathbf{q})$ about some point $\mathbf{q}_0$. For this, we approximate a power-law using a sum of exponentials \citep[see also, e.g.,][]{Bochud_2007}. 
Consider the function $f_{\rm pl}(q;z) = q^{-z}$. We will define an exponential basis function at a location $q_i$ as
\begin{align}
f_{\rm exp}(q;z,q_i) \equiv q_i^{-z} e^{-z \left(\frac{q}{q_i} - 1\right)}.  
\end{align}
Note that the basis function and its first derivative match the power-law at $q=q_i$: $f_{\rm exp}(q_i;z,q_i) = f_{\rm pl}(q_i;z)$ and $f_{\rm exp}'(q_i;z,q_i) = f_{\rm pl}'(q_i;z)$. To approximate $f_{\rm pl}(q;z)$, we sum exponential basis functions that are evenly spaced logarithmically over a desired range $\left\{ q_{\rm min}, q_{\rm max}\right\}$; we then normalize the result so that the approximate and exact forms match at some reference coordinate $q_{\rm ref}$. For instance, taking $q_{\rm ref} = 1$, we obtain
\begin{align}
 f_{\rm pl}(q;z) &\approx f_{\rm exp}(q;z,\{q_i\}) \equiv \frac{ \sum_{q_i} f_{\rm exp}(q;z,q_i)  }{ \sum_{q_i} f_{\rm exp}(1;z,q_i)  }\, .
\end{align}
Figure~\ref{fig::PowerLawApproximation} illustrates why this approximation is effective and shows the errors when 1, 2, and 5 exponentials are placed per decade in the range of interest. In essence, the exponential basis functions act as approximate step functions in log-log space; they are flat at values smaller than $q_i$ and quickly approach zero at larger values.

We can thus approximate the power spectrum near a point $\mathbf{q}_0$ as 
\begin{align}
\label{eq::Q_approx}
 Q_{\rm approx}(\mathbf{q};\mathbf{q}_0) &\equiv \frac{\bar{Q} \left(q_0^2 \rin^{2} \left[1+k_{\rm \zeta}\sin^2(\phi_{q_0}-\phi_{0,\rm PA})\right]   \right)^{-(\alpha+2)/2}}{2\pi \;_2F_1\left(1/2,1+\alpha/2;1;-k_{\rm \zeta}\right)} \, e^{-q^2\rin^2} f_{\rm exp}\left( \frac{q^2 \left[1+k_{\rm \zeta}\sin^2(\phi_q-\phi_{0,\rm PA})\right]}{q_0^2 \left[1+k_{\rm \zeta}\sin^2(\phi_{q_0}-\phi_{0,\rm PA})\right]},{-\frac{\alpha+2}{2}}\right)\, .
\end{align}
By normalizing the power-law in this way, the function $f_{\rm exp}$ can be referenced to a value $q_{\rm ref} = 1$, with basis functions that are independent of $\mathbf{q}$. 
Note that the argument of $f_{\rm exp}$ is quadratic in $\mathbf{q}$, and
\begin{align}
q^2 \left[1+k_{\rm \zeta}\sin^2(\phi_q-\phi_{0,\rm PA})\right] = \left( 1 + \frac{ k_{\rm \zeta} }{2} \left[ 1 - \cos\left(2\phi_{0,\rm PA}\right) \right] \right) \left( q_x^2 + q_y^2 \right) - k_{\rm \zeta} q_x q_y \sin\left(2\phi_{0,\rm PA}\right).
\end{align}
Thus, the exponential approximation for $Q(\mathbf{q})$ (Eq.~\ref{eq::Q_approx}) yields a sum of Gaussian functions in $\mathbf{q}$.

\end{appendix}

\ \\

\bibliography{SgrA_Kernel.bib}

\end{document}